\documentclass[fleqn,usenatbib]{mnras}

\usepackage{newtxtext,newtxmath}

\usepackage[T1]{fontenc}

\DeclareRobustCommand{\VAN}[3]{#2}
\let\VANthebibliography\thebibliography
\def\thebibliography{\DeclareRobustCommand{\VAN}[3]{##3}\VANthebibliography}

\usepackage{graphicx}	
\usepackage{amsmath}	

\usepackage{soul}
\usepackage{xcolor}

\def   \paul{\color{black}}

\newcommand{\teff}{T_{\rm eff}}
\newcommand{\logg}{\log g}
\newcommand{\mh}{\rm{[M/H]}}

\newcommand{\afe}{\rm{[\alpha/Fe]}}

\newcommand{\kms}{$\rm km\,s^{-1}$}

\newcommand{\vsini}{v\sin{i}}

\newcommand{\rsun}{{\rm R_{\sun}}}
\newcommand{\msun}{{\rm M_{\sun}}}

\usepackage{makecell} 
\usepackage{adjustbox}
\usepackage{tabu}

\title[\texttt{ZeeTurbo} to characterize M dwarfs with SPIRou]{Constraining atmospheric parameters and surface magnetic fields with \texttt{ZeeTurbo}: an application to SPIRou spectra}

\author[P. I. Cristofari et al.]{
P. I. Cristofari$^{1}$,\thanks{E-mail: paul.cristofari@irap.omp.eu (IRAP)}
J.-F. Donati$^{1}$,
C. P. Folsom$^{2,3}$,
T. Masseron$^{4}$,
P. Fouqu\'e$^{1,5}$,
C. Moutou$^{1}$,
\newauthor
E. Artigau$^{6}$,
A. Carmona$^{7}$,
P. Petit$^{1}$,
X. Delfosse$^{7}$,
E. Martioli$^{8,9}$,
and the SLS consortium
\\
$^{1}$Univ. de Toulouse, CNRS, IRAP, 14 av. Belin, 31400 Toulouse, France\\
$^{2}$Tartu Observatory, University of Tartu, Observatooriumi 1, Tõravere, 61602 Tartumaa, Estonia\\
$^{3}$University of Western Ontario, Department of Physics \& Astronomy, London, ON, N6A 3K7, Canada \\
$^{4}$Departamento de Astrofísica, Universidad de La Laguna, E-38206 La Laguna, Tenerife, Spain\\
$^{5}$Canada–France–Hawaii Telescope, CNRS, Kamuela, HI 96743, USA\\
$^{6}$Université de Montréal, Département de Physique, IREX, Montréal, QC, H3C 3J7, Canada \\
$^{7}$Univ. Grenoble Alpes, CNRS, IPAG, 38000 Grenoble, France
$^8$Institut d’Astrophysique de Paris, CNRS, UMR 7095, Sorbonne Université, F-75014 Paris, France \\
$^{9}$Laboratório National de Astrofíısica, Itajubá, MG 37504-364, Brazil \\
}

\date{Accepted 2023 March 16. Received 2023 March 15; in original form 2023 January 06}

\pubyear{2015}

\begin{document}
\label{firstpage}
\pagerange{\pageref{firstpage}--\pageref{lastpage}}
\maketitle

\begin{abstract}
We report first results on a method aimed at simultaneously characterising atmospheric parameters and magnetic properties of  M dwarfs from high-resolution nIR spectra recorded with SPIRou in the framework of the SPIRou Legacy Survey. Our analysis relies on fitting synthetic spectra computed from \texttt{MARCS} model atmospheres to selected spectral lines, both sensitive and insensitive to magnetic fields. We introduce a new code, \texttt{ZeeTurbo}, obtained by including the Zeeman effect and polarised radiative transfer capabilities to \texttt{Turbospectrum}. We compute a grid of synthetic spectra with \texttt{ZeeTurbo} for different magnetic field strengths and develop a process to simultaneously constrain $\teff$, $\logg$, $\mh$, $\afe$ and the  average surface magnetic flux. In this paper, we present our approach and assess its performance using simulations, before applying it to six targets observed in the context of the SPIRou Legacy Survey (SLS), namely AU~Mic, EV~Lac, AD~Leo, CN~Leo, PM~J18482+0741, and DS~Leo. Our method allows us to retrieve atmospheric parameters in good agreement with the literature, and simultaneously  yields surface magnetic fluxes in the range 2--4~kG with a typical precision of $0.05$~kG, in agreement with literature estimates, and consistent with the saturated dynamo regime in which most of these stars are.

\end{abstract}




\section{Introduction}
 
M dwarfs are known to harbour magnetic fields~\citep{saar_1985, johns-krull_1996, shulyak_2014, kochukhov_2021} and thus trigger activity that can impact the detection and characterisation of the planets they may host~\citep{hebrard_2016, dumusque_2021, bellotti_2022}. One direct consequence of magnetic fields in the stellar photosphere is the splitting of energy levels caused by the Zeeman effect, affecting the shape of spectral lines~\citep{landi_2004, reiners_2007, reiners_2012, shulyak_2014}. Some authors have estimated the surface magnetic flux of cool stars by modelling synthetic spectra including magnetic fields, and fitting them to   observed unpolarised near-infrared spectra, that are ideal for characterising the broadening impact of magnetic fields on spectral lines~\citep{valenti_1995, johns-krull_2004, shulyak_2014, lavail_2017, kochukhov_2020, reiners_2022}.

Several tools have been developed for the synthesis of magnetic stars spectra, such as COSSAM~\citep{stift_1985, stift_2003}, INVERS~\citep{piskunov_2002}, \texttt{Synmast}~\citep{kochukhov_2007}, \texttt{MOOGStokes}~\citep{deen_2013} or \texttt{Zeeman}~\citep{landstreet_1988, wade_2001, folsom_2016}. The latter, in particular, computes spectra from \texttt{MARCS} model atmospheres but does not consider molecules in the computed chemical equilibrium, which limits its application for cool stars. 
Given that \texttt{Turbospectrum}~\citep{plez_1998, plez_2012} allowed us to obtain good constraints on the stellar parameters of M dwarfs~\citep{cristofari_2022b}, we undertook to build  a new tool, called \texttt{ZeeTurbo}, by merging \texttt{Turbospectrum} and \texttt{Zeeman}, allowing us to synthesise spectra of magnetic M dwarfs.

With this paper, we report first results with an updated version of our tools to characterise M dwarfs~\citep{cristofari_2022a, cristofari_2022b} monitored with SPIRou~\citep{donati_2020}. Our goal is to provide the community with reliable constraints on the atmospheric parameters of targets observed in the context of the SPIRou Legacy Survey~\citep[SLS,][]{donati_2020}  and its follow-up program called SPICE, respectively allocated
310 and 174 nights on the 3.6-m Canada-France-Hawaii Telescope (CFHT). 
In the present work, we focus on a few very active M dwarfs already known to host strong magnetic fields (AU Mic = Gl~803, AD Leo = Gl~388, EV Lac = Gl~873, CN Leo = Gl~406, and PM~J18482+0741) thereby ideal targets for assessing the capabilities of our new atmospheric characterisation tool, and on one moderately active star (DS Leo = Gl~410), in order to confirm that our tool also performs adequately for such stars.
We use \texttt{ZeeTurbo} to compute synthetic spectra for different magnetic field strengths, in order to simultaneously constrain the atmospheric parameters and magnetic field strengths of our 6 targets.

In Sec~\ref{sec:observations} we describe the data used in this work, and introduce \texttt{ZeeTurbo} in Sec.~\ref{sec:zeeturbo}. We then discuss a revised procedure inspired by our previous work~\citep{cristofari_2022b} and assess its performance through simulations in Sec.~\ref{sec:method}, before presenting applications to SPIRou spectra in Sec.~\ref{sec:results}. In Sec.~\ref{sec:conclusions}, we discuss our results, and lay our conclusions and perspectives.  

\section{Observations and reduction}
\label{sec:observations}

In this paper, we analyse SPIRou spectra ~\citep[covering a domain of 0.95-2.5~$\mu$m at a resolving power of 70,000,][]{donati_2020} of AU Mic, AD Leo, EV Lac, DS Leo, CN Leo, and PM~J18482+0741 monitored in the context of the SLS. For these targets, spectra were acquired over 100 to 200 nights.
Data were processed through the SPIRou reduction pipeline \texttt{APERO}~\citep[version 0.7.254, ][]{cook_2022}. \texttt{APERO} provides a calibrated wavelength solution and blaze functions estimated from flat field exposure, used to correct observations. \texttt{APERO}  also performs the correction of telluric lines.

Each spectral order is normalised with a third-degree polynomial fitted on continuum points. For each star, we correct all observed spectra for the barycentric Earth radial velocity (BERV), use a cubic interpolation to bin all spectra on a common wavelength grid, and take the median of the telluric corrected spectra in the barycentric reference frame. These median spectra are referred to as templates in the rest of the paper and provide reference stellar spectra of typical signal-to-noise ratio (SNR) per 2~\kms pixel in the $H$ band reaching up to about 2000.

\section{\texttt{ZeeTurbo}, polarised radiative transfer with \texttt{Turbospectrum}}
\label{sec:zeeturbo}

\texttt{ZeeTurbo} was built directly from \texttt{Turbospectrum} and includes most of the its capabilities while solving the polarised radiative transfer equation with routines adapted or inspired from the \texttt{Zeeman} code.

\subsection{General description and functionalities}

\begin{figure}
    \centering
    \includegraphics[scale=.15]{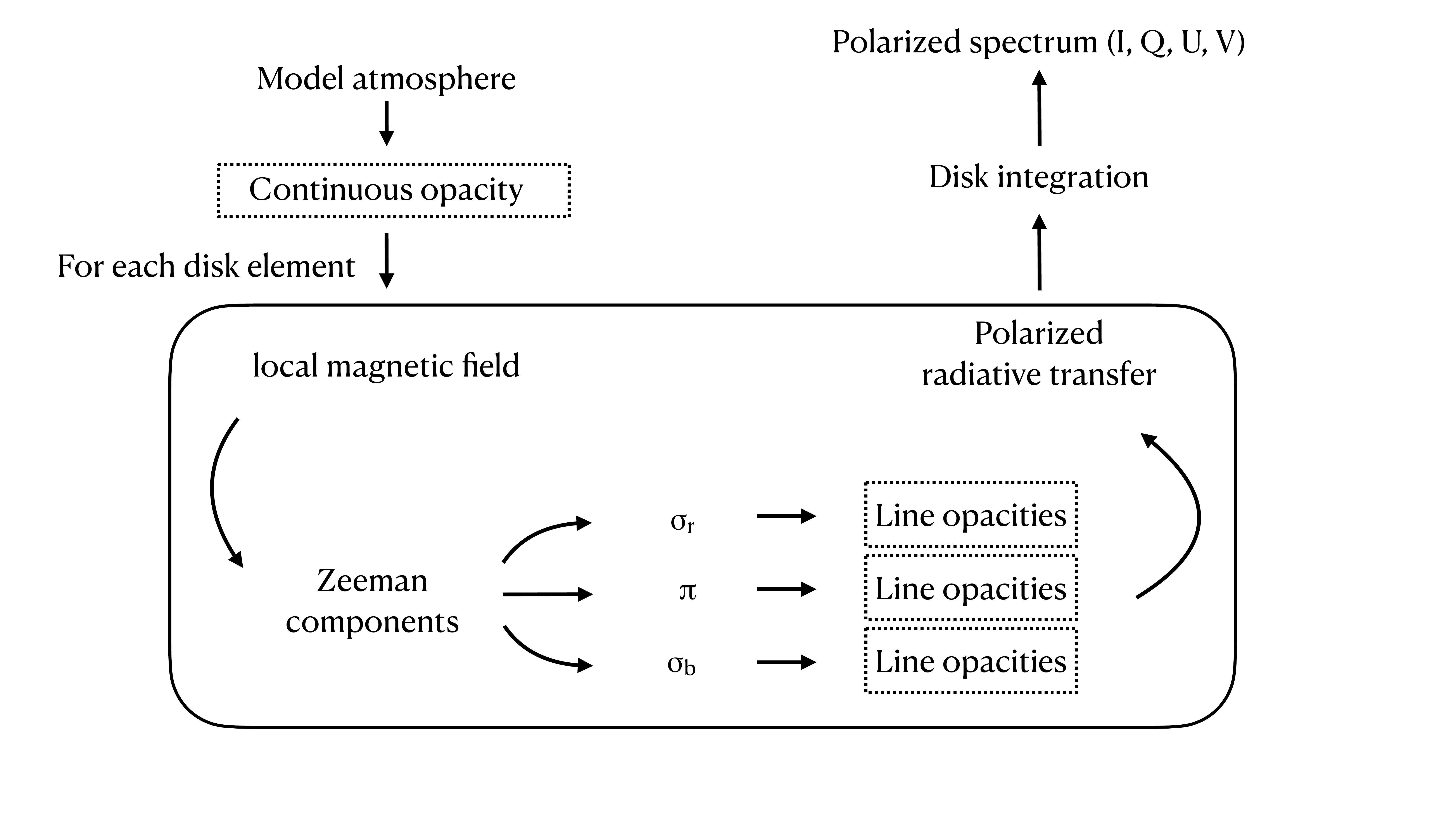}
    \caption{Schematic of the ZeeTurbo implementation.}
    \label{fig:zeeturbo_schematic}
\end{figure}

The general scheme of \texttt{ZeeTurbo} is described in Fig.~\ref{fig:zeeturbo_schematic}. For a given model atmosphere, the continuous opacities are computed by \texttt{Turbospectrum}.
The stellar disk is divided into concentric rings, each divided into cells.
For each disk element, we compute the local field strength, orientation with respect to the line-of-sight, and its projection on the line-of-sight. 
The computation of line opacities is also performed by \texttt{Turbospectrum}, but called for each $\sigma$ and $\pi$ Zeeman components, and adapted to support anomalous dispersion.
The line list format used by \texttt{ZeeTurbo} is inspired by that of  \texttt{Turbospectrum}, but also stores Land\'e factors for the lower and upper energy levels of each transition. For lines with no tabulated Land\'e factors, we compute the lower and upper Land\'e factors from the atomic structures assuming LS coupling.
The solution of the polarised radiative transfer equation is carried out by a routine adapted from that of the \texttt{Zeeman} code,  with the implementation of the quasi-analytic technique proposed by~\citet{martin_1979} and discussed in~\citet{wade_2001}. 

\texttt{ZeeTurbo} was implemented on the latest published version of \texttt{Turbospectrum}~\citep[version 20, with NLTE capabilities,][]{gerber_2022}. Most modifications of the \texttt{Turbospectrum} code where kept in separate routines and files whenever possible. Consequently, the modification to the code mostly affects the bsyn.f file of the \texttt{Turbospectrum} source code. We implemented a trigger to bypass any added feature and use the original \texttt{Turbospectrum} functions only. Currently, \texttt{ZeeTurbo} does not support NLTE computations for line list formatting reasons, but minor modifications to the code will allow us to implement this capability in the future.
  For the time being, rotation, and macroturbulence are applied as post-processing steps by convolving the spectra with rotation or macroturbulence profiles~\citep{gray_1975, gray_2005}. In this work, we focus on the analysis of Stokes $I$ spectra, although \texttt{ZeeTurbo} is also able to compute Stokes $Q$, $U$ and $V$ spectra. The analysis of polarised spectra will be treated in subsequent studies.

\subsection{Verification and validation}

\begin{figure*}
    \centering
    \includegraphics[width=\linewidth]{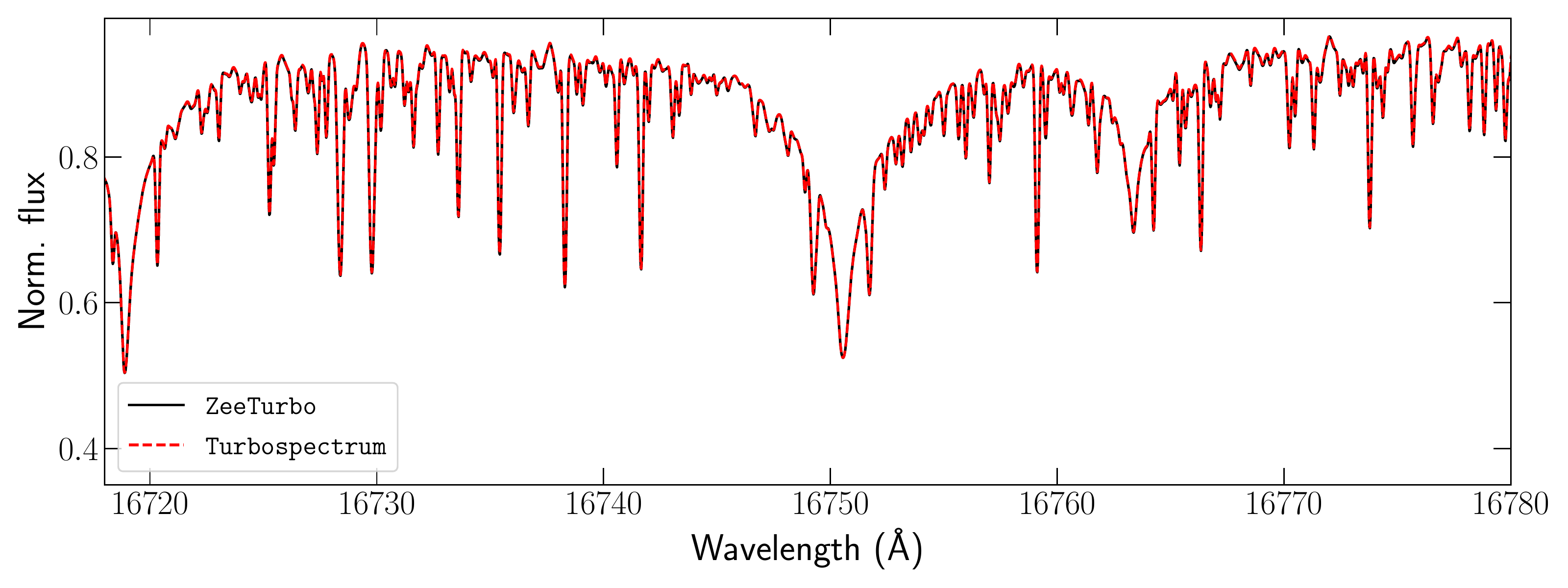}
    \includegraphics[width=\linewidth]{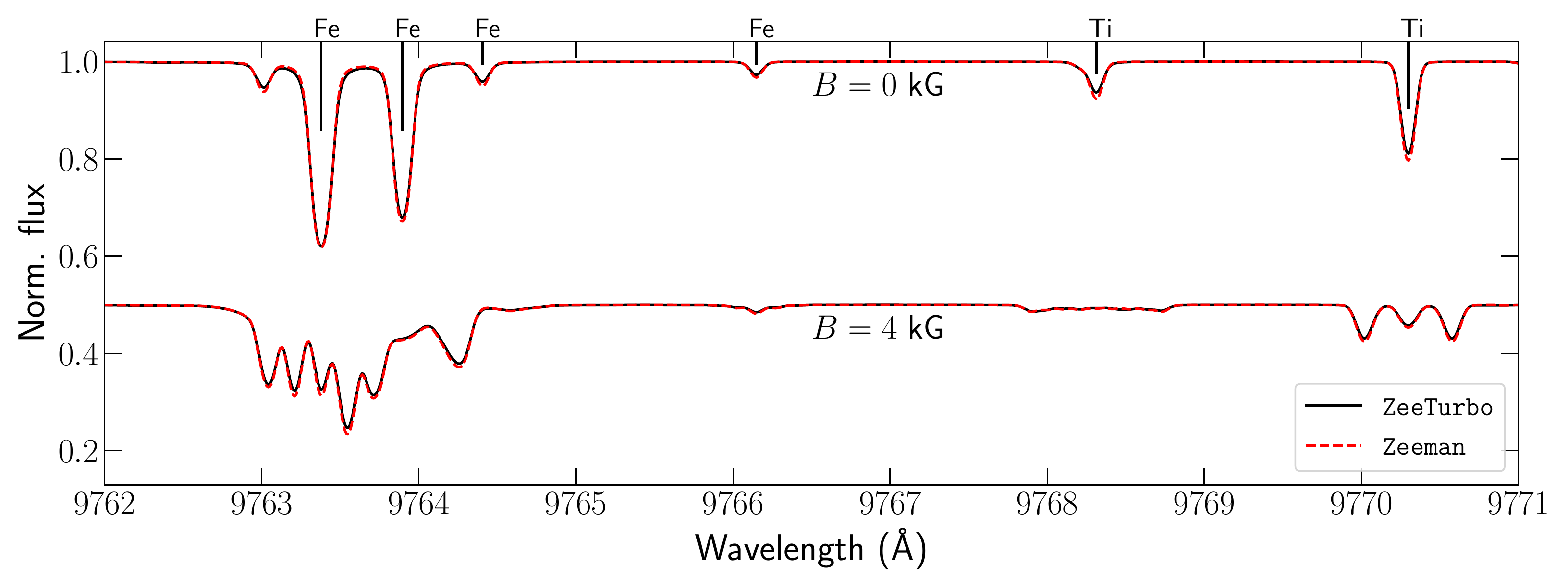}
    \caption{Top panel: comparison between spectra computed for a 0~kG field with \texttt{ZeeTurbo} and \texttt{Turbospectrum} for models with $\teff=3500$~K, $\logg=5.0$~dex, $\mh=0.0$~dex, and $\afe=0.0$~dex. Bottom panel: comparison between spectra computed with \texttt{ZeeTurbo} and \texttt{Zeeman} for $\teff=6000$~K, $\logg=5.0$~dex, $\mh=0.0$~dex, and $\afe=0.0$~dex. The spectra were computed assuming that the magnetic field is radial in all points of the photosphere.}
    \label{fig:example_comparisons}
\end{figure*}

In order to ensure that the spectra synthesised with \texttt{ZeeTurbo} are reliable, we compared them to those computed with \texttt{Turbospectrum} and \texttt{Zeeman}. We find that \texttt{ZeeTurbo} and \texttt{Turbospectrum} produce similar spectra when no magnetic field is considered.
The \texttt{Zeeman} and \texttt{Turbospectrum} codes, however, were found to produce significantly different outputs, both in the continuum levels and in the shape of spectral lines. These discrepancies are particularly obvious at temperatures lower than 3500~K.
In order to ensure that \texttt{ZeeTurbo} produces Zeeman patterns in agreement with the \texttt{Zeeman} code, we synthesised spectra at higher temperatures (e.g.\, 6000~K, see Fig.~\ref{fig:example_comparisons}) and compared the Zeeman patterns modelled by both codes. 
We found that the Zeeman patterns computed with \texttt{ZeeTurbo} are consistent with those computed with the \texttt{Zeeman} code.
Several comparisons allowed us to validate that \texttt{ZeeTurbo} behaves as expected (see Fig.~\ref{fig:example_comparisons} for an example).

\subsection{Computing a grid of synthetic spectra with \texttt{ZeeTurbo}}

We computed a new grid of synthetic spectra with \texttt{ZeeTurbo} for the analysis  of our 6 M dwarfs. The parameters covered by our grid are presented in Table~\ref{tab:parameter_space}. This grid was extended to cover lower temperatures than in our previous studies~\citep{cristofari_2022a, cristofari_2022b} in order to analyse cooler targets.
All models are computed assuming that the magnetic field is radial and of equal strength for all surface grid cells.
Our coverage in $\teff$, $\logg$, $\mh$, and $\afe$ is expected to be sufficient for most stars observed in the context of the SLS, and the step and span in magnetic field strengths are inspired from previous studies~\citep[e.g.,][]{kochukhov_2020, reiners_2022}.  

\begin{table}
    \centering
    \caption{Coverage and step size of the computed grid of \texttt{ZeeTurbo} spectra.}
    \begin{tabular}{cc}
        \hline
        $\teff$ (K) &  2700~--~4000 (100) \\
         $\logg$ (dex) & 4.0~--~5.5 (0.5) \\
         $\mh$ (dex) & $-1.0$~--~$+1.0$ (0.5) \\
         $\afe$ (dex) & $-0.25$~--~$+0.50$ (0.25) \\
         $B$ (kG) & 0~--~10 (2)  \\
         \hline
    \end{tabular}
    \label{tab:parameter_space}
\end{table}

\section{Characterising M dwarfs with \texttt{ZeeTurbo}}
\label{sec:method}

\subsection{Modelling magnetic activity -- filling factors}

Following the results of previous studies~\citep{shulyak_2010, shulyak_2014, kochukhov_2020, reiners_2022}, we choose to model the stellar spectra as a combination of spectra computed for various magnetic field strengths. 
This allows us to obtain better fits of the observed spectra by 
assuming a simple N-component model (with magnetic and non-magnetic regions at the surface of the star).
Considering the spectrum $S_X$ computed with a field of $X$~kG, the modelled spectrum $S$ is then 
\begin{equation}
    S = a_0 S_0 + a_2 S_2 + a_4 S_4 + a_6 S_6 + a_8 S_8 + a_{10} S_{10}
\end{equation} 
where $a_X$ is the filling factor for the field of $X$~kG, verifying that $a_0+a_2+a_4+a_6+a_8+a_{10}=1$ with all $a_X\geq0$.

Modelling the spectrum then amounts to finding the filling factors that lead to the best fit to our observations.

\subsection{Analysis}
\label{sec:analysis_process}

\subsubsection{Constraining atmospheric parameters and magnetic fields}
Our analysis is inspired by \citet{cristofari_2022a, cristofari_2022b}, searching for the model that provides the best fit to observations. However, unlike our previous studies, we now carry out a MCMC analysis, relying on the emcee package\footnote{https://emcee.readthedocs.io/en/stable/}~\citep{emcee} to estimate the atmospheric parameters from posterior distributions.
Prior to their comparison to observations, synthetic spectra are broadened to account for instrumental resolution, macroturbulence and rotation, and shifted to match the observed radial velocity. We then perform an adjustment of the continuum following the steps described in~\citet{cristofari_2022b}. Fixing both the instrumental width and either the macroturbulence or the rotation velocity to their known values, we end up with 11 parameters to be estimated with our MCMC process. Macroturbulent velocity and rotation are typically difficult to disentangle due to the similar effect they have on spectral lines~\citep[see e.g.][]{gray_2005, valenti_2005}. In the present paper, we chose to set the value of $\vsini$ and fit the macroturbulent velocity. We found that fixing macroturbulence and fitting the rotational velocity lead to very similar results in atmospheric parameters and magnetic field strengths.

Priors set on atmospheric parameters are meant to prevent walkers to run outside the boundaries of our grid. Priors are also set to ensure that the filling factors remain positive.
To ensure that the sum of the filling factors is one, we compute $a_0 = 1-(a_2+a_4+a_6+a_8+a_{10})$.
For each walker, if one of the filling factors differs from 1, or if the atmospheric parameters fall out of the grid, we set the likelihood value to -infinity.

\subsubsection{Deriving error bars}

Error bars on atmospheric parameters and filling factors are estimated from posterior distributions. In practice, we find that the minimum reduced $\chi^2$ ($\chi^2_{\rm r, min}$) derived from fitting the observed spectrum is larger than 1, because of systematic differences between the model and observations, which impacts the results of our MCMC analysis. In order to overcome the issue, we artificially expand the error bars on each pixel by $\sqrt{\chi^2_{\rm r, min}}$ before running our analysis, to ensure that the best fit corresponds to a unit $\chi^2_{\rm r}$. The factor used to expand the error bars is estimated after a preliminary run.

In our previous work~\citep{cristofari_2022a}, formal error bars on atmospheric parameters were found to be smaller than the dispersion between parameters derived using different grids of synthetic spectra. Following ~\citet{cristofari_2022a, cristofari_2022b} we chose to further enlarge these errors again by quadratically adding to our formal error bars 30~K for $\teff$, 0.05~dex for $\logg$, 0.10~dex for $\mh$ and 0.04~dex in $\afe$. The resulting error bars are referred to as `empirical error bars' in the rest of the paper.

\subsection{Line list}

For this analysis, we start from the same atomic and molecular line list used in~\citet{cristofari_2022b}, adding several Ti, K, and Mg lines included in previous studies~\citep{kochukhov_2020, reiners_2022}, and shown to be useful for estimating magnetic fields.
Atomic data, including Land\'e factors, were extracted from the VALD database~\citep{piskunov_1995, kupta_2000, ryabchikova_2015, pakhomov_2019}. For a few Ti lines, corrections to the Van Der Waals parameters were applied following~\citet{cristofari_2022b}.
Data for molecular lines was compiled from~\citet{burrows_2002},~\citet{barber_2006},~\citet{yadin_2012},~\citet{sneden_2014},~\citet{masseron_2014},~\citet{brooke_2016}, the ExoMol database~\citep{yadin_2012,barton_2013,yurchenko_2018, exomol_2020}, and the HITRAN database~\citep{rothman_2013, gordon_2017}.
Our line list also contains a number of OH and CO lines, assumed to be insensitive to magnetic fields. This assumption was supported by comparing the spectra of weakly and strongly magnetic targets, as well as by the results of previous studies~\citep[e.g.][]{lopez-valdivia_2021}.
The lines used in the present analysis are listed in Table~\ref{tab:selected_lines}.

\begin{table}
	\center
	\caption{List of lines used for our analysis. The identification relied on wavelengths and depths extracted from the VALD database. 
	Effective Land\'e factors are given for atomic lines. 
	 For each atomic line in the table, if two values of effective Landé factors are given, the first is that listed in the VALD database, and the second was computed assuming LS coupling. When effective Landé factors were missing from the VALD database, we display a single value computed assuming LS coupling.}
	\begin{tabular}[h]{cc}
	\hline
 		Species & Wavelength (Å) [effective Land\'e factor]\\
 		\hline
Ti I & 9678.20 [1.35~--~1.35], 
9708.33 [1.26~--~1.25], 9721.63 [0.95~--~1.00], \\
& 9731.07 [1.00~--~1.00], 9746.28 [0.00~--~0.00], 9785.99 [1.48~--~1.50], \\
& 9786.27 [1.49~--~1.50], 9790.37 [1.50~--~1.50], 22217.28 [2.08~--~2.00], \\
& 22238.91 [1.66~--~1.67], 22280.09 [1.58~--~1.58], \\
& 22316.70 [2.50~--~2.50], 22969.60 [1.11~--~1.10], \\
Fe I  & 10343.72 [0.68~--~0.67], \\
Mg I & 10968.42 [1.33], 15044.36 [1.75], 15051.83 [2.00], \\
K I & 12435.67 [1.33], 12525.56 [1.17], 15167.21 [1.07~--~1.07], \\
Mn I  & 12979.46 [1.21~--~1.21], \\
Al I  & 13127.00 [1.17], 16723.52 [0.83], 16755.14 [1.10], \\
Na I & 22062.42 [1.17], 22089.69 [1.33], \\
OH & 16073.91, 16539.10, 16708.92, 16712.08, \\
& 16753.83, 16756.30, 16907.35, 16908.89, \\
& 16910.25, \\
CO & 22935.23, 22935.29, 22935.58, 22935.75, \\
& 22936.34, 22936.63, 22937.51, 22937.90, \\
& 22939.09, 22939.58, 22941.09, 22943.49, \\
& 22944.16, 22946.31, 22947.06, 22949.54, \\
& 22950.36, 22953.19, 22954.06, 22957.26, \\
& 22958.16, 22961.74, 22962.67, 22966.65, \\
& 22967.58, 22971.97, 22972.88, 22977.72, \\
& 22978.60, 22983.89, 22984.71, 22990.49 \\
		\hline
\end{tabular}
\label{tab:selected_lines}
\end{table}

\subsection{Benchmarking \texttt{ZeeTurbo}}

\subsubsection{Building model templates}
We ran a benchmark, to ensure that our new tool is indeed capable of constraining atmospheric parameters and filling factors. To this end, we generated a set of model template spectra as follows. From a set of atmospheric parameters and filling factors, we computed a synthetic spectrum. We then broadened this spectrum with a Gaussian profile of full width at half maximum (FWHM) 4.3~\kms{} to account for the instrumental width of SPIRou, and optionally applied convolution with rotation and/or macroturbulence profiles. The spectrum was then convolved with a 2~\kms-wide rectangular function representing pixels, and re-sampled on a typical SPIRou wavelength solution. Noise was added to the spectrum, accounting for the typical SPIRou throughput~\citep{donati_2020} and the typical blaze function for a SPIRou observation. The modelled spectrum, therefore, resembles template spectra in that the noise varies throughout each order, and from order to order.

\subsubsection{Simulating the estimation of the atmospheric parameters and filling factors.}
\begin{figure}
    \centering
    \includegraphics[scale=.5]{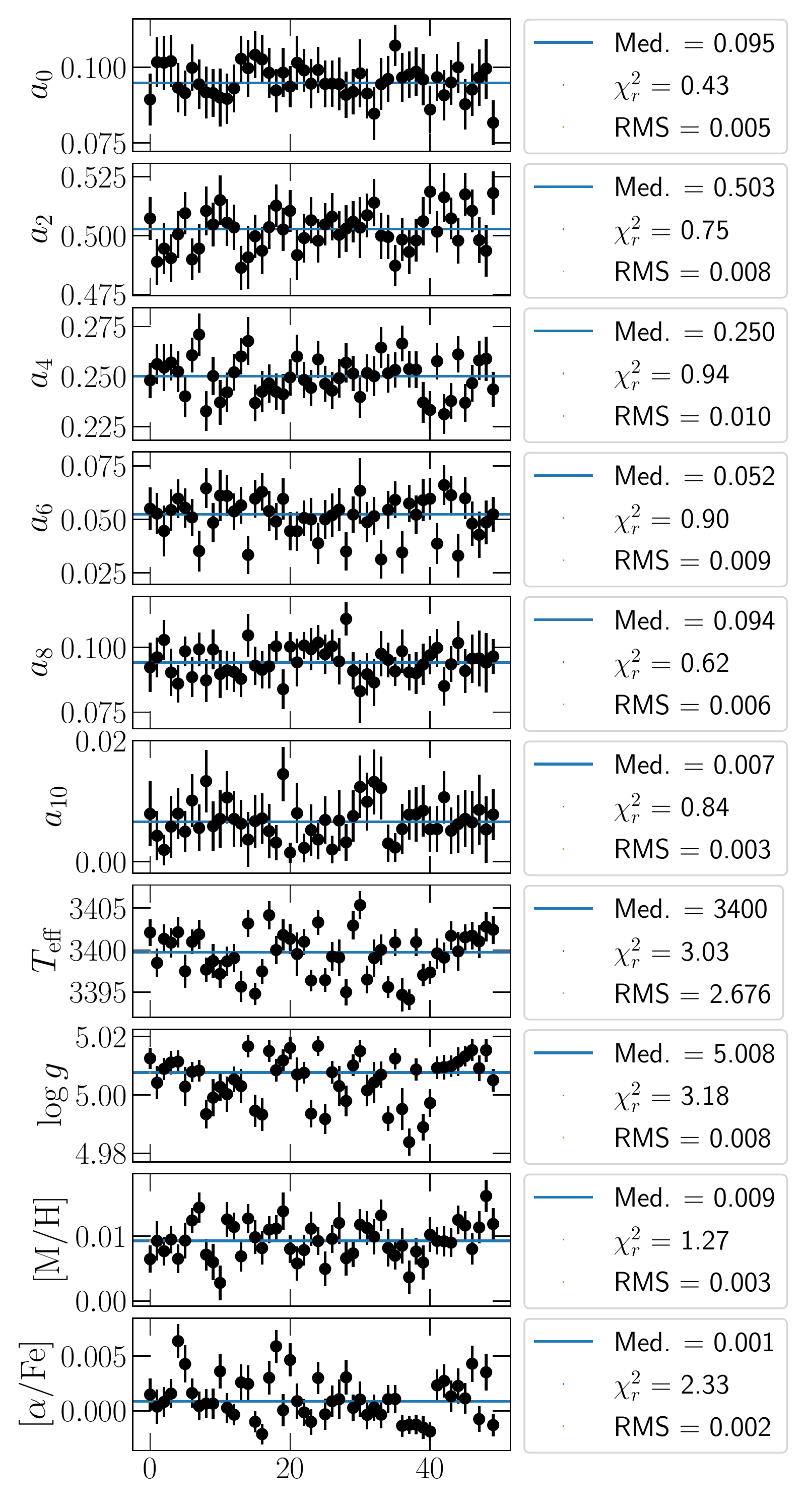}
    \caption{Example of comparison between input and output parameters. The blue horizontal solid line marks the median. For each parameter, we compute the $\chi^2$ of the series of points. The models were computed for $\teff=3400$~K, $\logg=5.0$~dex, $\mh=0.00$~dex and $\afe=0.00$~dex.}
    \label{fig:simu_results}
\end{figure}

We performed our analysis on 50 modelled templates computed for the same atmospheric parameters and filling factors but different noise realisations with the process described in Sec.~\ref{sec:analysis_process}. The modelled templates were computed assuming an SNR in the H band of $\sim$500, $\teff=3400$~K, $\logg=5.0$~dex, $\mh=0.0$~dex and $\afe=0.0$~dex. We set the filling factors of the models to $a_0$=0.10, $a_2$=0.50, $a_4$=0.25, $a_6$=0.05, $a_8$=0.10 and $a_{10}=0.00$ ( yielding an average magnetic field strength $<\!B\!>=3.1$~kG), thus adopting values consistent with typically observed targets (see Sec.~\ref{sec:results}). We simultaneously constrained atmospheric parameters and filling factors and analysed posterior distributions to find out potential correlations and estimate uncertainties. 
Figure~\ref{fig:simu_results} presents the results of our benchmark.
We find that the dispersion on the series of 50 points is not fully consistent with our formal error bars,  especially for the atmospheric parameters $\teff$, $\logg$, $\mh$ and $\afe$,
the reduced $\chi^2$ ($\chi^2_{\rm r}$) on the residuals (the retrieved parameters minus the median) reaching up to $3.2$. Subsequent tests showed that most of this excess dispersion can be attributed to the continuum adjustment step.
We also find that the effect of the continuum adjustment is sensitive to the SNR, and can introduce systematic offsets in the retrieved atmospheric parameters of up to 0.01~dex in $\logg$ or $\mh$ and up to 0.5~K in $\teff$ with a SNR~$\sim$500. These shifts reach up to 20~K in $\teff$ and 0.1~dex in $\logg$, $\mh$ and $\afe$ if we assume a SNR~$\sim$100. In practice, the SPIRou templates usually reach an SNR in the $H$ band of $\sim$2000, implying that our results should not be affected by such biases.

\begin{figure}
    \centering
    \includegraphics[scale=.5]{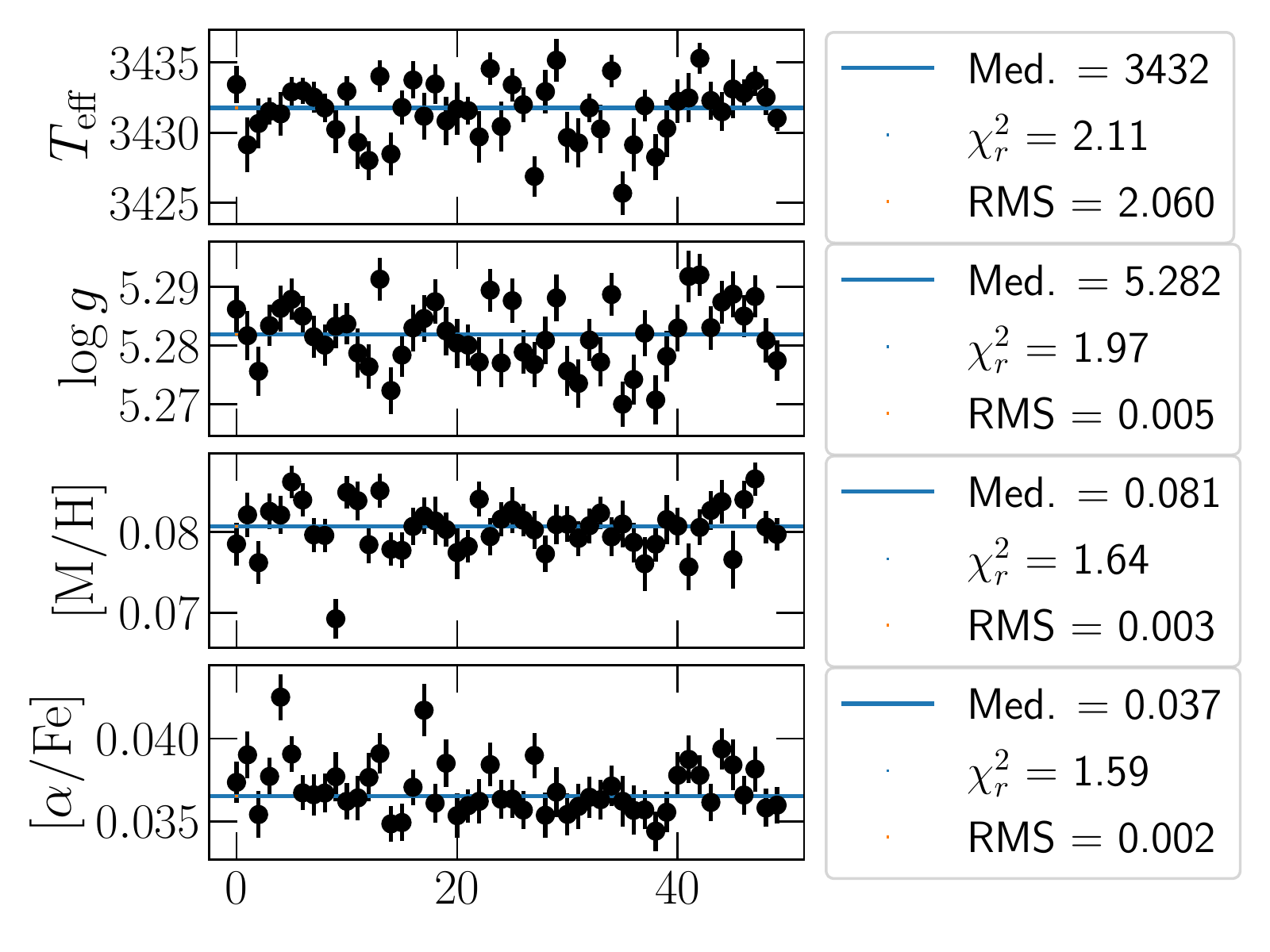}
    \caption{Same as Fig.~\ref{fig:simu_results} but with no magnetic field considered for the analysis.}
    \label{fig:simulations_nofield}
\end{figure}

With our benchmark, we explored the impact of magnetic fields on the estimation of atmospheric parameters. We generated templates for magnetic stars, and ran our analysis with non-magnetic models. The recovered atmospheric parameters deviate from the input parameters by up to 30~K in $\teff$ and 0.3~dex $\logg$ (see Fig.~\ref{fig:simulations_nofield}) for this particular magnetic configuration. Smaller  biases ($<0.1$~dex) are found on $\mh$ and $\afe$. These systematic shifts can be $\sim$10 times larger than our formal error bars for large values of the magnetic flux.\\

\begin{figure}
    \centering
    \includegraphics[width=\columnwidth]{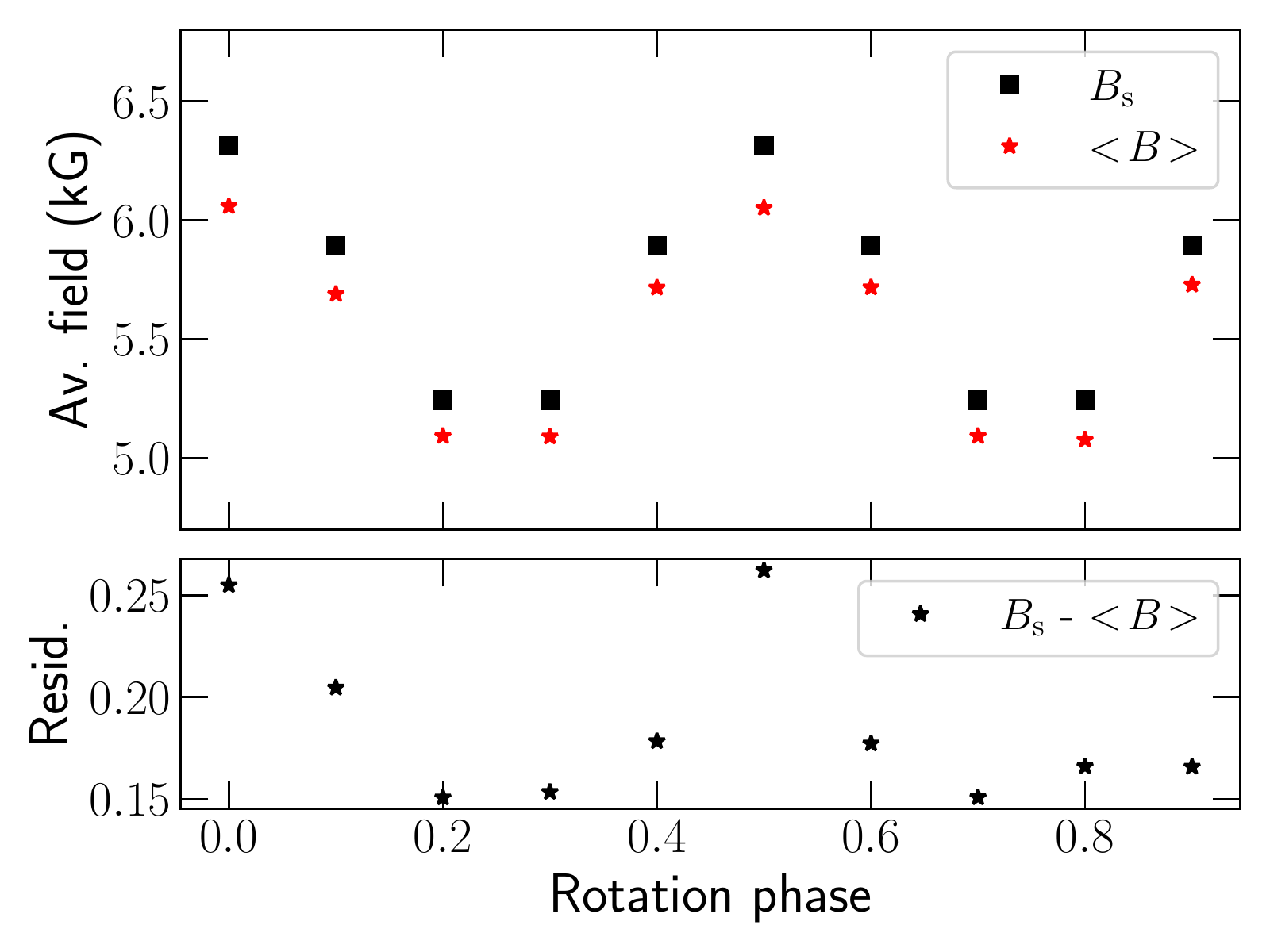}
    \caption{{\paul Magnetic field strengths for a rotating star hosting a 8~kG dipole whose axis is perpendicular to the rotation axis. For each rotation phase, we compare the average magnetic strength recovered with our analysis ($<\!B\!>$, red stars) to the true magnetic field averaged over the visible hemisphere of the star ($B_{\rm s}$, black squares). The bottom panel shows the residuals.}}
    \label{fig:rel_mag_phase}
\end{figure}

{\paul 
\subsubsection{Estimating field strengths from known magnetic configurations}
\label{sec:simu_dipole}

We carried out additional simulations to assess the precision at which field strengths are recovered given the a priori assumptions of our model, in particular on the field topology; we achieve this by running our tool on synthetic spectra of a star with a known magnetic configuration.

We consider a star hosting an 8~kG dipolar magnetic field inclined at 90$^{\circ}$ with respect to the rotation axis, for a star viewed equator on.
We computed synthetic spectra for 10 evenly spaced rotation phases. We added noise to each spectra and ran our analysis at each phase. We then compared the retrieved average magnetic field strengths (${<\!B\!>}$) to the true field strengths averaged over the visible hemisphere of the star ($B_{\rm s}$, see Fig.~\ref{fig:rel_mag_phase}).
We find that $<\!B\!>$ is in good agreement with $B_{\rm s}$, though slightly smaller by about 3-4\% (i.e. 0.15-0.25~kG). This slight difference comes from our modelling assumption that the field is radial over the whole surface.

We also performed our analysis on a spectrum obtained by taking the median of the spectra at all phases. The average magnetic field obtained with the median spectrum is 5.5~kG, consistent with the median of the retrieved $<\!B\!>$ values.

Altogether, it demonstrates that our modeling assumptions are quite reasonable and introduce only marginal biases in the measured field strengths.

}

\section{Application to SPIRou spectra}
\label{sec:results}

We applied our new tool to our template SPIRou spectra of AU~Mic, AD~Leo, EV~Lac, DS~Leo, CN Leo and PM~J18482+0741, relying on models computed for magnetic fields ranging from 0 to 10~kG in steps of 2~kG.

\begin{figure*}
	\includegraphics[width=\linewidth]{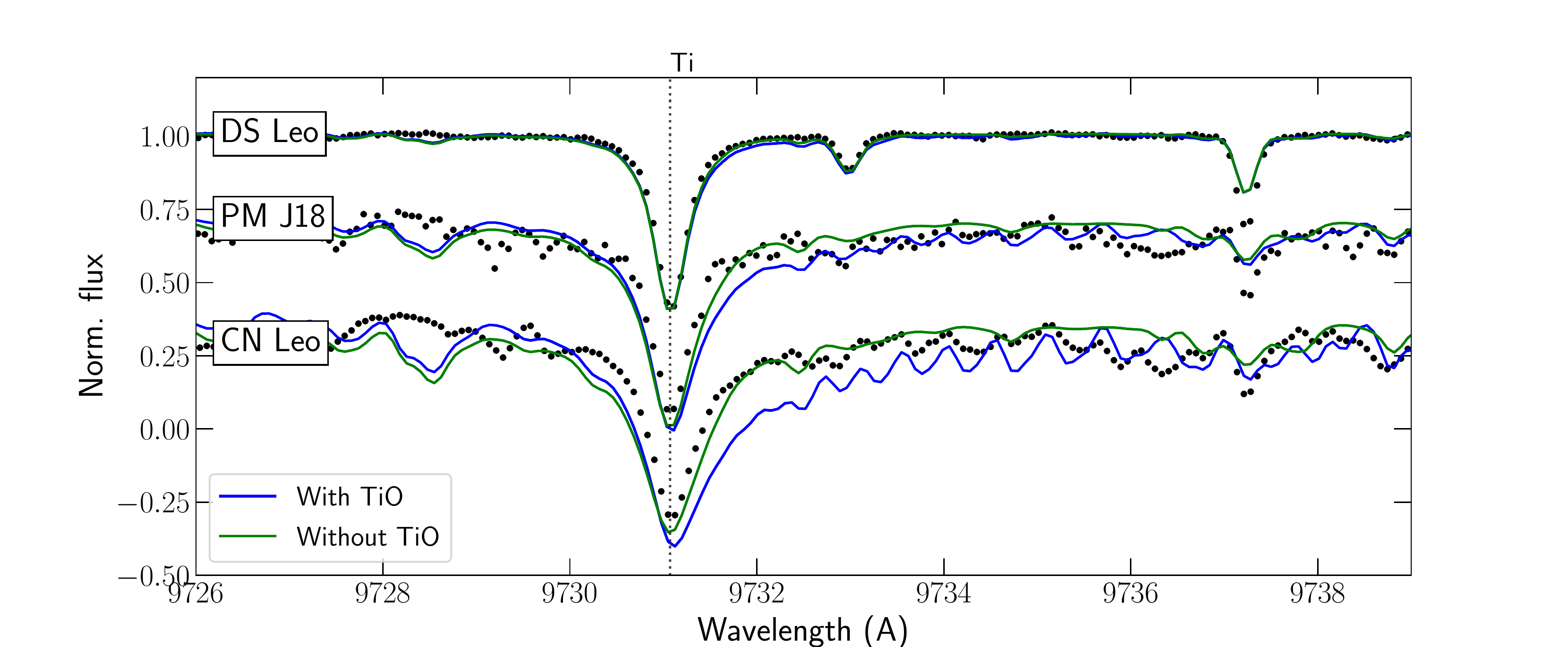}
	\caption{Best obtained fit with synthetic spectra computed with and without TiO for the two coolest stars in our sample (CN Leo and PM~J18482+0741) and DS~Leo. The black points show the SPIRou templates.} The label of PM~J18482+0741 was abbreviated PM~J18 for better readability.
	\label{fig:discrepent_line}
\end{figure*}

For the coolest targets in our sample (CN Leo and PM~J18482+0741), we found discrepancies between the best-fitted model and the SPIRou template for some lines, such as the Ti line at 9678~\r{A} (see Fig.~\ref{fig:discrepent_line}). We worked out that the presence of spurious TiO  lines in the synthetic spectra were responsible for some of these discrepancies, and that removing this molecule from the spectral synthesis improved the fit quality for the coolest stars in our sample. 
The results presented in this section were obtained with synthetic spectra computed without TiO, after checking that very similar results (and worse fits) were obtained when keeping TiO in.

\begin{table*}
	\caption{Literature parameters for the stars in our sample. For all stars but AU~Mic, masses were obtained from the mass-$K$ band magnitude proposed by~\citet{mann_2019}, and radii were computed from the mass with the models of~\citet{baraffe_2015} assuming an age of 5 Gyr for all stars but AU~Mic. $K$ band magnitudes were extracted from the SIMBAD database~\citep{wenger_2000}. For each star, we report $\vsini$ estimates from the literature, or our adopted values if rotation period and radius estimates suggested that the literature $\vsini$ was overestimated. The convective turnover time ($\tau$) is taken from~\citet{reiners_2022}. The Rossby number ($R_{\rm O}$) is computed from the rotation period ($P_{\rm rot}$) and $\tau$. Ref. -- 
		\textit{a}:~\citet{plavchan_2020}, 
			\textit{b}:~Donati et al., (submitted), 
		\textit{c}:~\citet{gallenne_2022},
		\textit{d}:~\citet{morin_2008}, 
		\textit{e}:~\citet{diez_alonso_2019}, 
		\textit{f}:~\citet{donati_2008}
		\textit{g}:~\citet{reiners_2018}
		\textit{h}:~\citet{reiners_2007}.}
	\label{tab:literature_params}
	\begin{tabular}{ccccccccc}
		\hline
		Star & GJ ID & Spectral type & $P_{\rm rot}$  (d) & $M$ ($M_{\sun}$) & $R$ ($R_{\sun}$)  & $\vsini$ (\kms{}) & $\tau$ (d) & $R_{\rm O}$ \\ 
		\hline
		AU Mic & Gl 803 & M1V & $4.863\pm0.010^{\textit{a}}$ & $0.60\pm0.03^{\textit{b}}$ & \makecell{$0.82\pm0.05^{\textit{b	}}$  } & $8.5\pm1.0^{\textit{b}}$  & 39 & $0.125$ \\
		EV Lac & Gl 873 &  M4.0V & $4.3715\pm0.0006^{\textit{d}}$ & $0.32 \pm0.02$ & $0.31\pm0.02$ & $3\pm1$ & 110 & $0.040$ \\
		AD Leo & Gl 388 & M3V & $2.2399\pm0.0006^{\textit{d}}$ & $0.42\pm0.02$ & $0.39\pm0.02$ & $3\pm1^\textit{d}$& 80 & 0.028 \\ 
		CN Leo & Gl 406 & M6V & $2.704\pm0.003^{\textit{e}}$ &  $0.11\pm0.02$  & $0.13\pm0.02$ & $2\pm1$ & 387 & 0.007 \\
		PM J18482+0741 & -- &  M5.0V &  $2.76\pm0.01^{\textit{e}}$ & $0.14\pm0.02$ & $0.17\pm0.02$ & $2.4\pm1.5^\textit{g}$ &  230 & 0.012 \\  
		DS Leo & Gl 410 & M1.0V & $14.0\pm0.1^\textit{f}$ & $0.57\pm0.02$ & $0.53\pm0.02$ & $1.5\pm1.0$ & 60 & 0.233 \\
		\hline
	\end{tabular}
\end{table*}

\subsection{AU Mic = Gl~803}
The young planetary system AU Mic attracted significant attention in the recent years~\citep{boccaletti_2018, kochukhov_2020, martioli_2020, martioli_2021, klein_2021, klein_2022} and has been monitored by several instruments.
\begin{figure*}
    \centering
    \includegraphics[scale=.27]{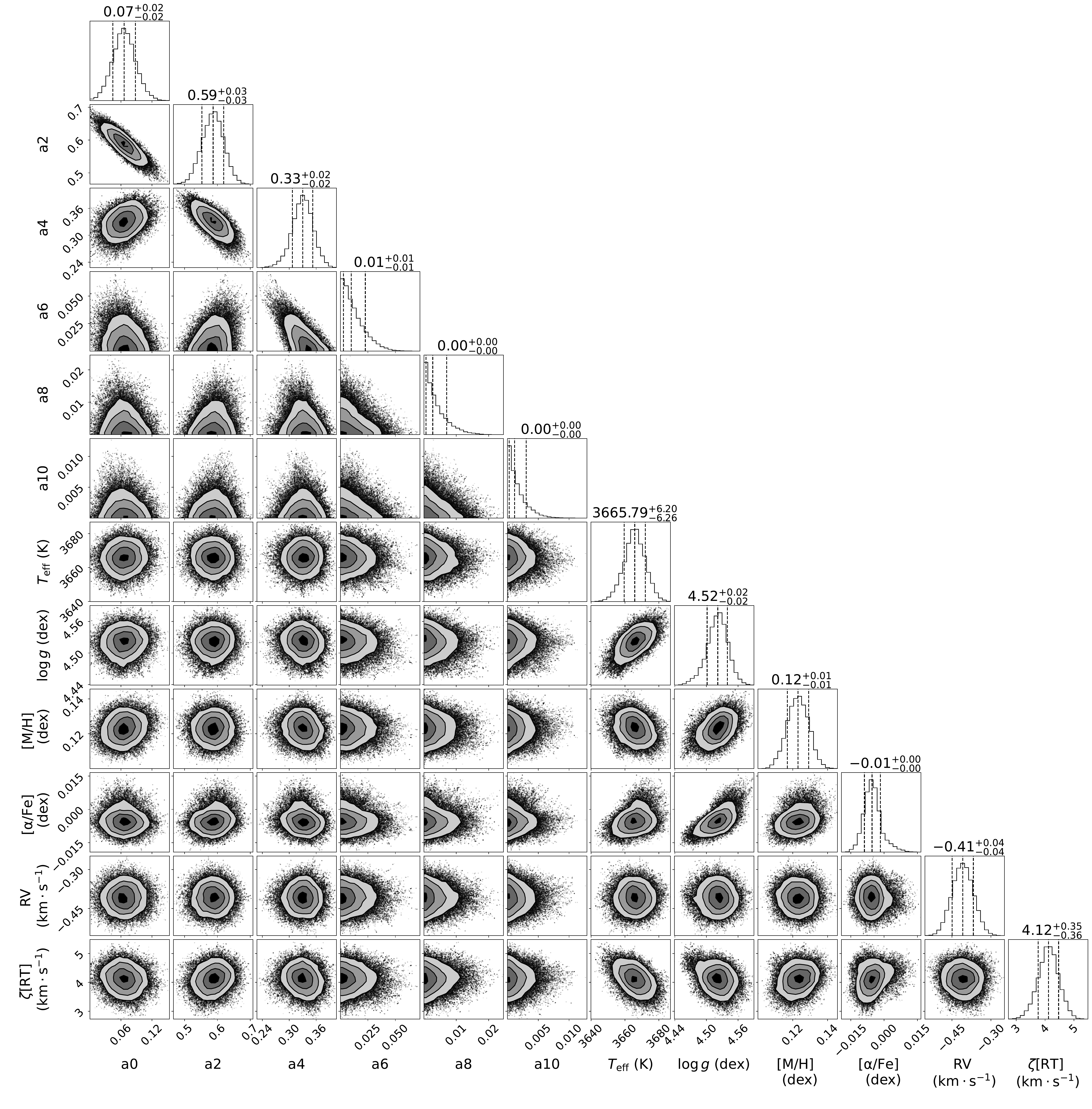}
    \caption{Corner plot presenting posterior distributions for filling factors and atmospheric parameters for AU Mic.}
    \label{fig:aumic_corner_full}
\end{figure*}
The rotation period of this star is ${P_{\rm rot}=4.863\pm0.010}$~d~\citep{plavchan_2020, klein_2021} with an angle between the line of sight and the rotation axis close to $90^{\circ}$, and its radius was estimated from interferometric measurements to $0.862\pm0.052~\rsun$~\citep{gallenne_2022}. For this star, we adopt a $\vsini=8.5\pm1.0$~\kms{}~(Donati et al., in prep), yielding a radius of $0.82~\rsun$. With a mass estimated at $0.60\pm0.03~\msun$~(Donati et al., in prep), the logarithmic surface gravity of AU~Mic is then equal to $\logg=4.39$~dex.

We performed an analysis of AU Mic fitting $\zeta_{\rm RT}$ and fixing $\vsini=8.5$~\kms{} (see Table~\ref{tab:results}).
From posterior distributions, {\paul we estimate a $\teff=3665\pm31$~K, $\logg=4.52\pm0.05$~dex, $\mh=0.12\pm0.10$~dex and $\afe=0.00\pm0.04$~dex.}
These estimates are listed Table~\ref{tab:results}.
The temperature is consistent with that estimated from SEDs~\citep{alfram_2019}.
Our $\logg$ is significantly larger than that estimated from mass and radius.
We attempted to perform the analysis by fixing the value of $\logg$ to 4.40~dex. In that case, we retrieve a $\teff=3641\pm31$~K and $\mh=0.06\pm0.10$~dex, and $\zeta_{\rm RT}=5.4\pm0.2$~\kms{}, still consistent with literature values.

\begin{figure*}
    \centering
    \includegraphics[scale=.55]{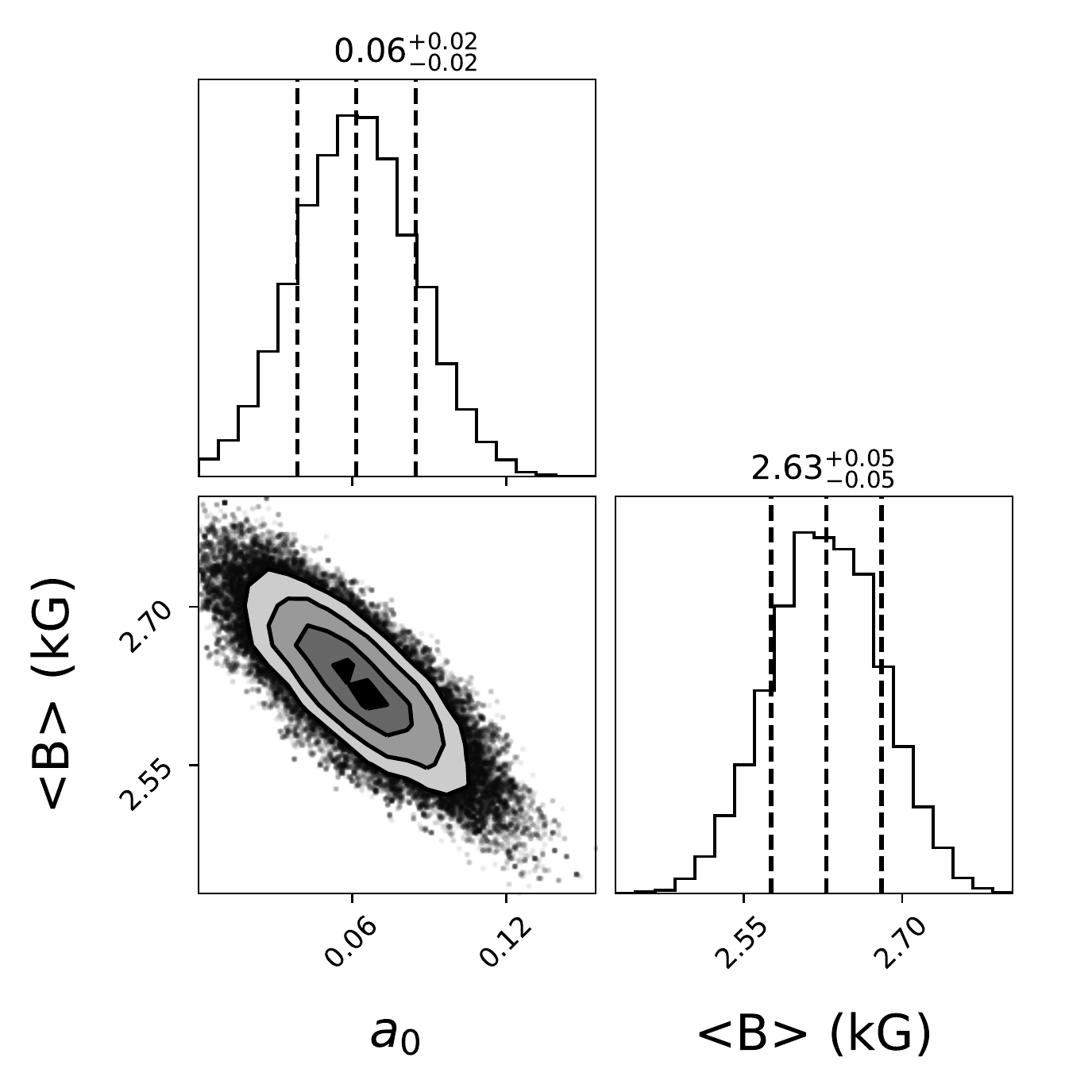}\includegraphics[scale=.5]{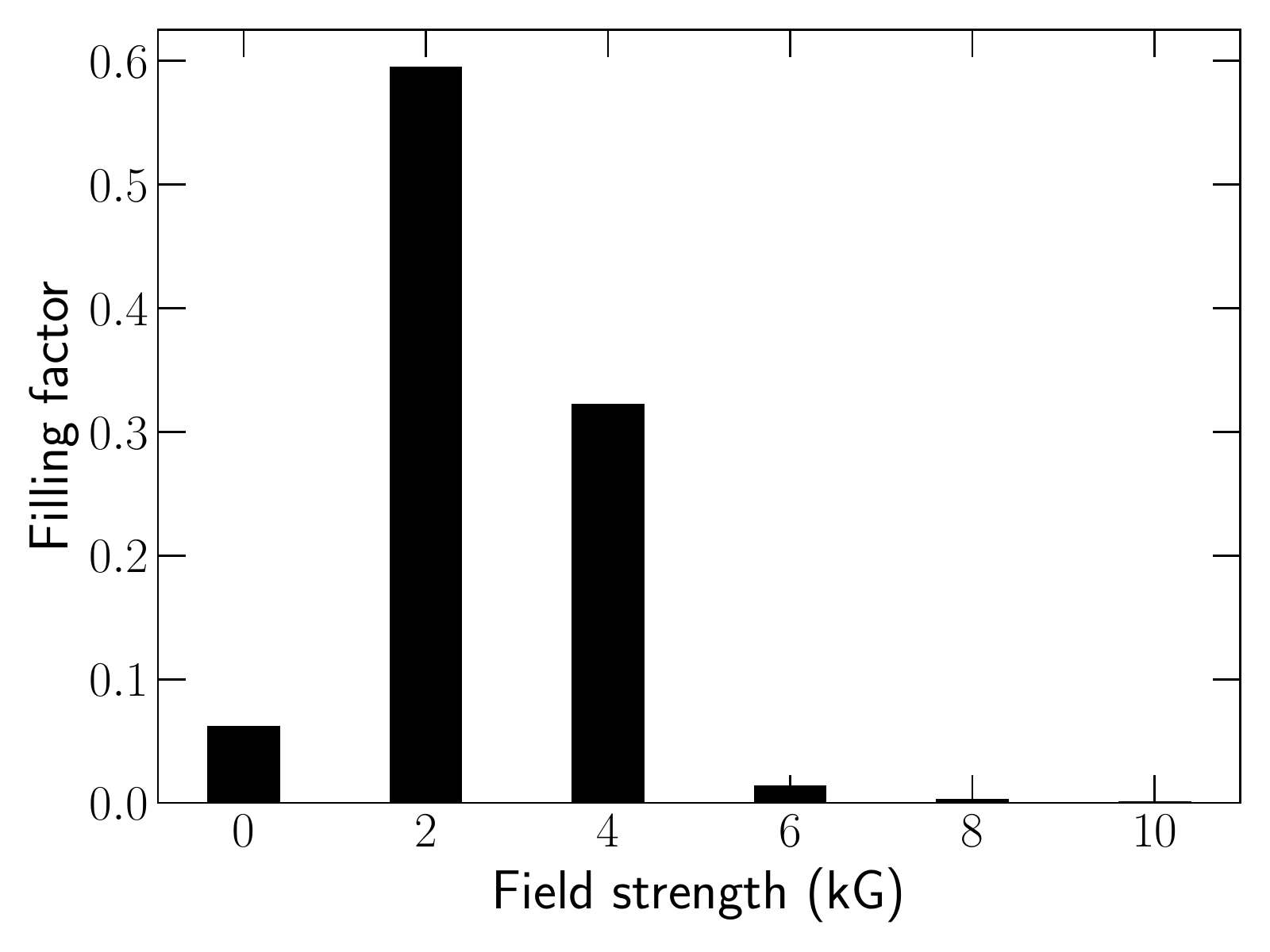}
    \caption{Left panel: non-magnetic component against average field strength for AU Mic. Right panel: distribution of the filling factors over magnetic field components for AU Mic.}
    \label{fig:b_a0_aumic}
\end{figure*}

The derived filling factors amount to an average field strength $<\!B\!>=2.61\pm0.05$~kG (see Fig.~\ref{fig:b_a0_aumic}), which compares well to values reported in the literature of, for example, 2.1--2.3~kG~\citep{kochukhov_2020} and $3.01\pm0.22$~kG~\citep{reiners_2022}.
When fixing $\logg$ to 4.40~dex, the average field strength rises up to ${<\!B\!>=2.68\pm0.05}$~kG, still consistent with the values reported in the literature.

\begin{figure*}
    \centering
    \includegraphics[scale=.47]{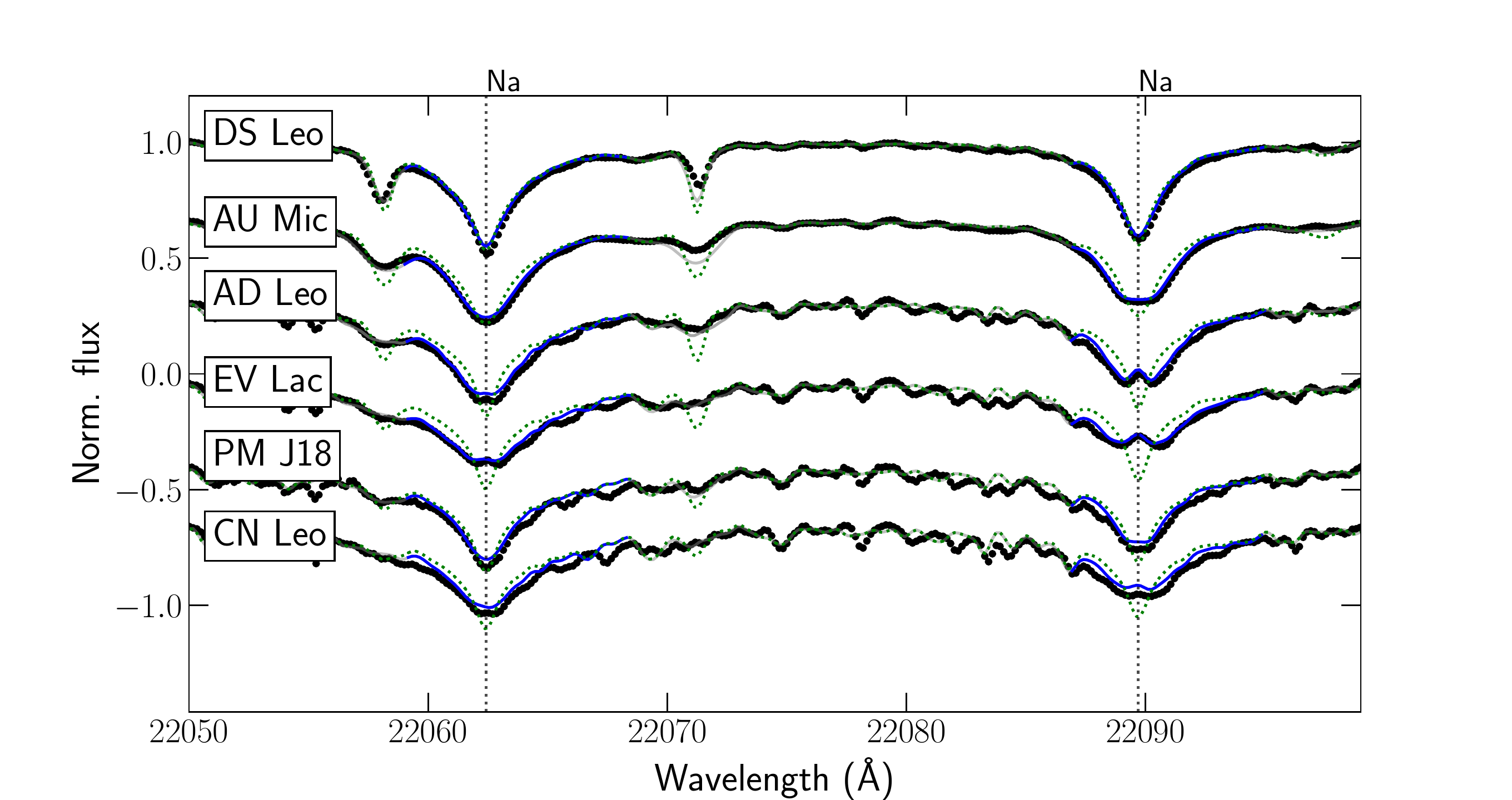}
    \caption{Best fit obtained for the seven stars included in our study with \texttt{ZeeTurbo} for two Na lines (22062.4 and 22089.7~\r{A}). Black points present the data. The grey solid line shows the best fit, and the blue solid blue line presents the part of the windows used for the fit. The green dotted line shows the model obtained for the same atmospheric parameters but with a zero magnetic field. The name PM~J18482+0741 was replaced by PM~J18 for better readability.}
    \label{fig:example_lines}
\end{figure*}

We repeated our analysis assuming a Gaussian macroturbulence.
With this kernel, we recovered $\teff$, $\logg$, $\mh$, $\afe$ and $<\!B\!>$ very close to those assuming a radial-tangential macroturbulence profile (see Table~\ref{tab:results}). 
The strong constraint derived for $<\!B\!>$ can be explained by the dependence of line shapes on the magnetic field, as illustrated in Fig.~\ref{fig:example_lines} and Fig.~B1 (available as supplementary material).
We note that the filling factors $a_2$ and $a_4$ associated with the 2 and 4~kG components account for most of the surface field of AU Mic (see Fig.~\ref{fig:b_a0_aumic}). 
To diagnose the influence of the higher-field components on the results, we performed a second analysis, omitting the 8 and 10~kG models.
We find no change in the atmospheric parameters, and that the average magnetic field is lowered by a negligible amount, with a difference of 0.01~kG on $<\!B\!>$, thus confirming that keeping the 8 and 10~kG components do not generate additional errors when characterizing the surface magnetic field of AU~Mic.

{\paul We also applied our analysis on spectra recorded for each night (Donati et al., submitted). As in our simulation (see Sec.~\ref{sec:simu_dipole}), we find that the average magnetic field strength derived from the SPIRou template is consistent with the median of the field strengths derived for each night. These results provide further evidence that analysing median spectra does not introduce biases in the derived magnetic field strengths.}

\subsection{AD Leo = Gl~388}

We performed a similar analysis on AD Leo (Gl~388). This star was included in the sample of previous studies studies~\citep{morin_2008, reiners_2022}, and its projected rotational velocity was estimated to $\vsini=3\pm1$~\kms~\citep{morin_2008}. The mass and radius of this star were estimated from the mass-K band magnitude relation of~\citep{mann_2019} and the models of~\citep{baraffe_2015} (see Table~\ref{tab:literature_params}). These yield a $\logg\sim4.88\pm0.05$~dex. The  rotation period  of AD~Leo is $P_{\rm rot}=2.2399\pm0.0006$~d~\citep[][see Table~\ref{tab:literature_params}]{morin_2008}.

 We chose to fix the value of $\vsini$ and fit a radial-tangential macroturbulence in this analysis. With these constraints, we derive an average magnetic field of $3.03\pm0.06$~kG, consistent with some previous estimates, e.g. $2.9\pm0.2$~kG~\citep{reiners_2007} and $3.57\pm0.09$~kG~\citep{reiners_2022}. Just like for AU Mic, we find the largest filling factors for the 2 and 4~kG components for this star.
The retrieved atmospheric parameters, i.e. $\teff=3467\pm31$, $\logg=4.90\pm0.05$, $\mh=0.24\pm0.10$ and $\afe=0.00\pm0.04$ compare well with previous estimates~\citep{mann_2015}, with the exception of a few recent studies suggesting that this star may be metal-poor~\citep{marfil_2021}. 
Our $\logg$ is in good agreement with the mass and radius estimates.

With $\vsini=3$~\kms, we retrieve a radial-tangential macroturbulence $\zeta_{\rm{RT}}=1.7\pm0.2$~\kms. We repeat the analysis, this time with a Gaussian macroturbulence, and retrieve a FWHM of $\zeta_{\rm G}=2.0\pm0.2$~\kms. We find that changing the macroturbulence model has a negligible impact on the derived atmospheric parameters and filling factors (see Table~\ref{tab:results}).

\subsection{EV Lac = Gl~873}

EV Lac  (Gl~873) is another very well-known magnetic M dwarf observed in the context of the SLS, with a rotation period of ${P_{\rm rot}=4.3715\pm0.0006}$~d~\citep{morin_2008}. We estimated its mass and radius to $M=0.32\pm0.02~\msun$ and $R=0.31\pm0.02~\rsun$ (see Table~\ref{tab:literature_params}), thus implying $\logg=4.96\pm0.06$~dex. The projected rotational velocity of this star was estimated to about $4\pm1$~\kms~\citep{morin_2008}. The radius and rotation period of this star would suggest that this value is slightly over-estimated, and we therefore choose to fix its value to $\vsini=3$~\kms{}.

We fixed $\vsini$, and fitted the radial-tangential macroturbulent velocity $\zeta_{\rm RT}$. For this star we derive ${\teff=3340\pm31}$~K, ${\logg=4.87\pm0.05}$~dex, ${\mh=0.04\pm0.10}$~dex and ${\afe=0.01\pm0.04}$~dex. These atmospheric parameters are in good agreement with those reported by previous studies~\citep{maldonado_2020}.  Our $\logg$ estimate is also in good agreement with the estimated mass and radius for this star. 
We compute an average magnetic field of $<\!B\!>=4.53\pm0.07$~kG, consistent with estimates reported in the literature, of $3.8\pm0.5$~kG~\citep{johns-krull_2000} or $4.32\pm0.11$~kG~\citep{reiners_2022}.
For this star, we note that the filling factors associated to the 6, 8 and 10~kG components are not close to 0, but rather account for 30~\% of the total magnetic flux (see Fig.~\ref{fig:b_a0_evlac}).

We retrieved a macroturbulent velocity $\zeta_{\rm RT}=4.2\pm0.2$~\kms. We repeat the analysis, this time fitting a Gaussian macroturbulence model, and retrieve a FWHM of $\zeta_{\rm G}=4.6\pm0.3$~\kms. 
We further checked that changing the adopted value of $\vsini$ by 1~\kms{} had negligible impact on the retrieved atmospheric parameters and magnetic field strength.
Here again, the choice of model for the macroturbulence  profile has negligible impact on the derived atmospheric parameters and magnetic field strength (see Table~\ref{tab:results}).

\subsection{CN Leo = Gl~406}

We then performed our analysis on the SPIRou template of CN Leo (Gl~406), an active late-type M dwarf. The rotation period of this star is ${P_{\rm rot} = 2.704\pm0.003}$~d~\citep{diez_alonso_2019}, and we estimate its mass and radius to $M=0.11\pm0.02~\msun$ and $R=0.13\pm0.02~\rsun$ (see Table~\ref{tab:literature_params}). The projected rotational velocity of CN~Leo was previously estimated to $\vsini=3\pm1$~\kms{}~\citep{reiners_2007}. Given the rotation period and radius for this star, we find that $\vsini=3$~\kms{} is likely overestimated, and we chose to fix its value to $\vsini=2$~\kms{}.

We retrieve $\teff=2898\pm31$~K, in good agreement with previously reported estimates~\citep{mann_2015}. Our estimate of $\logg=4.82\pm0.08$~dex is significantly lower than that computed from mass and radius, of $\logg=5.25\pm0.17$~dex.
 We derive $\mh=0.17\pm0.12$~dex,  consistent with previous estimates~\citep{rojas-ayala_2012, mann_2015}. 

We recover an average magnetic field strength $<\!B\!>=3.45\pm0.20$~kG, consistent with previously reported values~\citep[$<\!B\!>=3.01\pm0.16$~kG, ][]{reiners_2022}. 
Again, we find that fitting the data with a Gaussian macroturbulence profile yields almost identical parameters (see Table~\ref{tab:results}).

\subsection{PM~J18482+0741}

PM J18482+0741 is another cool M dwarf observed in the context of the SLS, with a projected rotational velocity estimated to $\vsini=2.4\pm1.5$~\kms{}~\citep{reiners_2018}, and a mass and radius estimated to ${M=0.14\pm0.02~\msun}$~\&~${R=0.17\pm0.02~\rsun}$ (see Table~\ref{tab:literature_params}), yielding $\logg=5.12\pm0.13$~dex. The rotation period of this star was estimated by~\citep{diez-alonso_2019} to $2.76\pm0.01$~d. 
 
 For this target, we retrieve $\teff=3078\pm32$~K, consistent with previously reported effective temperatures for this target~\citep{gaidos_2014, passegger_2019}. Our recovered $\logg=4.72\pm0.06$ is lower than that reported by~\citep{passegger_2019} and that implied by our radius and mass estimates.
With our process, we retrieve an average magnetic field $<\!B\!>=1.99\pm0.15$~kG, almost twice that of~\citet[][$<\!B\!>=1.19\pm0.23$~kG]{reiners_2022}. We find that for this star too, fitting the data with a Gaussian instead of a radial-tangential macroturbulence profile has negligible impact on the results (see Table~\ref{tab:results}).

\subsection*{DS Leo = Gl~410}

Finally, we run our process on the moderately active DS Leo (Gl~410). The rotation period of this star, of $P_{\rm rot}=14\pm0.1$~d~\citep{donati_2008}, is the largest in our sample. The mass and radius of DS~Leo, estimated to $M=0.57\pm0.02~\msun$ and $R=0.53\pm0.02~\rsun$ (see Table~\ref{tab:literature_params}), implies a surface gravity of $\logg=4.74\pm0.04$~dex. 
For this star, $\vsini$ was estimated to $2\pm1$~\kms{} by~\citet{morin_2008}.

With a fixed value of $\vsini$, we retrieved ${\teff=3818\pm30}$~K, $\logg=4.79\pm0.05$~dex, $\mh=0.01\pm0.10$~dex and ${\afe=0.03\pm0.04}$~dex (see Table~\ref{tab:results}). These values are in good agreement with previous estimates, including ours~\citep{mann_2015, cristofari_2022b}. Our $\logg$ is also comparable to that implied by previous mass and radius estimates.
We derive $<\!B\!>=0.73\pm0.03$~kG, lower than that reported by~\citet{reiners_2022}, of $<\!B\!>=1.04\pm0.06$~kG. For DS Leo, we find $a_4$, $a_6$ and $a_8$ to be close to 0. We repeat our analysis process, only using models computed for 0 and 2~kG, and find that removing the 4, 6, 8 and 10~kG components has negligible impact on the estimation of atmospheric parameters and filling factors.

\subsection{Comparison with the literature}

\begin{figure}
	\includegraphics[width=\linewidth]{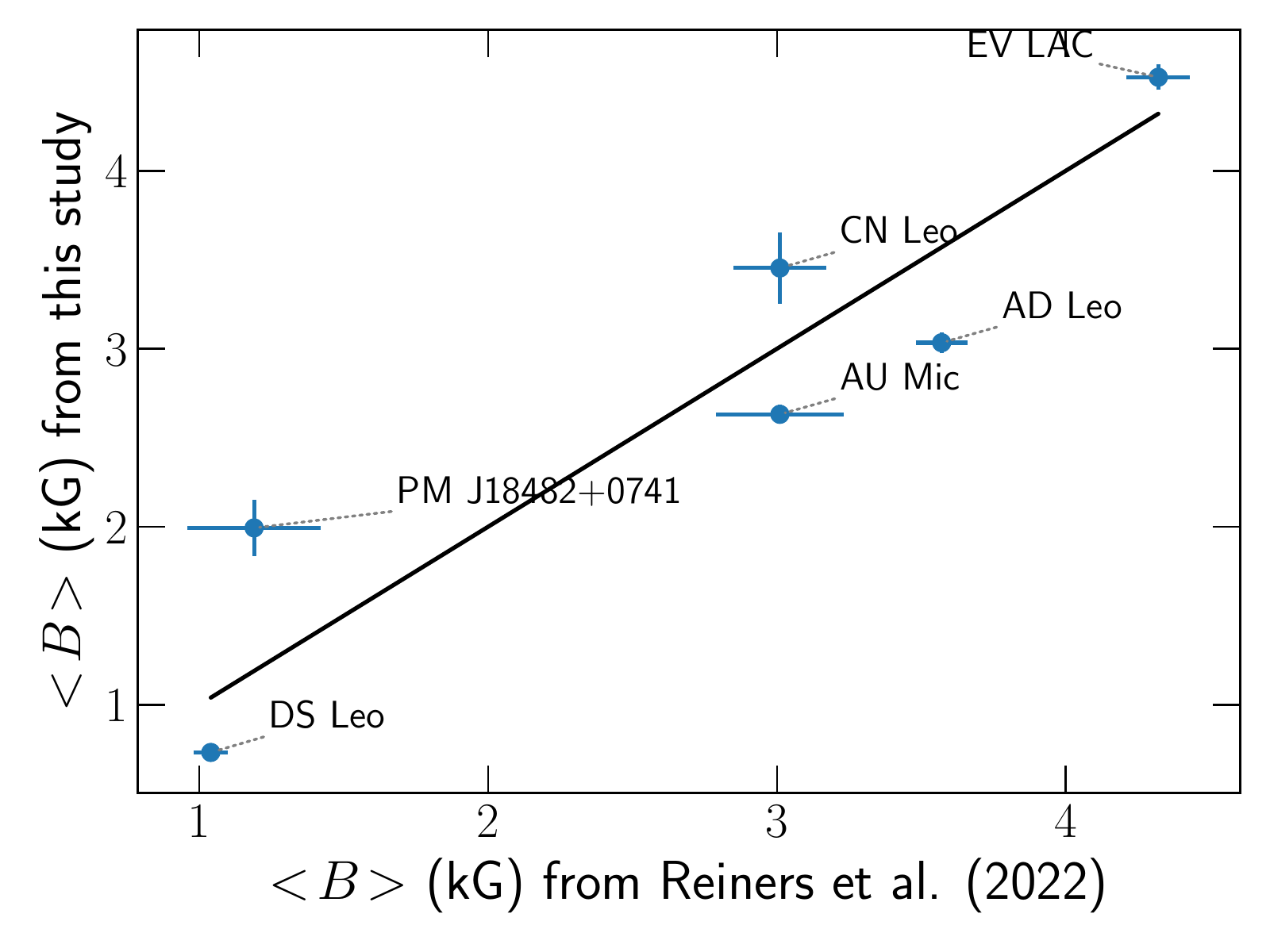}
	\caption{Comparison between our derived $<\!B\!>$ values and those of~\citet{reiners_2022}. The black line marks the equality}
	\label{fig:b_b}
\end{figure}

\begin{figure}
	\includegraphics[width=\linewidth]{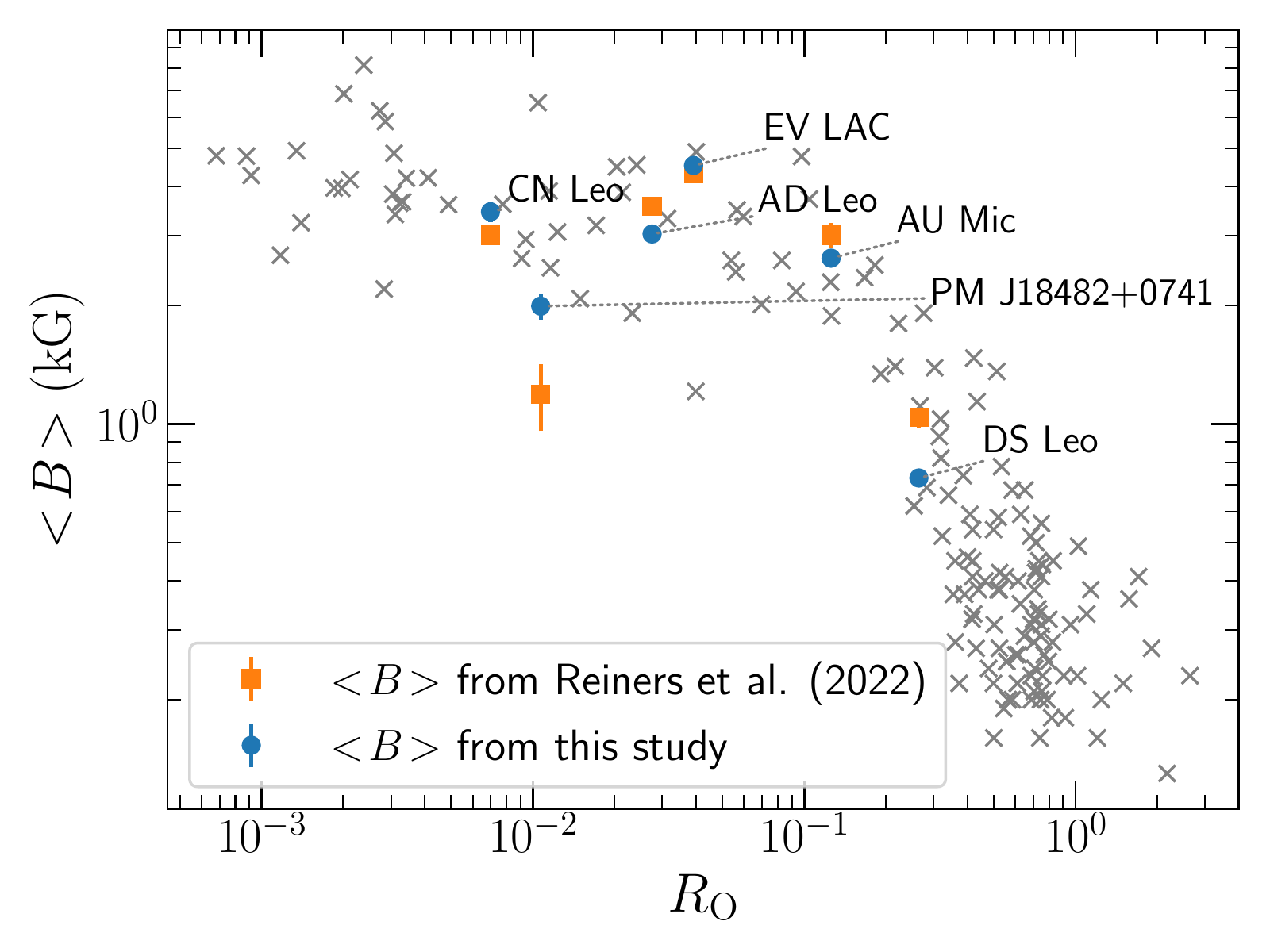}
	\caption{Comparison Rossby number ($R_{\rm O}$) and $<\!B\!>$ values derived from our study (blue dots) and reported by~\citet[][orange squares]{reiners_2022}. The grey symbols mark the position of all the stars studied by~\citet{reiners_2022}.}
	\label{fig:rossby_b}
\end{figure}

Figure~\ref{fig:b_b} presents a comparison between our retrieved $<\!B\!>$ estimates and those reported by~\citet{reiners_2022}. We find an overall good agreement between the two sets of values. $<\!B\!>$ is expected to evolve with time, which can at least partly account for some of the observed differences. Figure~\ref{fig:rossby_b} presents the position of the stars in a $<\!B\!>$ vs Rossby number ($R_{\rm O}$) diagram. Most active M dwarfs included in our sample fall within the saturated dynamo regime, with the exception of DS~Leo. These results are also in good agreement with those reported in~\citet{reiners_2022}.
Comparisons between our retrieved $\teff$, $\logg$ and literature estimates are presented in Figs.~\ref{fig:comparison_lit_teffs}~\&~\ref{fig:comparison_lit_loggs}.

\begin{table*}
	\centering
	\caption{Retrieved stellar parameter and magnetic fields for our sample of targets. Values given with no associated uncertainties were fixed. For each star, multiple lines present the results obtained when fixing different parameters.}
	\begin{adjustbox}{angle=90}   
		\resizebox{!}{1.1\height}{
  {
			\begin{tabu}{cccccccccc}
				\hline
				Star  (GJ ID) & $\teff$ & $\logg$ & $\mh$ & $\afe$ & $\vsini$ & $\zeta_{\rm RT}$ & $\zeta_{\rm G}$ & $<\!B\!>$  & \makecell{$f_0$, $f_2$, $f_4$, \\ $f_6$, $f_8$, $f_{10}$}\\ 
				\hline
\rowfont{\color{black}}
AU Mic (Gl 803) & $3665\pm31$ & $4.52\pm0.05$ & $0.12\pm0.10$ & $0.00\pm0.04$ & $8.5$ & $4.1\pm0.3$ & -- & $2.61\pm0.05$& $0.07\pm0.02$, $0.59\pm0.03$, $0.33\pm0.03$, \\  \rowfont{\color{black}} & & & & & & & & & $0.01\pm0.01$, $0.00\pm0.00$, $0.00\pm0.00$\\
\rowfont{\color{black}}
AU Mic (Gl 803) & $3666\pm31$ & $4.52\pm0.05$ & $0.12\pm0.10$ & $0.00\pm0.04$ & $8.5$ & -- & $4.7\pm0.4$ & $2.62\pm0.05$& $0.07\pm0.02$, $0.59\pm0.03$, $0.32\pm0.03$, \\ \rowfont{\color{black}} & & & & & & & & & $0.01\pm0.01$, $0.00\pm0.00$, $0.00\pm0.00$\\
\rowfont{\color{black}}
AU Mic (Gl 803) & $3640\pm31$ & $4.40$ & $0.10\pm0.10$ & $-0.02\pm0.04$ & $8.5$ & $5.1\pm0.2$ & -- & $2.68\pm0.05$& $0.06\pm0.02$, $0.58\pm0.03$, $0.33\pm0.03$, \\ \rowfont{\color{black}} & & & & & & & & & $0.01\pm0.01$, $0.00\pm0.00$, $0.00\pm0.00$\\
\rowfont{\color{black}}
AU Mic (Gl 803) & $3641\pm31$ & $4.40$ & $0.10\pm0.10$ & $-0.02\pm0.04$ & $8.5$ & -- & $6.4\pm0.4$ & $2.68\pm0.05$& $0.06\pm0.02$, $0.59\pm0.03$, $0.33\pm0.03$, \\ & & & & & & & & & $0.01\pm0.01$, $0.00\pm0.00$, $0.00\pm0.00$\\
EV~LAC (Gl 873) & $3340\pm31$ & $4.87\pm0.05$ & $0.04\pm0.10$ & $0.01\pm0.04$ & $3.0$ & $4.2\pm0.2$ & -- & $4.53\pm0.07$& $0.02\pm0.04$, $0.29\pm0.03$, $0.38\pm0.04$, \\ & & & & & & & & & $0.09\pm0.04$, $0.13\pm0.04$, $0.08\pm0.02$\\
EV~LAC (Gl 873) & $3342\pm31$ & $4.88\pm0.05$ & $0.03\pm0.10$ & $0.01\pm0.04$ & $3.0$ & -- & $4.6\pm0.2$ & $4.52\pm0.07$& $0.02\pm0.04$, $0.29\pm0.03$, $0.39\pm0.04$, \\ & & & & & & & & & $0.09\pm0.05$, $0.13\pm0.04$, $0.08\pm0.02$\\
AD Leo (Gl 388) & $3467\pm31$ & $4.90\pm0.05$ & $0.24\pm0.10$ & $-0.00\pm0.04$ & $3.0$ & $1.7\pm0.2$ & -- & $3.03\pm0.06$& $0.04\pm0.02$, $0.50\pm0.03$, $0.41\pm0.03$, \\ & & & & & & & & & $0.04\pm0.02$, $0.01\pm0.01$, $0.00\pm0.00$\\
AD Leo (Gl 388) & $3467\pm31$ & $4.89\pm0.05$ & $0.24\pm0.10$ & $-0.01\pm0.04$ & $3.0$ & -- & $2.0\pm0.3$ & $3.04\pm0.06$& $0.03\pm0.02$, $0.50\pm0.03$, $0.41\pm0.03$, \\ & & & & & & & & & $0.04\pm0.02$, $0.01\pm0.01$, $0.00\pm0.00$\\
CN Leo (Gl 406) & $2898\pm31$ & $4.82\pm0.08$ & $0.17\pm0.12$ & $-0.04\pm0.04$ & $2.0$ & $4.3\pm0.3$ & -- & $3.45\pm0.20$& $0.10\pm0.07$, $0.44\pm0.11$, $0.29\pm0.10$, \\ & & & & & & & & & $0.07\pm0.06$, $0.07\pm0.05$, $0.03\pm0.03$\\
CN Leo (Gl 406) & $2899\pm31$ & $4.82\pm0.07$ & $0.16\pm0.11$ & $-0.04\pm0.04$ & $2.0$ & -- & $4.8\pm0.3$ & $3.45\pm0.19$& $0.10\pm0.07$, $0.44\pm0.11$, $0.29\pm0.10$, \\ & & & & & & & & & $0.07\pm0.06$, $0.07\pm0.05$, $0.03\pm0.03$\\
DS Leo (Gl 410) & $3818\pm30$ & $4.79\pm0.05$ & $0.01\pm0.10$ & $0.03\pm0.04$ & $1.5$ & $2.7\pm0.1$ & -- & $0.73\pm0.03$& $0.65\pm0.00$, $0.34\pm0.02$, $0.00\pm0.00$, \\ & & & & & & & & & $0.00\pm0.00$, $0.00\pm0.00$, $0.00\pm0.00$\\
DS Leo (Gl 410) & $3818\pm33$ & $4.79\pm0.06$ & $0.01\pm0.10$ & $0.03\pm0.04$ & $1.5$ & -- & $3.0\pm0.3$ & $0.82\pm0.10$& $0.66\pm0.01$, $0.32\pm0.04$, $0.01\pm0.01$, \\ & & & & & & & & & $0.01\pm0.01$, $0.00\pm0.00$, $0.00\pm0.00$\\
PM~J18482+0741  & $3078\pm32$ & $4.72\pm0.06$ & $-0.02\pm0.10$ & $-0.05\pm0.04$ & $2.4$ & $4.8\pm0.4$ & -- & $1.99\pm0.16$& $0.23\pm0.04$, $0.65\pm0.09$, $0.08\pm0.05$, \\ & & & & & & & & & $0.02\pm0.02$, $0.01\pm0.01$, $0.01\pm0.01$\\
PM~J18482+0741  & $3081\pm34$ & $4.72\pm0.08$ & $-0.03\pm0.11$ & $-0.05\pm0.04$ & $2.4$ & -- & $5.2\pm0.5$ & $2.08\pm0.24$& $0.25\pm0.05$, $0.63\pm0.13$, $0.07\pm0.06$, \\ & & & & & & & & & $0.02\pm0.02$, $0.02\pm0.02$, $0.01\pm0.01$\\

\hline 

\end{tabu}
}
}
\end{adjustbox}
\label{tab:results}
\end{table*}

\section{Discussion and conclusions}
\label{sec:conclusions}

In this paper, we present our first results with our new tools aimed at characterising M dwarfs from SPIRou spectra.
Our process relies on the comparison of high-resolution synthetic spectra computed from state-of-the-art \texttt{MARCS} model atmospheres to data, and is used to constrain $\teff$, $\logg$, $\mh$, $\afe$ and the average magnetic field strengths for 4 targets observed in the context of the SLS. 

We introduce a new code, \texttt{ZeeTurbo}, built from the \texttt{Turbospectrum} and \texttt{Zeeman} codes, allowing us to synthesise spectra of magnetic stars from \texttt{MARCS} model atmospheres. We compared the output spectra computed with \texttt{ZeeTurbo}, \texttt{Zeeman} and \texttt{Turbospectrum} and found that our new code allows us to properly synthesise spectra for magnetic M dwarfs. Our code also allowed us to synthesise molecular lines, assumed to be insensitive to magnetic fields in the present work. This assumption holds for the lines our analysis relies on, namely the few OH lines and the CO lines at 2.28~$\mu$m. Modelling molecular lines is particularly critical to the analysis of M dwarfs spectra since they are numerous and blend with atomic features.

With our newly implemented code, we computed a grid of synthetic spectra assuming a constant magnetic field, radial in all points of the photosphere.
We modelled the spectra by a linear combination of profiles computed for different magnetic strengths, and fitted our model to SPIRou templates to constrain $\teff$, $\logg$, $\mh$, $\afe$, the filling factors and thereby the surface magnetic flux. Our analysis relies on a MCMC process, and the atmospheric parameters and filling factors are estimated from posterior distributions.
We performed a benchmark, designed to assess the performances of our new tool, and found that it was capable of simultaneously constraining magnetic fields and atmospheric parameters. 
{\paul We also show that our modeling assumptions, e.g. on the field topology, introduce only small biases in the measured field strengths.}
We then applied our tool to a few well-known magnetic stars observed in the context of the SLS (AU~Mic, AD~Leo, EV~Lac, DS~Leo, CN~Leo and PM~J18482+0741). Our recovered atmospheric parameters and magnetic field estimates are found in good agreement with the literature for most stars.
The largest discrepancies between our results and the literature are found for the two coolest stars in our sample (CN~Leo and PM~J18482+0741), with $\logg$ estimates significantly lower than those computed from masses and radii.

The average surface magnetic flux retrieved with our process for the six targets in our sample are in good agreement with previous estimates reported by~\citet[][see Fig.~\ref{fig:b_b}]{reiners_2022}. Our estimates are also consistent with most of our stars being in the saturated dynamo regime, with the exception of DS~Leo, whose rotation period is significantly  longer than that of the other stars. The differences in the values reported in the literature and those derived in this study may partly arise from the evolution of the surface magnetic flux with time.

We find that the way the surface magnetic flux is distributed across the magnetic field strengths differs from star to star. In particular, we find significantly larger contributions of the 6, 8 and 10~kG components for EV~Lac of CN~Leo, than for the other targets of our sample. For the quietest star in our sample, DS~Leo, the best fit relies almost entirely on the 0 and 2~kG components. Moreover, the contribution of the 0~kG component also differs from star to star, and is not necessarily smallest for the most magnetic targets (e.g. the case of CN Leo, where $a_0=0.10$, see Fig.~\ref{fig:b_a0_cnleo}). These results illustrate the variety of magnetic topologies encountered in our sample, and the possibility to distinguish them using unpolarised spectra.
{\paul Besides, we find that applying our tool to spectra collected on individual nights (e.g., in the case of AU~Mic, Donati et al., submitted) yields field strengths whose median value is consistent with the field strength derived from the median spectrum.}

{\paul In this work, spectra computed with \texttt{ZeeTurbo} relied on \texttt{MARCS} models that neglect the impact of the additional pressure from magnetic fields on the structure of the atmosphere. Recent works have attempted to obtain improved model atmospheres of magnetic stars~\citep{valyavin_2004, kochukhov_2005, shulyak_2010, stift_2016, jarvinen_2020}. Future studies using \texttt{ZeeTurbo} may build up on such improvements.}

\texttt{ZeeTurbo} will allow us to analyse all stars observed in the context of the SLS in a self-consistent way. In particular, we will reprocess the M dwarfs included in our previous studies, to measure their surface magnetic fluxes and assess their impact on the atmospheric characterisation. 
We will also look for temporal evolution in the average magnetic flux of stars monitored over several years, in order to find new means of constraining rotation, activity cycles, and help disentangle activity jitters from radial velocity signals~\citep[][Donati et al., in prep]{haywood_2016, suarez_2020}.
We will also expand our analysis to PMS stars, whose modelling may require to account for veiling and starspots, and whose characterisation is essential to the study of stars and planets formation~\citep{flores_2021, lopez-valdivia_2021}.

\section*{Acknowledgements}

This project received funding from the European Research Council (ERC) under the H2020 research and innovation programme (grant \#740651, NewWorlds).
TM acknowledges financial support from the Spanish Ministry of Science and Innovation (MICINN) through the Spanish State Research Agency, under the Severo Ochoa Program 2020-2023(CEX2019-000920-S) as well as support from the ACIISI, Consejer\'{i}a de Econom\'{i}a, Conocimiento y Empleo del Gobiernode Canarias and the European Regional Development Fund (ERDF) under grant with reference  PROID2021010128

This work is based on observations obtained at the Canada– France–Hawaii Telescope (CFHT), operated by the National Research Council (NRC) of Canada, the Institut National des Sciences de l’Univers of the Centre National de la Recherche Scientifique (CNRS) of France, and the University of Hawaii. The observations at the CFHT were performed with care and respect from the summit of Maunakea, which is a significant cultural and historic site.

We acknowledge B. Plez for his implication in developing the freely available \texttt{Turbospectrum} code which allowed us to develop \texttt{ZeeTurbo}.

We acknowledge funding from the French ANR under contract number ANR\-18\-CE31\-0019 (SPlaSH).

\section*{Data Availability}

The data used in this work were recorded in the context of the SLS, and will be available to the public at the Canadian Astronomy Data Center one year after completion of the program.



\bibliographystyle{mnras}
\bibliography{BibPaper3} 

\begin{thebibliography}{}
\makeatletter
\relax
\def\mn@urlcharsother{\let\do\@makeother \do\$\do\&\do\#\do\^\do\_\do\%\do\~}
\def\mn@doi{\begingroup\mn@urlcharsother \@ifnextchar [ {\mn@doi@}
  {\mn@doi@[]}}
\def\mn@doi@[#1]#2{\def\@tempa{#1}\ifx\@tempa\@empty \href
  {http://dx.doi.org/#2} {doi:#2}\else \href {http://dx.doi.org/#2} {#1}\fi
  \endgroup}
\def\mn@eprint#1#2{\mn@eprint@#1:#2::\@nil}
\def\mn@eprint@arXiv#1{\href {http://arxiv.org/abs/#1} {{\tt arXiv:#1}}}
\def\mn@eprint@dblp#1{\href {http://dblp.uni-trier.de/rec/bibtex/#1.xml}
  {dblp:#1}}
\def\mn@eprint@#1:#2:#3:#4\@nil{\def\@tempa {#1}\def\@tempb {#2}\def\@tempc
  {#3}\ifx \@tempc \@empty \let \@tempc \@tempb \let \@tempb \@tempa \fi \ifx
  \@tempb \@empty \def\@tempb {arXiv}\fi \@ifundefined
  {mn@eprint@\@tempb}{\@tempb:\@tempc}{\expandafter \expandafter \csname
  mn@eprint@\@tempb\endcsname \expandafter{\@tempc}}}

\bibitem[\protect\citeauthoryear{{Afram} \& {Berdyugina}}{{Afram} \&
  {Berdyugina}}{2019}]{alfram_2019}
{Afram} N.,  {Berdyugina} S.~V.,  2019, \mn@doi [\aap]
  {10.1051/0004-6361/201935793}, \href
  {https://ui.adsabs.harvard.edu/abs/2019A&A...629A..83A} {629, A83}

\bibitem[\protect\citeauthoryear{{Alvarez} \& {Plez}}{{Alvarez} \&
  {Plez}}{1998}]{plez_1998}
{Alvarez} R.,  {Plez} B.,  1998, \aap, \href
  {https://ui.adsabs.harvard.edu/abs/1998A&A...330.1109A} {330, 1109}

\bibitem[\protect\citeauthoryear{{Baraffe}, {Homeier}, {Allard}  \&
  {Chabrier}}{{Baraffe} et~al.}{2015}]{baraffe_2015}
{Baraffe} I.,  {Homeier} D.,  {Allard} F.,   {Chabrier} G.,  2015, \mn@doi
  [\aap] {10.1051/0004-6361/201425481}, \href
  {https://ui.adsabs.harvard.edu/abs/2015A&A...577A..42B} {577, A42}

\bibitem[\protect\citeauthoryear{{Barber}, {Tennyson}, {Harris}  \&
  {Tolchenov}}{{Barber} et~al.}{2006}]{barber_2006}
{Barber} R.~J.,  {Tennyson} J.,  {Harris} G.~J.,   {Tolchenov} R.~N.,  2006,
  \mn@doi [\mnras] {10.1111/j.1365-2966.2006.10184.x}, \href
  {https://ui.adsabs.harvard.edu/abs/2006MNRAS.368.1087B} {368, 1087}

\bibitem[\protect\citeauthoryear{{Barton}, {Yurchenko}  \& {Tennyson}}{{Barton}
  et~al.}{2013}]{barton_2013}
{Barton} E.~J.,  {Yurchenko} S.~N.,   {Tennyson} J.,  2013, \mn@doi [\mnras]
  {10.1093/mnras/stt1105}, \href
  {https://ui.adsabs.harvard.edu/abs/2013MNRAS.434.1469B} {434, 1469}

\bibitem[\protect\citeauthoryear{{Bellotti}, {Petit}, {Morin}, {Hussain},
  {Folsom}, {Carmona}, {Delfosse}  \& {Moutou}}{{Bellotti}
  et~al.}{2022}]{bellotti_2022}
{Bellotti} S.,  {Petit} P.,  {Morin} J.,  {Hussain} G.~A.~J.,  {Folsom} C.~P.,
  {Carmona} A.,  {Delfosse} X.,   {Moutou} C.,  2022, \mn@doi [\aap]
  {10.1051/0004-6361/202141812}, \href
  {https://ui.adsabs.harvard.edu/abs/2022A&A...657A.107B} {657, A107}

\bibitem[\protect\citeauthoryear{{Boccaletti} et~al.,}{{Boccaletti}
  et~al.}{2018}]{boccaletti_2018}
{Boccaletti} A.,  et~al., 2018, \mn@doi [\aap] {10.1051/0004-6361/201732462},
  \href {https://ui.adsabs.harvard.edu/abs/2018A&A...614A..52B} {614, A52}

\bibitem[\protect\citeauthoryear{{Brooke}, {Bernath}, {Western}, {Sneden},
  {Af{\c{s}}ar}, {Li}  \& {Gordon}}{{Brooke} et~al.}{2016}]{brooke_2016}
{Brooke} J. S.~A.,  {Bernath} P.~F.,  {Western} C.~M.,  {Sneden} C.,
  {Af{\c{s}}ar} M.,  {Li} G.,   {Gordon} I.~E.,  2016, \mn@doi [\jqsrt]
  {10.1016/j.jqsrt.2015.07.021}, \href
  {https://ui.adsabs.harvard.edu/abs/2016JQSRT.168..142B} {168, 142}

\bibitem[\protect\citeauthoryear{{Burrows}, {Ram}, {Bernath}, {Sharp}  \&
  {Milsom}}{{Burrows} et~al.}{2002}]{burrows_2002}
{Burrows} A.,  {Ram} R.~S.,  {Bernath} P.,  {Sharp} C.~M.,   {Milsom} J.~A.,
  2002, \mn@doi [\apj] {10.1086/342242}, \href
  {https://ui.adsabs.harvard.edu/abs/2002ApJ...577..986B} {577, 986}

\bibitem[\protect\citeauthoryear{{Cook} et~al.,}{{Cook}
  et~al.}{2022}]{cook_2022}
{Cook} N.~J.,  et~al., 2022, arXiv e-prints, \href
  {https://ui.adsabs.harvard.edu/abs/2022arXiv221101358C} {p. arXiv:2211.01358}

\bibitem[\protect\citeauthoryear{{Cristofari} et~al.,}{{Cristofari}
  et~al.}{2022a}]{cristofari_2022a}
{Cristofari} P.~I.,  et~al., 2022a, \mn@doi [\mnras] {10.1093/mnras/stab3679},
  \href {https://ui.adsabs.harvard.edu/abs/2022MNRAS.511.1893C} {511, 1893}

\bibitem[\protect\citeauthoryear{{Cristofari} et~al.,}{{Cristofari}
  et~al.}{2022b}]{cristofari_2022b}
{Cristofari} P.~I.,  et~al., 2022b, \mn@doi [\mnras] {10.1093/mnras/stac2364},
  \href {https://ui.adsabs.harvard.edu/abs/2022MNRAS.516.3802C} {516, 3802}

\bibitem[\protect\citeauthoryear{{Deen}}{{Deen}}{2013}]{deen_2013}
{Deen} C.~P.,  2013, \mn@doi [\aj] {10.1088/0004-6256/146/3/51}, \href
  {https://ui.adsabs.harvard.edu/abs/2013AJ....146...51D} {146, 51}

\bibitem[\protect\citeauthoryear{{D{\'\i}ez Alonso} et~al.,}{{D{\'\i}ez Alonso}
  et~al.}{2019a}]{diez_alonso_2019}
{D{\'\i}ez Alonso} E.,  et~al., 2019a, \mn@doi [\aap]
  {10.1051/0004-6361/201833316}, \href
  {https://ui.adsabs.harvard.edu/abs/2019A&A...621A.126D} {621, A126}

\bibitem[\protect\citeauthoryear{{D{\'\i}ez Alonso} et~al.,}{{D{\'\i}ez Alonso}
  et~al.}{2019b}]{diez-alonso_2019}
{D{\'\i}ez Alonso} E.,  et~al., 2019b, \mn@doi [\aap]
  {10.1051/0004-6361/201833316}, \href
  {https://ui.adsabs.harvard.edu/abs/2019A&A...621A.126D} {621, A126}

\bibitem[\protect\citeauthoryear{{Donati} et~al.,}{{Donati}
  et~al.}{2008}]{donati_2008}
{Donati} J.~F.,  et~al., 2008, \mn@doi [\mnras]
  {10.1111/j.1365-2966.2008.13799.x}, \href
  {https://ui.adsabs.harvard.edu/abs/2008MNRAS.390..545D} {390, 545}

\bibitem[\protect\citeauthoryear{{Donati} et~al.,}{{Donati}
  et~al.}{2020}]{donati_2020}
{Donati} J.~F.,  et~al., 2020, \mn@doi [\mnras] {10.1093/mnras/staa2569}, \href
  {https://ui.adsabs.harvard.edu/abs/2020MNRAS.498.5684D} {498, 5684}

\bibitem[\protect\citeauthoryear{{Dumusque} et~al.,}{{Dumusque}
  et~al.}{2021}]{dumusque_2021}
{Dumusque} X.,  et~al., 2021, \mn@doi [\aap] {10.1051/0004-6361/202039350},
  \href {https://ui.adsabs.harvard.edu/abs/2021A&A...648A.103D} {648, A103}

\bibitem[\protect\citeauthoryear{{Flores}, {Connelley}, {Reipurth}  \&
  {Duch{\^e}ne}}{{Flores} et~al.}{2021}]{flores_2021}
{Flores} C.,  {Connelley} M.~S.,  {Reipurth} B.,   {Duch{\^e}ne} G.,  2021,
  arXiv e-prints, \href {https://ui.adsabs.harvard.edu/abs/2021arXiv211103957F}
  {p. arXiv:2111.03957}

\bibitem[\protect\citeauthoryear{{Folsom} et~al.,}{{Folsom}
  et~al.}{2016}]{folsom_2016}
{Folsom} C.~P.,  et~al., 2016, \mn@doi [\mnras] {10.1093/mnras/stv2924}, \href
  {https://ui.adsabs.harvard.edu/abs/2016MNRAS.457..580F} {457, 580}

\bibitem[\protect\citeauthoryear{{Foreman-Mackey}, {Hogg}, {Lang}  \&
  {Goodman}}{{Foreman-Mackey} et~al.}{2013}]{emcee}
{Foreman-Mackey} D.,  {Hogg} D.~W.,  {Lang} D.,   {Goodman} J.,  2013, \mn@doi
  [\pasp] {10.1086/670067}, \href
  {https://ui.adsabs.harvard.edu/abs/2013PASP..125..306F} {125, 306}

\bibitem[\protect\citeauthoryear{{Gaidos} et~al.,}{{Gaidos}
  et~al.}{2014}]{gaidos_2014}
{Gaidos} E.,  et~al., 2014, \mn@doi [\mnras] {10.1093/mnras/stu1313}, \href
  {https://ui.adsabs.harvard.edu/abs/2014MNRAS.443.2561G} {443, 2561}

\bibitem[\protect\citeauthoryear{{Gallenne}, {Desgrange}, {Milli},
  {Sanchez-Bermudez}, {Chauvin}, {Kraus}, {Girard}  \& {Boccaletti}}{{Gallenne}
  et~al.}{2022}]{gallenne_2022}
{Gallenne} A.,  {Desgrange} C.,  {Milli} J.,  {Sanchez-Bermudez} J.,  {Chauvin}
  G.,  {Kraus} S.,  {Girard} J.~H.,   {Boccaletti} A.,  2022, arXiv e-prints,
  \href {https://ui.adsabs.harvard.edu/abs/2022arXiv220704116G} {p.
  arXiv:2207.04116}

\bibitem[\protect\citeauthoryear{{Gerber}, {Magg}, {Plez}, {Bergemann},
  {Heiter}, {Olander}  \& {Hoppe}}{{Gerber} et~al.}{2022}]{gerber_2022}
{Gerber} J.~M.,  {Magg} E.,  {Plez} B.,  {Bergemann} M.,  {Heiter} U.,
  {Olander} T.,   {Hoppe} R.,  2022, arXiv e-prints, \href
  {https://ui.adsabs.harvard.edu/abs/2022arXiv220600967G} {p. arXiv:2206.00967}

\bibitem[\protect\citeauthoryear{{Gordon} et~al.,}{{Gordon}
  et~al.}{2017}]{gordon_2017}
{Gordon} I.~E.,  et~al., 2017, \mn@doi [\jqsrt] {10.1016/j.jqsrt.2017.06.038},
  \href {https://ui.adsabs.harvard.edu/abs/2017JQSRT.203....3G} {203, 3}

\bibitem[\protect\citeauthoryear{{Gray}}{{Gray}}{1975}]{gray_1975}
{Gray} D.~F.,  1975, \mn@doi [\apj] {10.1086/153960}, \href
  {https://ui.adsabs.harvard.edu/abs/1975ApJ...202..148G} {202, 148}

\bibitem[\protect\citeauthoryear{Gray}{Gray}{2005}]{gray_2005}
Gray D.~F.,  2005, The Observation and Analysis of Stellar Photospheres, 3 edn.
Cambridge University Press, \mn@doi{10.1017/CBO9781316036570}

\bibitem[\protect\citeauthoryear{{Haywood} et~al.,}{{Haywood}
  et~al.}{2016}]{haywood_2016}
{Haywood} R.~D.,  et~al., 2016, \mn@doi [\mnras] {10.1093/mnras/stw187}, \href
  {https://ui.adsabs.harvard.edu/abs/2016MNRAS.457.3637H} {457, 3637}

\bibitem[\protect\citeauthoryear{{H{\'e}brard}, {Donati}, {Delfosse}, {Morin},
  {Moutou}  \& {Boisse}}{{H{\'e}brard} et~al.}{2016}]{hebrard_2016}
{H{\'e}brard} {\'E}.~M.,  {Donati} J.~F.,  {Delfosse} X.,  {Morin} J.,
  {Moutou} C.,   {Boisse} I.,  2016, \mn@doi [\mnras] {10.1093/mnras/stw1346},
  \href {https://ui.adsabs.harvard.edu/abs/2016MNRAS.461.1465H} {461, 1465}

\bibitem[\protect\citeauthoryear{{Johns-Krull} \& {Valenti}}{{Johns-Krull} \&
  {Valenti}}{1996}]{johns-krull_1996}
{Johns-Krull} C.~M.,  {Valenti} J.~A.,  1996, \mn@doi [\apjl] {10.1086/309954},
  \href {https://ui.adsabs.harvard.edu/abs/1996ApJ...459L..95J} {459, L95}

\bibitem[\protect\citeauthoryear{{Johns-Krull} \& {Valenti}}{{Johns-Krull} \&
  {Valenti}}{2000}]{johns-krull_2000}
{Johns-Krull} C.~M.,  {Valenti} J.~A.,  2000, in {Pallavicini} R.,  {Micela}
  G.,   {Sciortino} S.,  eds,  Astronomical Society of the Pacific Conference
  Series Vol. 198, Stellar Clusters and Associations: Convection, Rotation, and
  Dynamos. p.~371

\bibitem[\protect\citeauthoryear{{Johns-Krull}, {Valenti}  \&
  {Saar}}{{Johns-Krull} et~al.}{2004}]{johns-krull_2004}
{Johns-Krull} C.~M.,  {Valenti} J.~A.,   {Saar} S.~H.,  2004, \mn@doi [\apj]
  {10.1086/425652}, \href
  {https://ui.adsabs.harvard.edu/abs/2004ApJ...617.1204J} {617, 1204}

\bibitem[\protect\citeauthoryear{Järvinen, Hubrig, Mathys, Khalack, Ilyin  \&
  Adigozalzade}{Järvinen et~al.}{2020}]{jarvinen_2020}
Järvinen S.~P.,  Hubrig S.,  Mathys G.,  Khalack V.,  Ilyin I.,   Adigozalzade
  H.,  2020, \mn@doi [Monthly Notices of the Royal Astronomical Society]
  {10.1093/mnras/staa2887}, 499, 2734

\bibitem[\protect\citeauthoryear{{Klein} et~al.,}{{Klein}
  et~al.}{2021}]{klein_2021}
{Klein} B.,  et~al., 2021, \mn@doi [\mnras] {10.1093/mnras/staa3702}, \href
  {https://ui.adsabs.harvard.edu/abs/2021MNRAS.502..188K} {502, 188}

\bibitem[\protect\citeauthoryear{{Klein} et~al.,}{{Klein}
  et~al.}{2022}]{klein_2022}
{Klein} B.,  et~al., 2022, \mn@doi [\mnras] {10.1093/mnras/stac761}, \href
  {https://ui.adsabs.harvard.edu/abs/2022MNRAS.512.5067K} {512, 5067}

\bibitem[\protect\citeauthoryear{{Kochukhov}}{{Kochukhov}}{2007}]{kochukhov_2007}
{Kochukhov} O.~P.,  2007, in {Romanyuk} I.~I.,  {Kudryavtsev} D.~O.,
  {Neizvestnaya} O.~M.,   {Shapoval} V.~M.,  eds, Physics of Magnetic Stars. pp
  109--118 (\mn@eprint {arXiv} {astro-ph/0701084})

\bibitem[\protect\citeauthoryear{{Kochukhov}}{{Kochukhov}}{2021}]{kochukhov_2021}
{Kochukhov} O.,  2021, \mn@doi [\aapr] {10.1007/s00159-020-00130-3}, \href
  {https://ui.adsabs.harvard.edu/abs/2021A&ARv..29....1K} {29, 1}

\bibitem[\protect\citeauthoryear{{Kochukhov} \& {Reiners}}{{Kochukhov} \&
  {Reiners}}{2020}]{kochukhov_2020}
{Kochukhov} O.,  {Reiners} A.,  2020, \mn@doi [\apj]
  {10.3847/1538-4357/abb2a2}, \href
  {https://ui.adsabs.harvard.edu/abs/2020ApJ...902...43K} {902, 43}

\bibitem[\protect\citeauthoryear{{Kochukhov}, {Khan}  \& {Shulyak}}{{Kochukhov}
  et~al.}{2005}]{kochukhov_2005}
{Kochukhov} O.,  {Khan} S.,   {Shulyak} D.,  2005, \mn@doi [\aap]
  {10.1051/0004-6361:20042300}, \href
  {https://ui.adsabs.harvard.edu/abs/2005A&A...433..671K} {433, 671}

\bibitem[\protect\citeauthoryear{{Kupka}, {Ryabchikova}, {Piskunov}, {Stempels}
   \& {Weiss}}{{Kupka} et~al.}{2000}]{kupta_2000}
{Kupka} F.~G.,  {Ryabchikova} T.~A.,  {Piskunov} N.~E.,  {Stempels} H.~C.,
  {Weiss} W.~W.,  2000, \mn@doi [Baltic Astronomy] {10.1515/astro-2000-0420},
  \href {https://ui.adsabs.harvard.edu/abs/2000BaltA...9..590K} {9, 590}

\bibitem[\protect\citeauthoryear{{Landi Degl'Innocenti} \& {Landolfi}}{{Landi
  Degl'Innocenti} \& {Landolfi}}{2004}]{landi_2004}
{Landi Degl'Innocenti} E.,  {Landolfi} M.,  2004, {Polarization in Spectral
  Lines}.
~ Vol. 307, \mn@doi{10.1007/978-1-4020-2415-3, }

\bibitem[\protect\citeauthoryear{{Landstreet}}{{Landstreet}}{1988}]{landstreet_1988}
{Landstreet} J.~D.,  1988, \mn@doi [\apj] {10.1086/166155}, \href
  {https://ui.adsabs.harvard.edu/abs/1988ApJ...326..967L} {326, 967}

\bibitem[\protect\citeauthoryear{{Lavail}, {Kochukhov}, {Hussain}, {Alecian},
  {Herczeg}  \& {Johns-Krull}}{{Lavail} et~al.}{2017}]{lavail_2017}
{Lavail} A.,  {Kochukhov} O.,  {Hussain} G.~A.~J.,  {Alecian} E.,  {Herczeg}
  G.~J.,   {Johns-Krull} C.,  2017, \mn@doi [\aap]
  {10.1051/0004-6361/201731889}, \href
  {https://ui.adsabs.harvard.edu/abs/2017A&A...608A..77L} {608, A77}

\bibitem[\protect\citeauthoryear{{L{\'o}pez-Valdivia}
  et~al.,}{{L{\'o}pez-Valdivia} et~al.}{2021}]{lopez-valdivia_2021}
{L{\'o}pez-Valdivia} R.,  et~al., 2021, \mn@doi [\apj]
  {10.3847/1538-4357/ac1a7b}, \href
  {https://ui.adsabs.harvard.edu/abs/2021ApJ...921...53L} {921, 53}

\bibitem[\protect\citeauthoryear{{Maldonado} et~al.,}{{Maldonado}
  et~al.}{2020}]{maldonado_2020}
{Maldonado} J.,  et~al., 2020, \mn@doi [\aap] {10.1051/0004-6361/202039478},
  \href {https://ui.adsabs.harvard.edu/abs/2020A&A...644A..68M} {644, A68}

\bibitem[\protect\citeauthoryear{{Mann}, {Feiden}, {Gaidos}, {Boyajian}  \&
  {von Braun}}{{Mann} et~al.}{2015}]{mann_2015}
{Mann} A.~W.,  {Feiden} G.~A.,  {Gaidos} E.,  {Boyajian} T.,   {von Braun} K.,
  2015, \mn@doi [\apj] {10.1088/0004-637X/804/1/64}, \href
  {https://ui.adsabs.harvard.edu/abs/2015ApJ...804...64M} {804, 64}

\bibitem[\protect\citeauthoryear{{Mann} et~al.,}{{Mann}
  et~al.}{2019}]{mann_2019}
{Mann} A.~W.,  et~al., 2019, \mn@doi [\apj] {10.3847/1538-4357/aaf3bc}, \href
  {https://ui.adsabs.harvard.edu/abs/2019ApJ...871...63M} {871, 63}

\bibitem[\protect\citeauthoryear{{Marfil} et~al.,}{{Marfil}
  et~al.}{2021}]{marfil_2021}
{Marfil} E.,  et~al., 2021, arXiv e-prints, \href
  {https://ui.adsabs.harvard.edu/abs/2021arXiv211007329M} {p. arXiv:2110.07329}

\bibitem[\protect\citeauthoryear{{Martin} \& {Wickramasinghe}}{{Martin} \&
  {Wickramasinghe}}{1979}]{martin_1979}
{Martin} B.,  {Wickramasinghe} D.~T.,  1979, \mn@doi [\mnras]
  {10.1093/mnras/189.4.883}, \href
  {https://ui.adsabs.harvard.edu/abs/1979MNRAS.189..883M} {189, 883}

\bibitem[\protect\citeauthoryear{{Martioli} et~al.,}{{Martioli}
  et~al.}{2020}]{martioli_2020}
{Martioli} E.,  et~al., 2020, \mn@doi [\aap] {10.1051/0004-6361/202038695},
  \href {https://ui.adsabs.harvard.edu/abs/2020A&A...641L...1M} {641, L1}

\bibitem[\protect\citeauthoryear{{Martioli}, {H{\'e}brard}, {Correia}, {Laskar}
   \& {Lecavelier des Etangs}}{{Martioli} et~al.}{2021}]{martioli_2021}
{Martioli} E.,  {H{\'e}brard} G.,  {Correia} A.~C.~M.,  {Laskar} J.,
  {Lecavelier des Etangs} A.,  2021, \mn@doi [\aap]
  {10.1051/0004-6361/202040235}, \href
  {https://ui.adsabs.harvard.edu/abs/2021A&A...649A.177M} {649, A177}

\bibitem[\protect\citeauthoryear{{Masseron} et~al.,}{{Masseron}
  et~al.}{2014}]{masseron_2014}
{Masseron} T.,  et~al., 2014, \mn@doi [\aap] {10.1051/0004-6361/201423956},
  \href {https://ui.adsabs.harvard.edu/abs/2014A&A...571A..47M} {571, A47}

\bibitem[\protect\citeauthoryear{{Morin} et~al.,}{{Morin}
  et~al.}{2008}]{morin_2008}
{Morin} J.,  et~al., 2008, \mn@doi [\mnras] {10.1111/j.1365-2966.2008.13809.x},
  \href {https://ui.adsabs.harvard.edu/abs/2008MNRAS.390..567M} {390, 567}

\bibitem[\protect\citeauthoryear{{Pakhomov}, {Ryabchikova}  \&
  {Piskunov}}{{Pakhomov} et~al.}{2019}]{pakhomov_2019}
{Pakhomov} Y.~V.,  {Ryabchikova} T.~A.,   {Piskunov} N.~E.,  2019, \mn@doi
  [Astronomy Reports] {10.1134/S1063772919120047}, \href
  {https://ui.adsabs.harvard.edu/abs/2019ARep...63.1010P} {63, 1010}

\bibitem[\protect\citeauthoryear{{Passegger} et~al.,}{{Passegger}
  et~al.}{2019}]{passegger_2019}
{Passegger} V.~M.,  et~al., 2019, \mn@doi [\aap] {10.1051/0004-6361/201935679},
  \href {https://ui.adsabs.harvard.edu/abs/2019A&A...627A.161P} {627, A161}

\bibitem[\protect\citeauthoryear{{Piskunov} \& {Kochukhov}}{{Piskunov} \&
  {Kochukhov}}{2002}]{piskunov_2002}
{Piskunov} N.,  {Kochukhov} O.,  2002, \mn@doi [\aap]
  {10.1051/0004-6361:20011517}, \href
  {https://ui.adsabs.harvard.edu/abs/2002A&A...381..736P} {381, 736}

\bibitem[\protect\citeauthoryear{{Piskunov}, {Kupka}, {Ryabchikova}, {Weiss}
  \& {Jeffery}}{{Piskunov} et~al.}{1995}]{piskunov_1995}
{Piskunov} N.~E.,  {Kupka} F.,  {Ryabchikova} T.~A.,  {Weiss} W.~W.,
  {Jeffery} C.~S.,  1995, \aaps, \href
  {https://ui.adsabs.harvard.edu/abs/1995A&AS..112..525P} {112, 525}

\bibitem[\protect\citeauthoryear{{Plavchan} et~al.,}{{Plavchan}
  et~al.}{2020}]{plavchan_2020}
{Plavchan} P.,  et~al., 2020, \mn@doi [\nat] {10.1038/s41586-020-2400-z}, \href
  {https://ui.adsabs.harvard.edu/abs/2020Natur.582..497P} {582, 497}

\bibitem[\protect\citeauthoryear{{Plez}}{{Plez}}{2012}]{plez_2012}
{Plez} B.,  2012, {Turbospectrum: Code for spectral synthesis} (\mn@eprint
  {ascl} {1205.004})

\bibitem[\protect\citeauthoryear{{Reiners}}{{Reiners}}{2012}]{reiners_2012}
{Reiners} A.,  2012, \mn@doi [Living Reviews in Solar Physics]
  {10.12942/lrsp-2012-1}, \href
  {https://ui.adsabs.harvard.edu/abs/2012LRSP....9....1R} {9, 1}

\bibitem[\protect\citeauthoryear{{Reiners} \& {Basri}}{{Reiners} \&
  {Basri}}{2007}]{reiners_2007}
{Reiners} A.,  {Basri} G.,  2007, \mn@doi [\apj] {10.1086/510304}, \href
  {https://ui.adsabs.harvard.edu/abs/2007ApJ...656.1121R} {656, 1121}

\bibitem[\protect\citeauthoryear{{Reiners} et~al.,}{{Reiners}
  et~al.}{2018}]{reiners_2018}
{Reiners} A.,  et~al., 2018, \mn@doi [\aap] {10.1051/0004-6361/201732054},
  \href {https://ui.adsabs.harvard.edu/abs/2018A&A...612A..49R} {612, A49}

\bibitem[\protect\citeauthoryear{{Reiners} et~al.,}{{Reiners}
  et~al.}{2022}]{reiners_2022}
{Reiners} A.,  et~al., 2022, arXiv e-prints, \href
  {https://ui.adsabs.harvard.edu/abs/2022arXiv220400342R} {p. arXiv:2204.00342}

\bibitem[\protect\citeauthoryear{{Rojas-Ayala}, {Covey}, {Muirhead}  \&
  {Lloyd}}{{Rojas-Ayala} et~al.}{2012}]{rojas-ayala_2012}
{Rojas-Ayala} B.,  {Covey} K.~R.,  {Muirhead} P.~S.,   {Lloyd} J.~P.,  2012,
  \mn@doi [\apj] {10.1088/0004-637X/748/2/93}, \href
  {https://ui.adsabs.harvard.edu/abs/2012ApJ...748...93R} {748, 93}

\bibitem[\protect\citeauthoryear{{Rothman} et~al.,}{{Rothman}
  et~al.}{2013}]{rothman_2013}
{Rothman} L.~S.,  et~al., 2013, \mn@doi [\jqsrt] {10.1016/j.jqsrt.2013.07.002},
  \href {https://ui.adsabs.harvard.edu/abs/2013JQSRT.130....4R} {130, 4}

\bibitem[\protect\citeauthoryear{{Ryabchikova}, {Piskunov}, {Kurucz},
  {Stempels}, {Heiter}, {Pakhomov}  \& {Barklem}}{{Ryabchikova}
  et~al.}{2015}]{ryabchikova_2015}
{Ryabchikova} T.,  {Piskunov} N.,  {Kurucz} R.~L.,  {Stempels} H.~C.,  {Heiter}
  U.,  {Pakhomov} Y.,   {Barklem} P.~S.,  2015, \mn@doi [\physscr]
  {10.1088/0031-8949/90/5/054005}, \href
  {https://ui.adsabs.harvard.edu/abs/2015PhyS...90e4005R} {90, 054005}

\bibitem[\protect\citeauthoryear{{Saar} \& {Linsky}}{{Saar} \&
  {Linsky}}{1985}]{saar_1985}
{Saar} S.~H.,  {Linsky} J.~L.,  1985, \mn@doi [\apjl] {10.1086/184578}, \href
  {https://ui.adsabs.harvard.edu/abs/1985ApJ...299L..47S} {299, L47}

\bibitem[\protect\citeauthoryear{{Shulyak}, {Reiners}, {Wende}, {Kochukhov},
  {Piskunov}  \& {Seifahrt}}{{Shulyak} et~al.}{2010}]{shulyak_2010}
{Shulyak} D.,  {Reiners} A.,  {Wende} S.,  {Kochukhov} O.,  {Piskunov} N.,
  {Seifahrt} A.,  2010, \mn@doi [\aap] {10.1051/0004-6361/201015229}, \href
  {https://ui.adsabs.harvard.edu/abs/2010A&A...523A..37S} {523, A37}

\bibitem[\protect\citeauthoryear{{Shulyak}, {Reiners}, {Seemann}, {Kochukhov}
  \& {Piskunov}}{{Shulyak} et~al.}{2014}]{shulyak_2014}
{Shulyak} D.,  {Reiners} A.,  {Seemann} U.,  {Kochukhov} O.,   {Piskunov} N.,
  2014, \mn@doi [\aap] {10.1051/0004-6361/201322136}, \href
  {https://ui.adsabs.harvard.edu/abs/2014A&A...563A..35S} {563, A35}

\bibitem[\protect\citeauthoryear{Sneden, Lucatello, Ram, Brooke  \&
  Bernath}{Sneden et~al.}{2014}]{sneden_2014}
Sneden C.,  Lucatello S.,  Ram R.~S.,  Brooke J. S.~A.,   Bernath P.,  2014,
  \mn@doi [The Astrophysical Journal Supplement Series]
  {10.1088/0067-0049/214/2/26}, 214, 26

\bibitem[\protect\citeauthoryear{{Stift}}{{Stift}}{1985}]{stift_1985}
{Stift} M.~J.,  1985, \mn@doi [\mnras] {10.1093/mnras/217.1.55}, \href
  {https://ui.adsabs.harvard.edu/abs/1985MNRAS.217...55S} {217, 55}

\bibitem[\protect\citeauthoryear{{Stift} \& {Alecian}}{{Stift} \&
  {Alecian}}{2016}]{stift_2016}
{Stift} M.~J.,  {Alecian} G.,  2016, \mn@doi [\mnras] {10.1093/mnras/stv2962},
  \href {https://ui.adsabs.harvard.edu/abs/2016MNRAS.457...74S} {457, 74}

\bibitem[\protect\citeauthoryear{{Stift} \& {Leone}}{{Stift} \&
  {Leone}}{2003}]{stift_2003}
{Stift} M.~J.,  {Leone} F.,  2003, \mn@doi [\aap] {10.1051/0004-6361:20021605},
  \href {https://ui.adsabs.harvard.edu/abs/2003A&A...398..411S} {398, 411}

\bibitem[\protect\citeauthoryear{{Su{\'a}rez Mascare{\~n}o}
  et~al.,}{{Su{\'a}rez Mascare{\~n}o} et~al.}{2020}]{suarez_2020}
{Su{\'a}rez Mascare{\~n}o} A.,  et~al., 2020, \mn@doi [\aap]
  {10.1051/0004-6361/202037745}, \href
  {https://ui.adsabs.harvard.edu/abs/2020A&A...639A..77S} {639, A77}

\bibitem[\protect\citeauthoryear{{Tennyson} et~al.,}{{Tennyson}
  et~al.}{2020}]{exomol_2020}
{Tennyson} J.,  et~al., 2020, \mn@doi [\jqsrt] {10.1016/j.jqsrt.2020.107228},
  \href {https://ui.adsabs.harvard.edu/abs/2020JQSRT.25507228T} {255, 107228}

\bibitem[\protect\citeauthoryear{{Valenti} \& {Fischer}}{{Valenti} \&
  {Fischer}}{2005}]{valenti_2005}
{Valenti} J.~A.,  {Fischer} D.~A.,  2005, \mn@doi [\apjs] {10.1086/430500},
  \href {https://ui.adsabs.harvard.edu/abs/2005ApJS..159..141V} {159, 141}

\bibitem[\protect\citeauthoryear{{Valenti}, {Marcy}  \& {Basri}}{{Valenti}
  et~al.}{1995}]{valenti_1995}
{Valenti} J.~A.,  {Marcy} G.~W.,   {Basri} G.,  1995, \mn@doi [\apj]
  {10.1086/175231}, \href
  {https://ui.adsabs.harvard.edu/abs/1995ApJ...439..939V} {439, 939}

\bibitem[\protect\citeauthoryear{{Valyavin}, {Kochukhov}  \&
  {Piskunov}}{{Valyavin} et~al.}{2004}]{valyavin_2004}
{Valyavin} G.,  {Kochukhov} O.,   {Piskunov} N.,  2004, \mn@doi [\aap]
  {10.1051/0004-6361:20034345}, \href
  {https://ui.adsabs.harvard.edu/abs/2004A&A...420..993V} {420, 993}

\bibitem[\protect\citeauthoryear{{Wade}, {Bagnulo}, {Kochukhov}, {Landstreet},
  {Piskunov}  \& {Stift}}{{Wade} et~al.}{2001}]{wade_2001}
{Wade} G.~A.,  {Bagnulo} S.,  {Kochukhov} O.,  {Landstreet} J.~D.,  {Piskunov}
  N.,   {Stift} M.~J.,  2001, \mn@doi [\aap] {10.1051/0004-6361:20010735},
  \href {https://ui.adsabs.harvard.edu/abs/2001A&A...374..265W} {374, 265}

\bibitem[\protect\citeauthoryear{{Wenger} et~al.,}{{Wenger}
  et~al.}{2000}]{wenger_2000}
{Wenger} M.,  et~al., 2000, \mn@doi [\aaps] {10.1051/aas:2000332}, \href
  {https://ui.adsabs.harvard.edu/abs/2000A&AS..143....9W} {143, 9}

\bibitem[\protect\citeauthoryear{{Yadin}, {Veness}, {Conti}, {Hill},
  {Yurchenko}  \& {Tennyson}}{{Yadin} et~al.}{2012}]{yadin_2012}
{Yadin} B.,  {Veness} T.,  {Conti} P.,  {Hill} C.,  {Yurchenko} S.~N.,
  {Tennyson} J.,  2012, \mn@doi [\mnras] {10.1111/j.1365-2966.2012.21367.x},
  \href {https://ui.adsabs.harvard.edu/abs/2012MNRAS.425...34Y} {425, 34}

\bibitem[\protect\citeauthoryear{{Yurchenko}, {Sinden}, {Lodi}, {Hill},
  {Gorman}  \& {Tennyson}}{{Yurchenko} et~al.}{2018}]{yurchenko_2018}
{Yurchenko} S.~N.,  {Sinden} F.,  {Lodi} L.,  {Hill} C.,  {Gorman} M.~N.,
  {Tennyson} J.,  2018, \mn@doi [\mnras] {10.1093/mnras/stx2738}, \href
  {https://ui.adsabs.harvard.edu/abs/2018MNRAS.473.5324Y} {473, 5324}

\makeatother
\end{thebibliography}




\appendix

\section{Additional figures}

\begin{figure*}
    \centering
    \includegraphics[scale=.55]{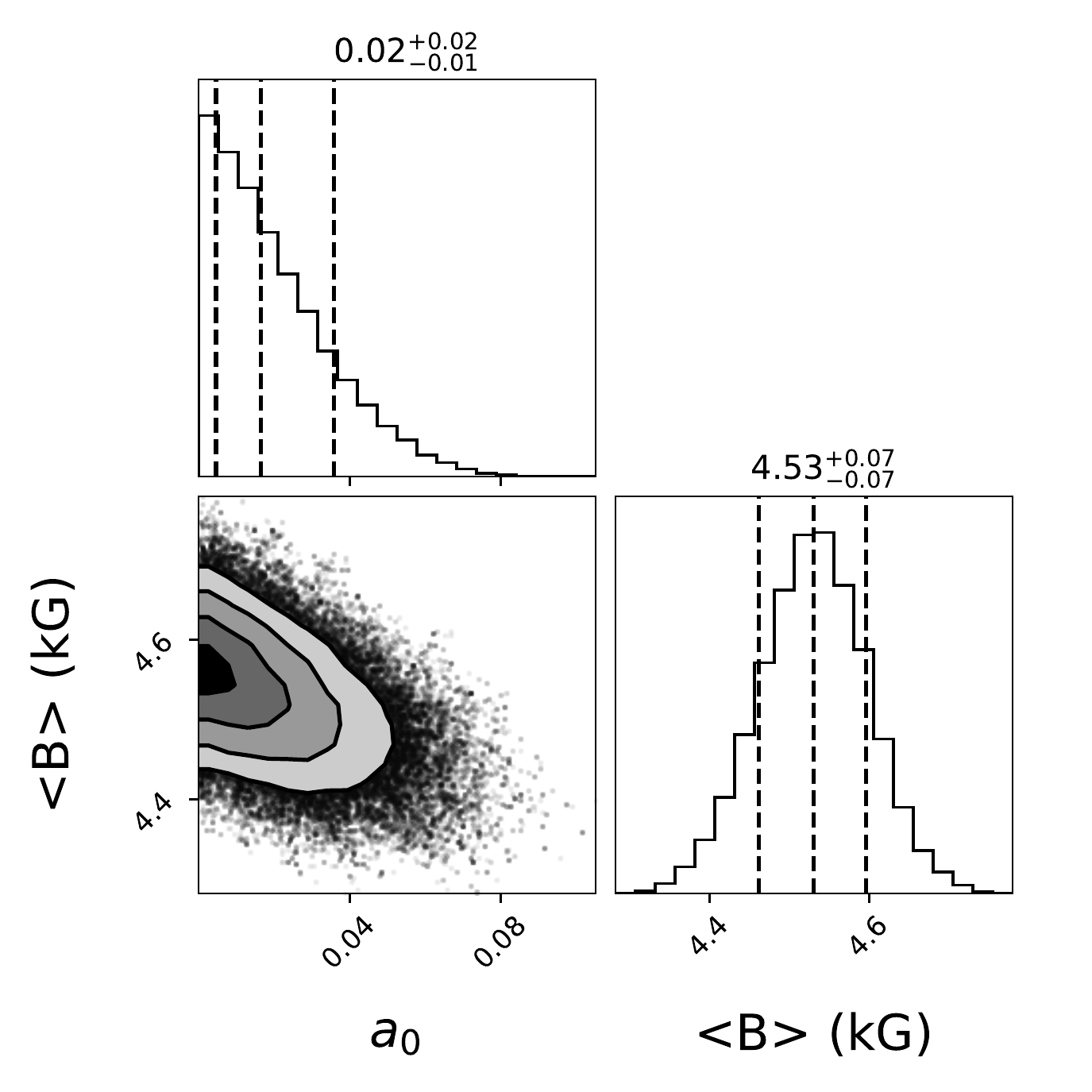}
    \includegraphics[scale=.5]{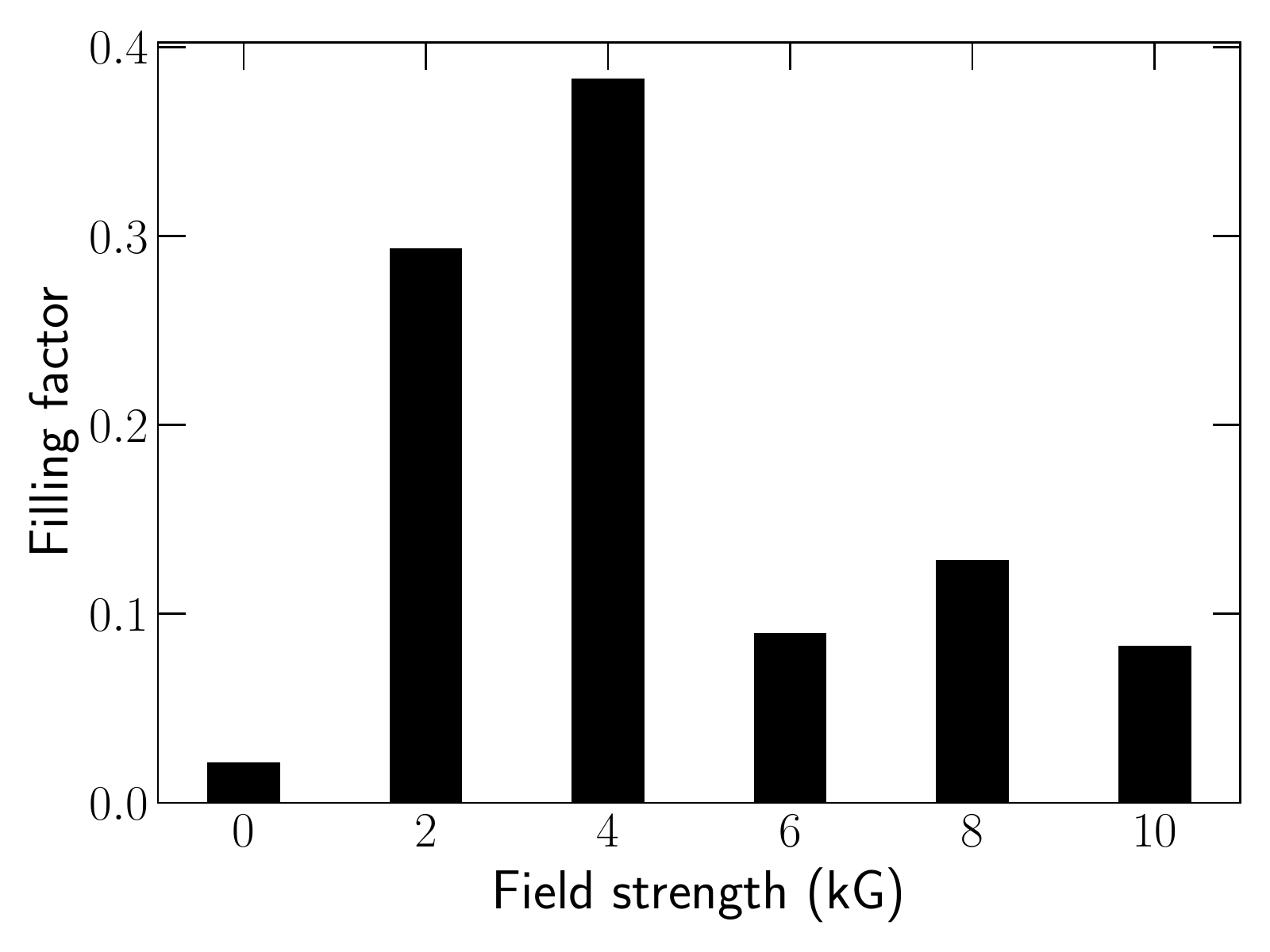}
    \caption{Same as~\ref{fig:b_a0_aumic} for EV Lac.}
    \label{fig:b_a0_evlac}
\end{figure*}

\begin{figure*}
    \centering
    \includegraphics[scale=.55]{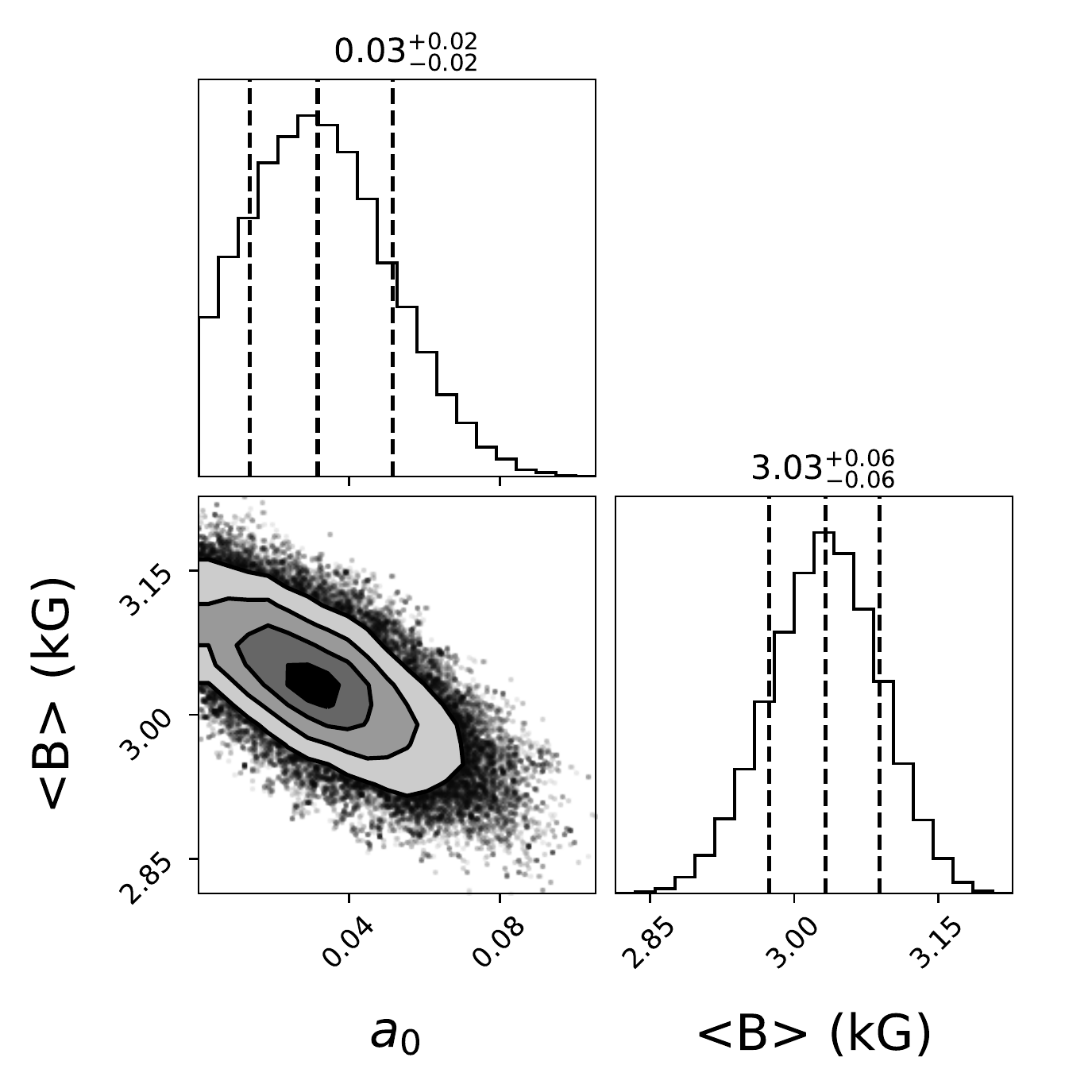}
    \includegraphics[scale=.5]{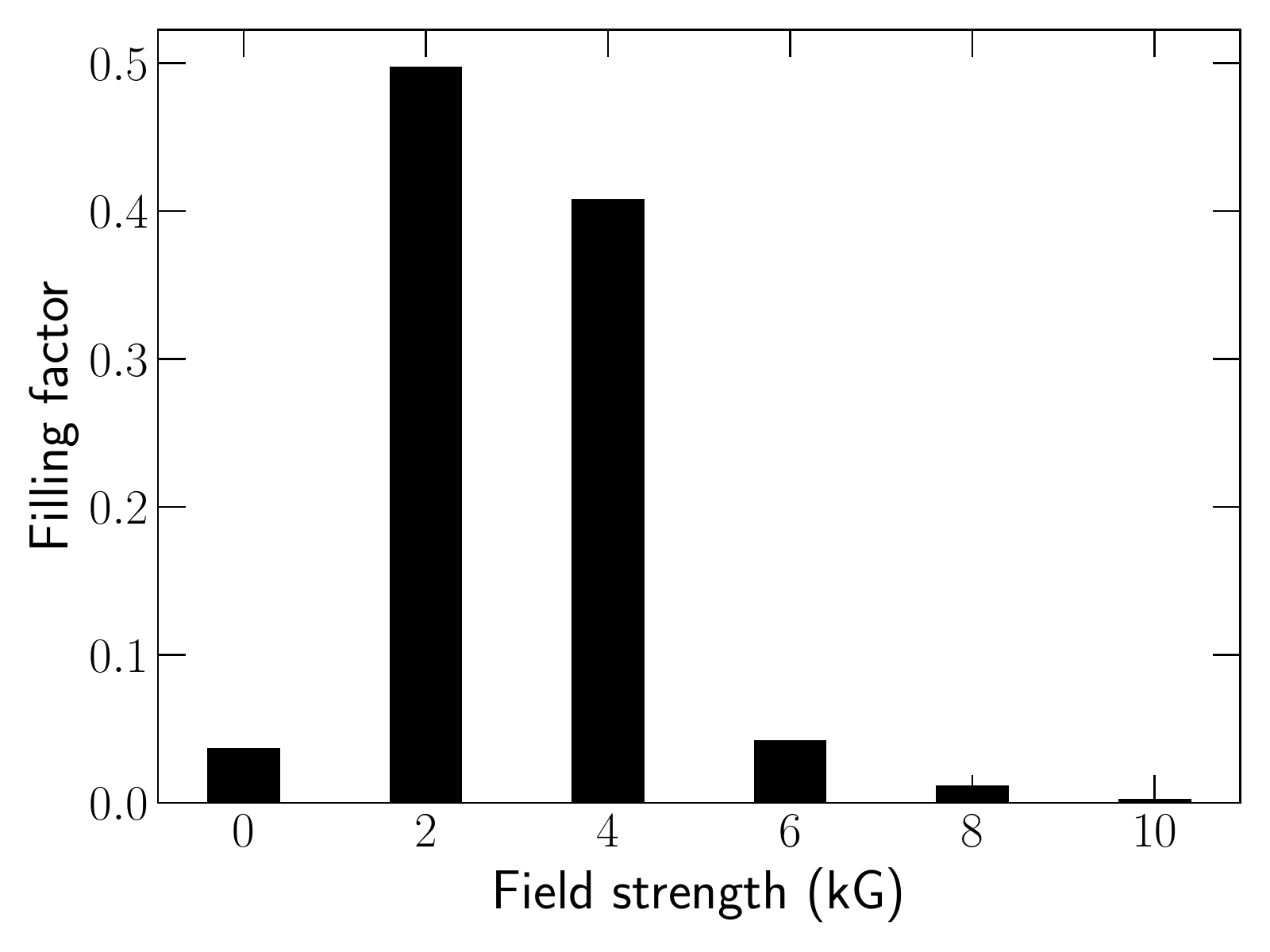}
    \caption{Same as~\ref{fig:b_a0_aumic} for AD Leo.}
    \label{fig:b_a0_adleo}
\end{figure*}

\begin{figure*}
	\centering
	\includegraphics[scale=.55]{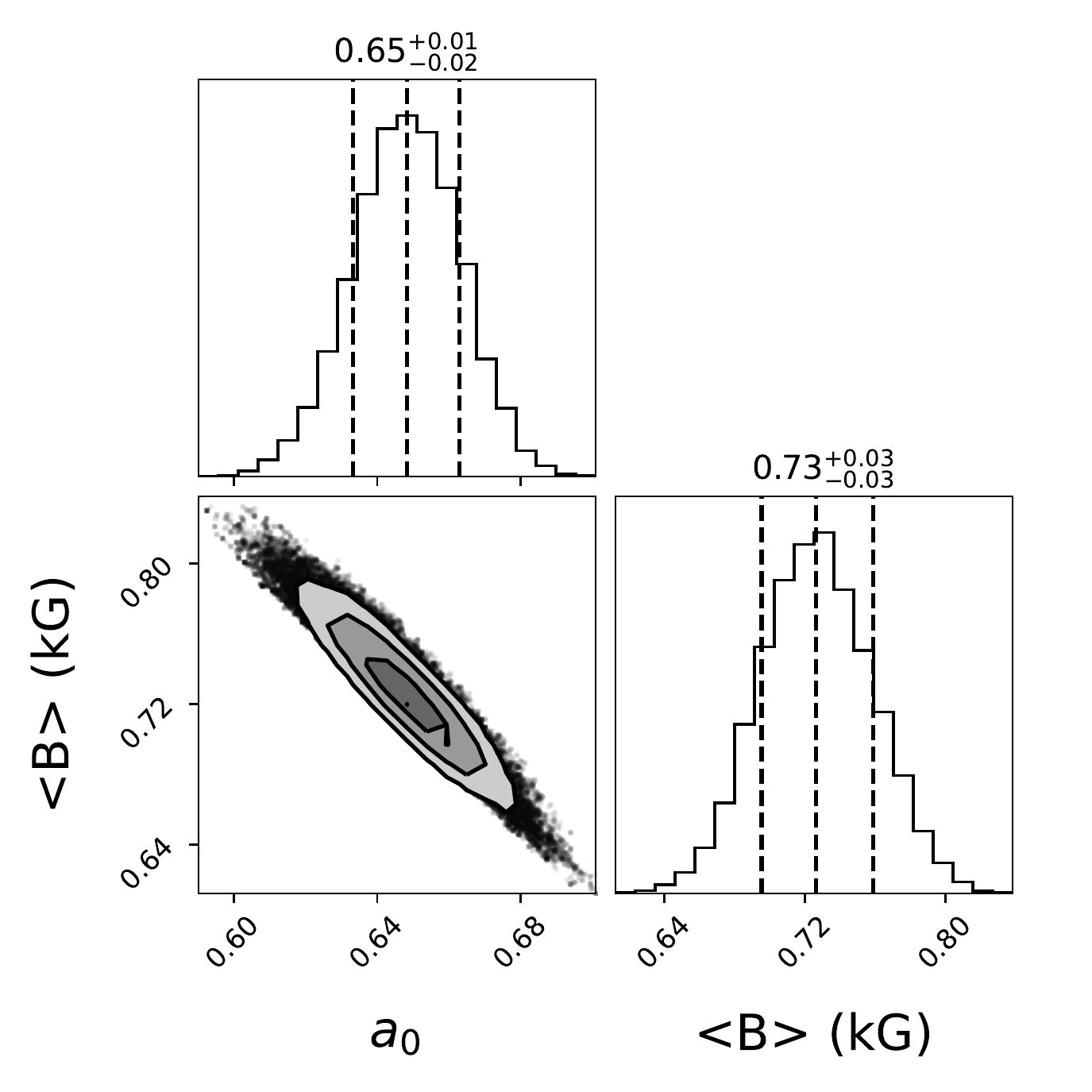}
	\includegraphics[scale=.5]{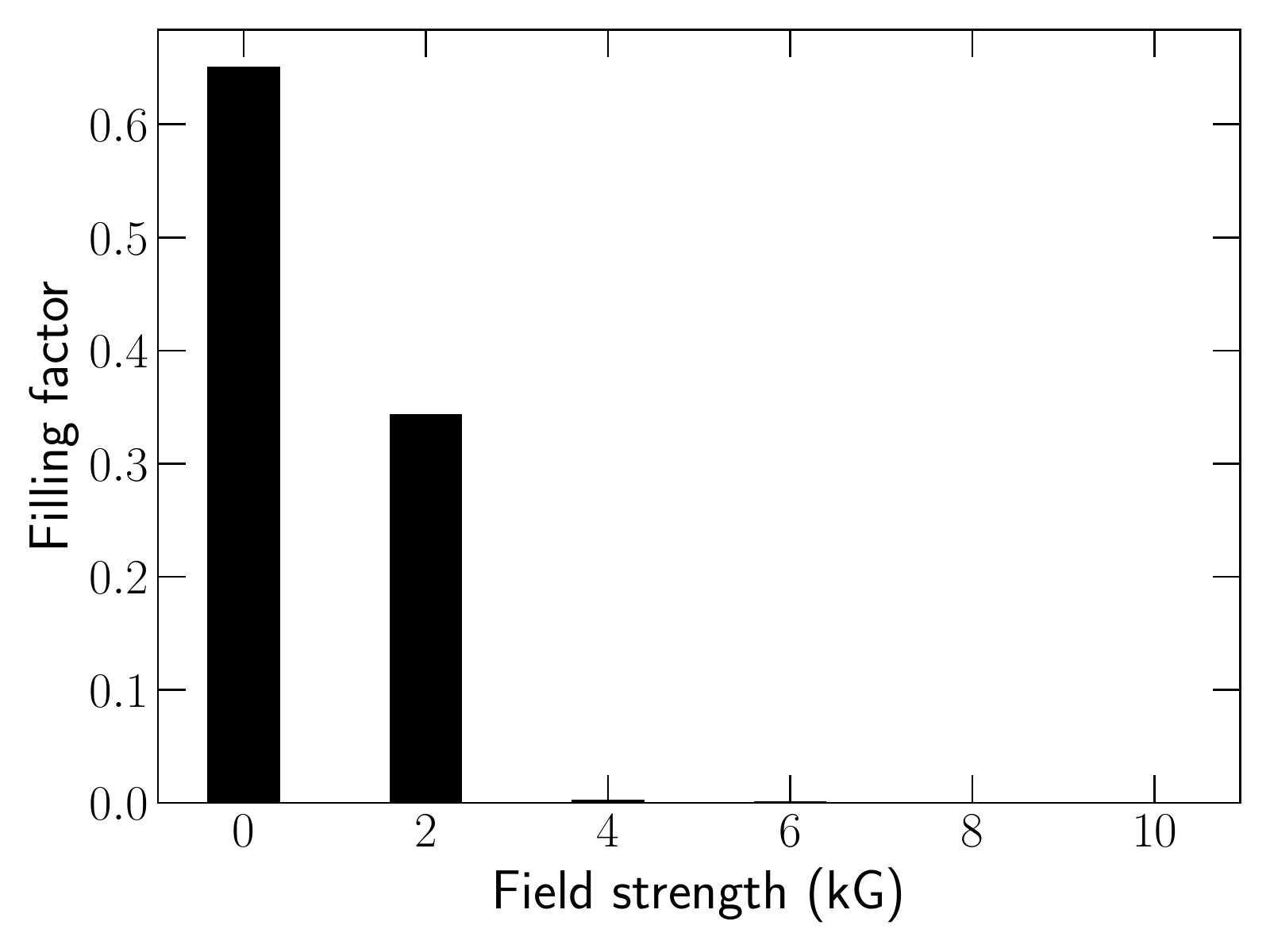}
	\caption{Same as~\ref{fig:b_a0_aumic} for DS Leo.}
	\label{fig:b_a0_dsleo}
\end{figure*}

Figures~\ref{fig:b_a0_evlac},~\ref{fig:b_a0_adleo} and~\ref{fig:b_a0_dsleo} present the same plots as Fig.~\ref{fig:b_a0_aumic} for EV Lac, AD Leo, DS~Leo, CN~Leo and PM~J18482+0741 respectively.

\begin{figure*}
    \centering
    \includegraphics[scale=.55]{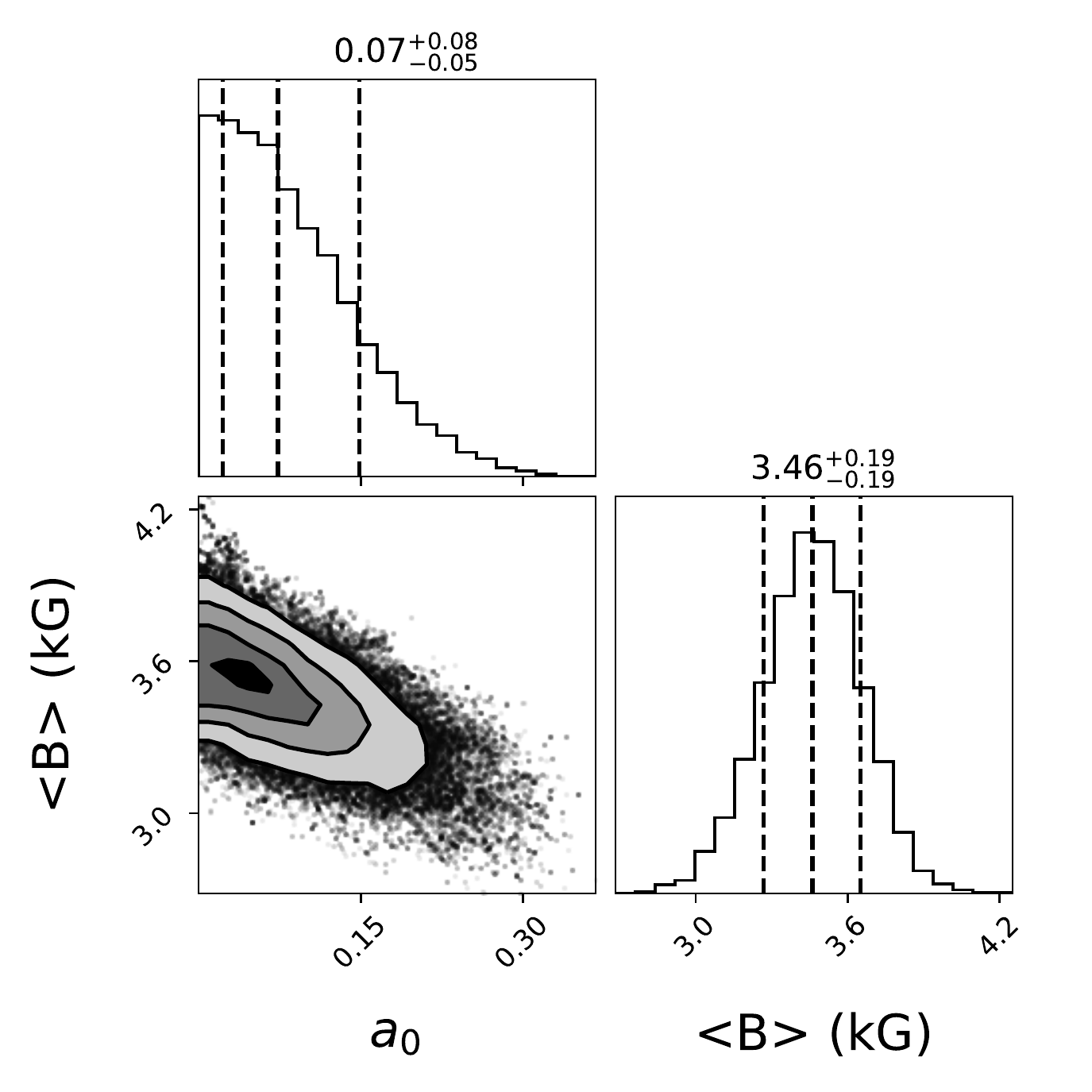}
    \includegraphics[scale=.5]{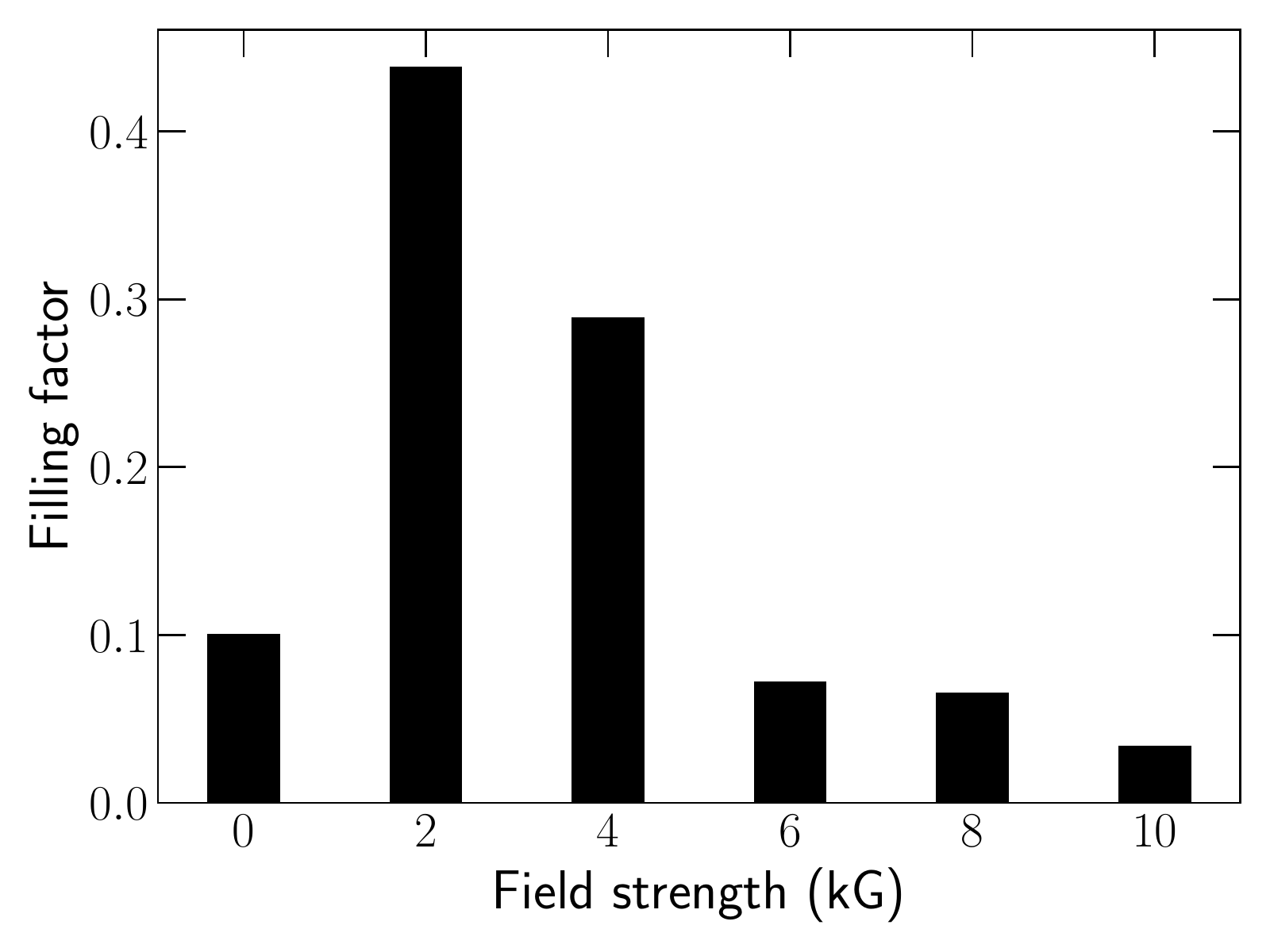}
    \caption{Same as~\ref{fig:b_a0_aumic} for CN Leo.}
    \label{fig:b_a0_cnleo}
\end{figure*}

\begin{figure*}
	\centering
	\includegraphics[scale=.55]{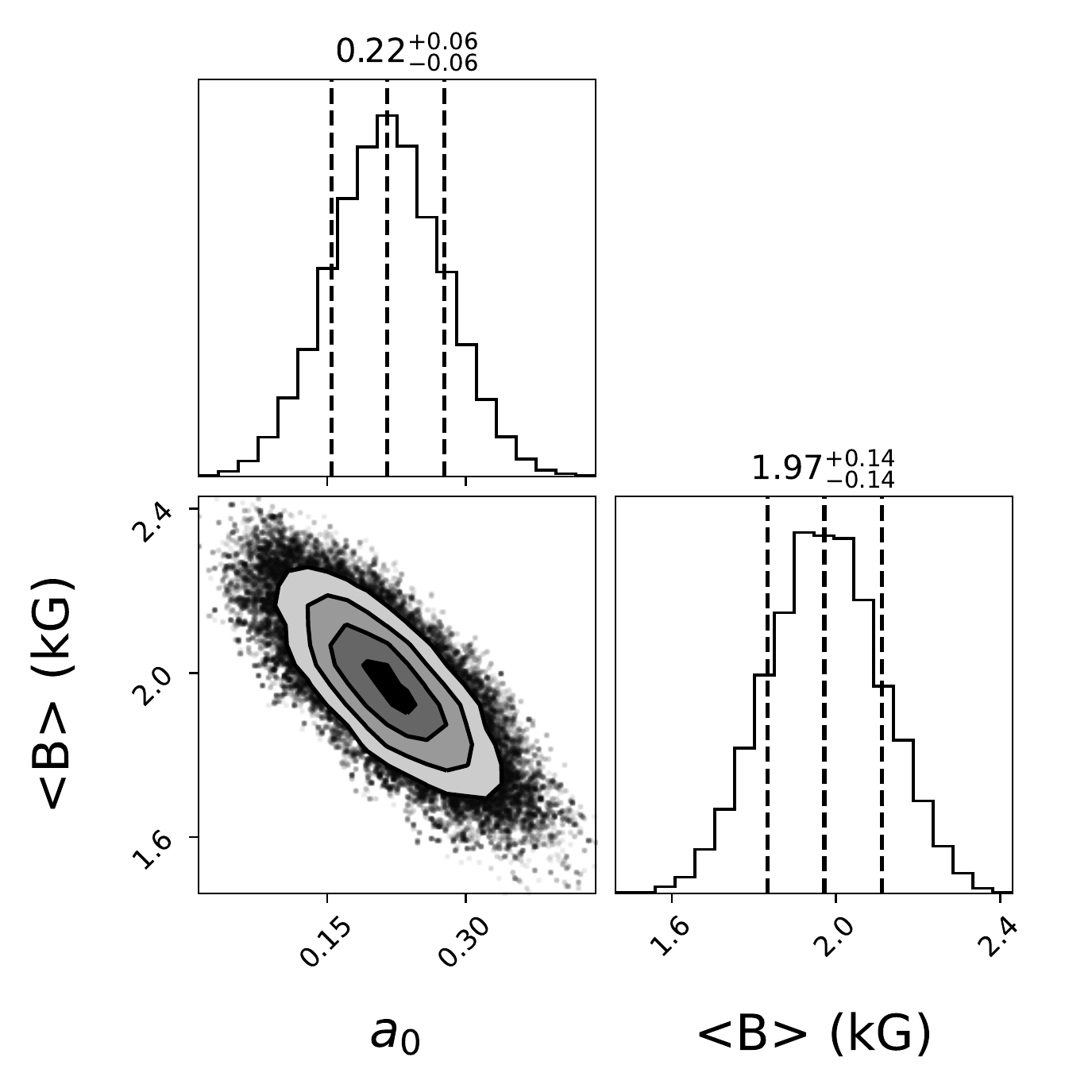}
	\includegraphics[scale=.5]{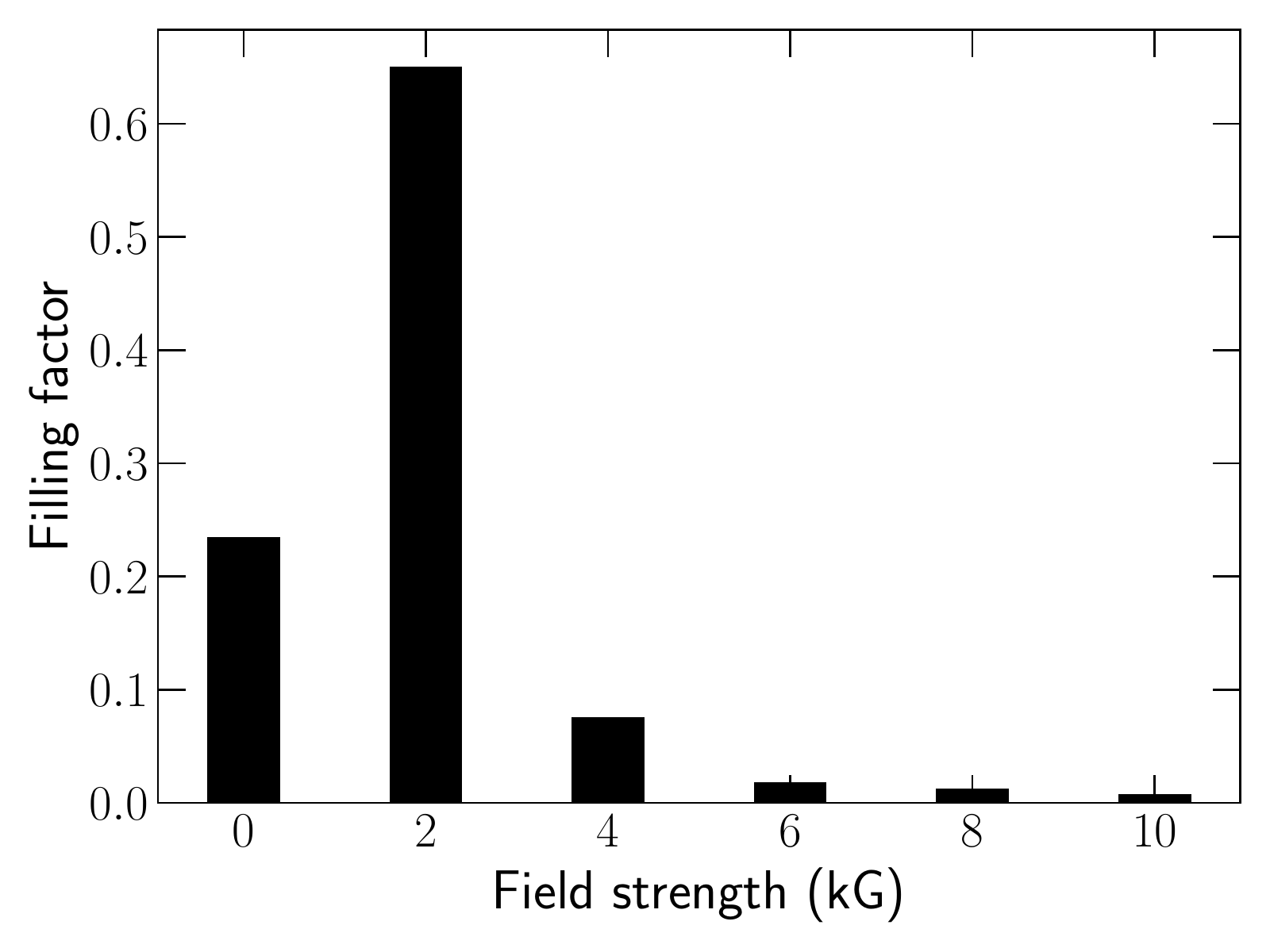}
	\caption{Same as~\ref{fig:b_a0_aumic} for PM~J18482+0741.}
	\label{fig:b_a0_pm}
\end{figure*}

\section{Best fits}

Figure~B1 available as supplementary material presents the best fits obtained for AU Mic, AD Leo, EV Lac and DS Leo for all lines used in our analysis.

{\paul 
\section{Corner plots}

Figures~C1-C14, available as supplementary material, present the corner plots obtained for the 6 stars in our sample.
}

\section{Comparison of $\teff$ and $\logg$ with the literature}

\begin{figure}
	\includegraphics[width=\columnwidth]{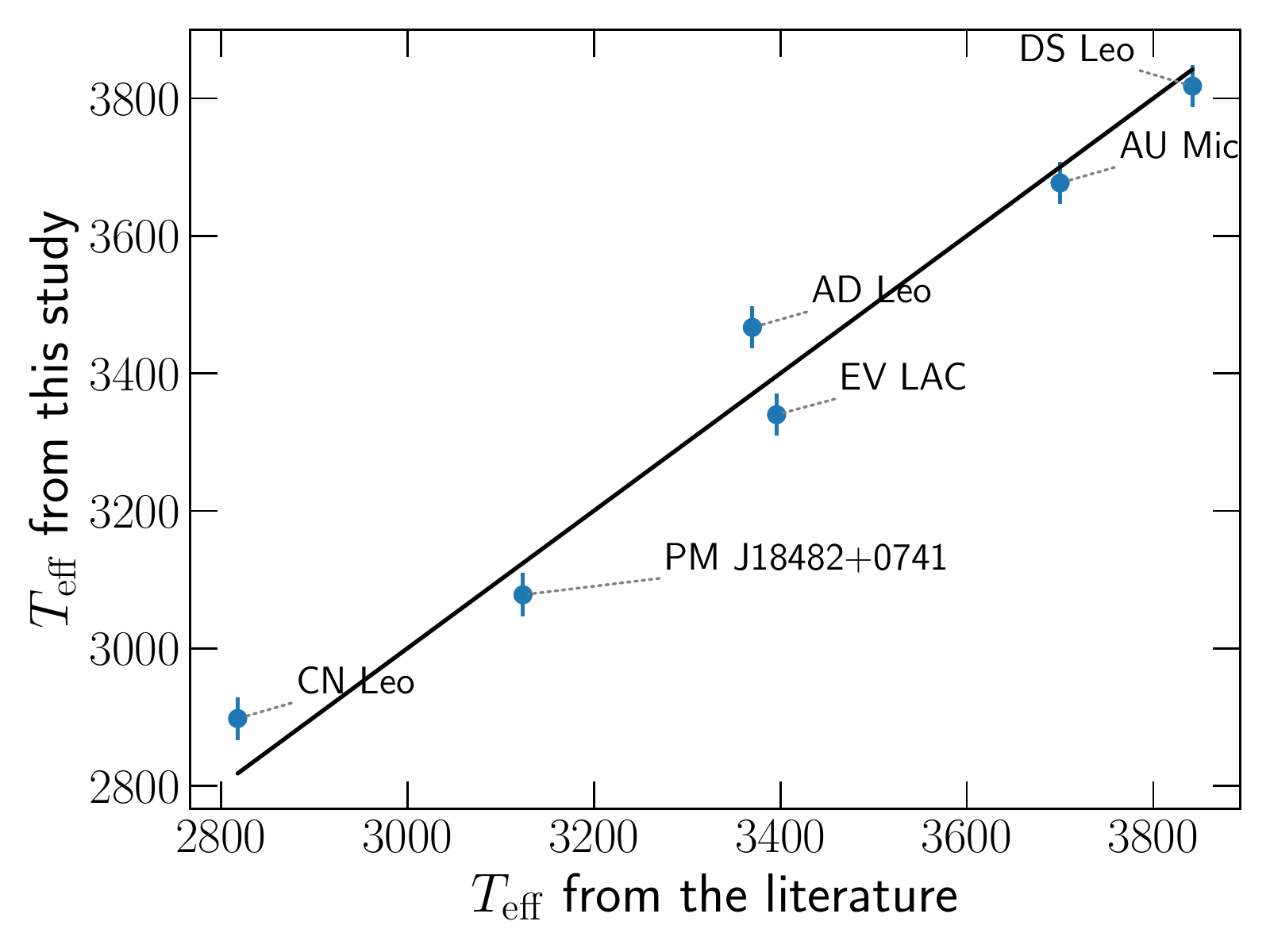}
	\caption{Retrieved $\teff$ compared to literature estimates taken from~\citet[][AU~Mic]{alfram_2019}, \citet[][AD~Leo, CN~Leo]{mann_2015}, \citet[][EV~Lac]{maldonado_2020},~\citet[][DS~Leo]{cristofari_2022b} and~\citet[][PM~J18482+0741]{passegger_2019}.}
	\label{fig:comparison_lit_teffs}
\end{figure}

\begin{figure}
	\includegraphics[width=\columnwidth]{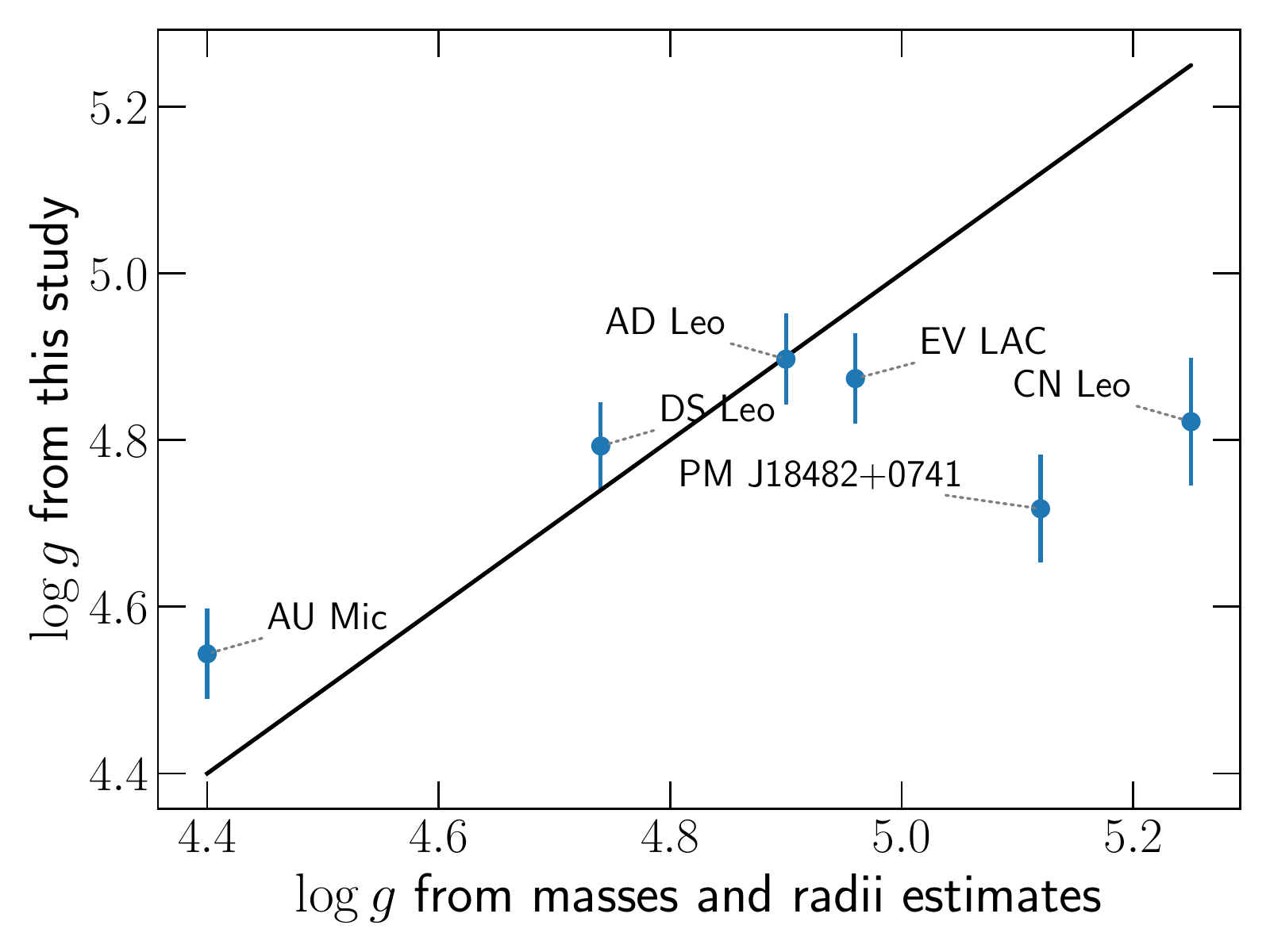}
	\caption{Retrieved $\logg$ compared to estimates derived from masses and radii (see Table~\ref{tab:literature_params}).}
	\label{fig:comparison_lit_loggs}
\end{figure}





\bsp	
\label{lastpage}
\end{document}


\maketitle

\section*{APPENDIX B: BEST FITS}
Figure~B1 presents the best fits obtained for AU Mic, AD Leo, EV Lac, CN Leo, PM~J18482+0741 and DS Leo for all lines used in our analysis.

\begin{figure*}
    \centering
    \includegraphics[scale=.47]{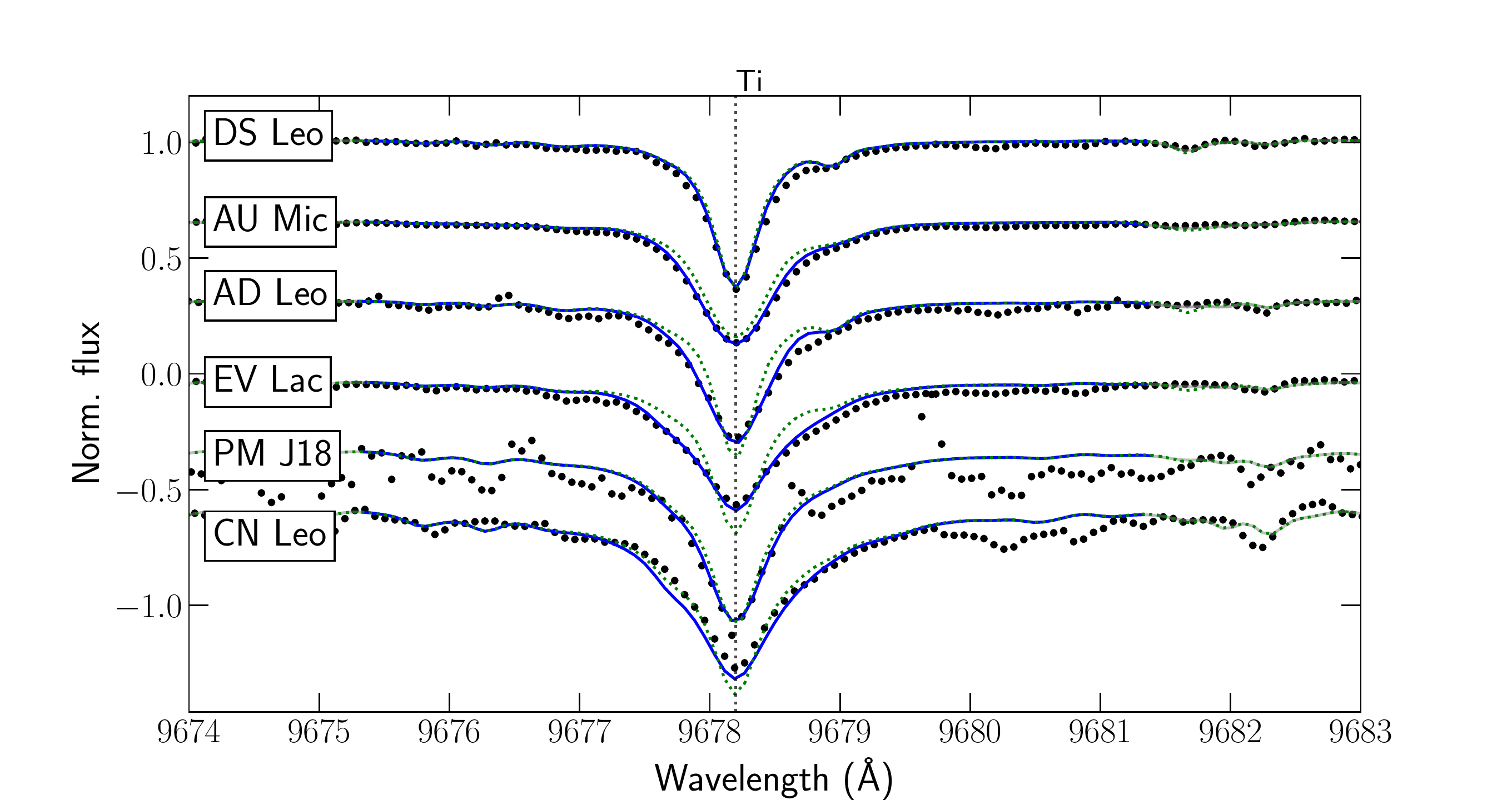}
    \caption*{\textbf{Figure B1.} Best fit obtained for the seven stars included in our study with \texttt{ZeeTurbo}. Black points present the data. The grey solid line shows the best fit, and the blue solid blue line presents the part of the windows used for the fit. The green dotted line shows the model obtained for the same atmospheric parameters but with a zero magnetic field. The name PM~J18482+0741 was replaced by PM~J18 for better readability.}
    \label{fig:example_lines_annex}
\end{figure*}

\begin{figure*}
    \centering
    \includegraphics[scale=.47]{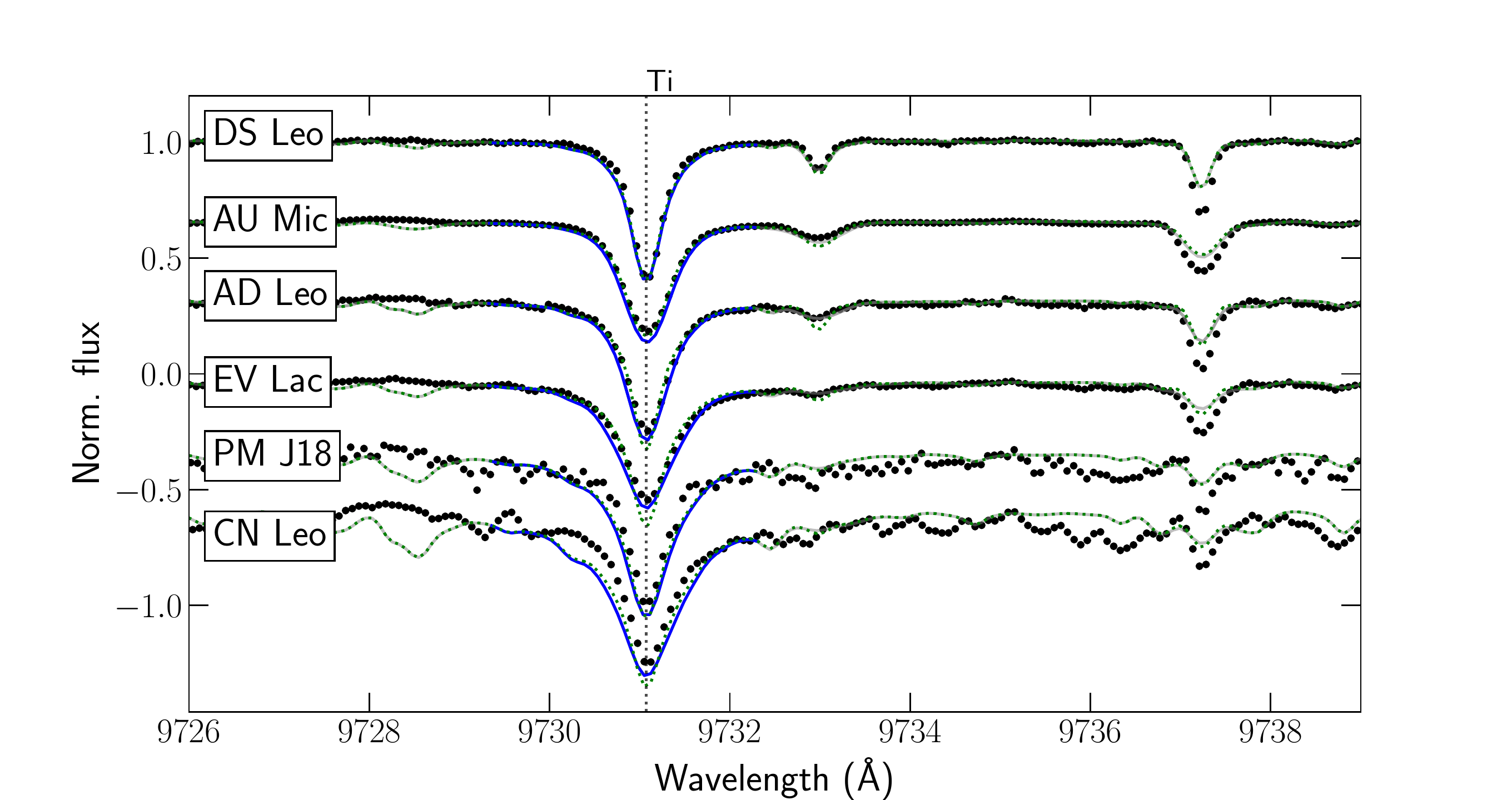}
    \caption*{\textbf{Figure B1} -- \textit{continued}}
\end{figure*}


\begin{figure*}
    \centering
    \includegraphics[scale=.47]{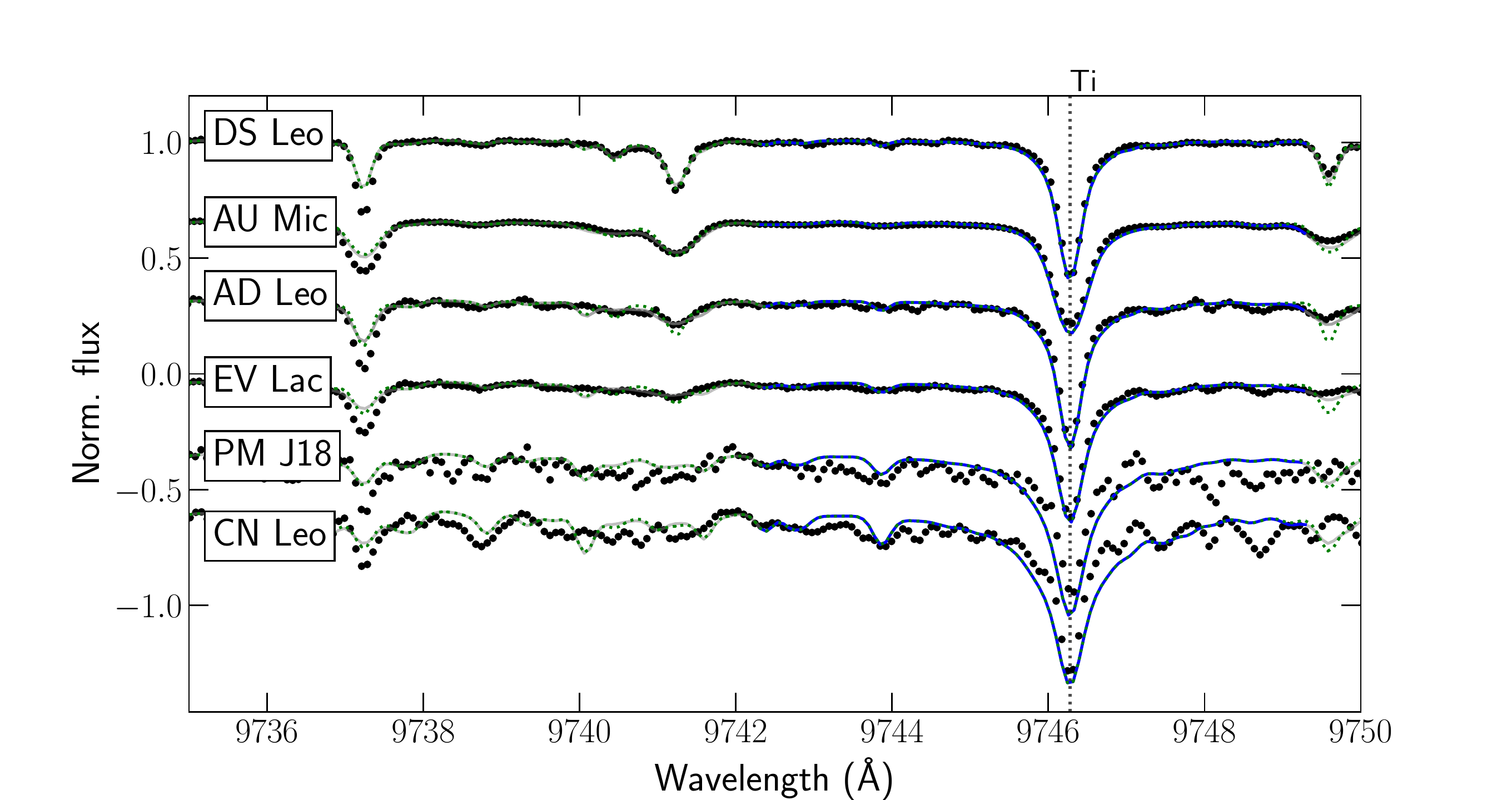}
    \caption*{\textbf{Figure B1} -- \textit{continued}}
\end{figure*}

\begin{figure*}
    \centering
    \includegraphics[scale=.47]{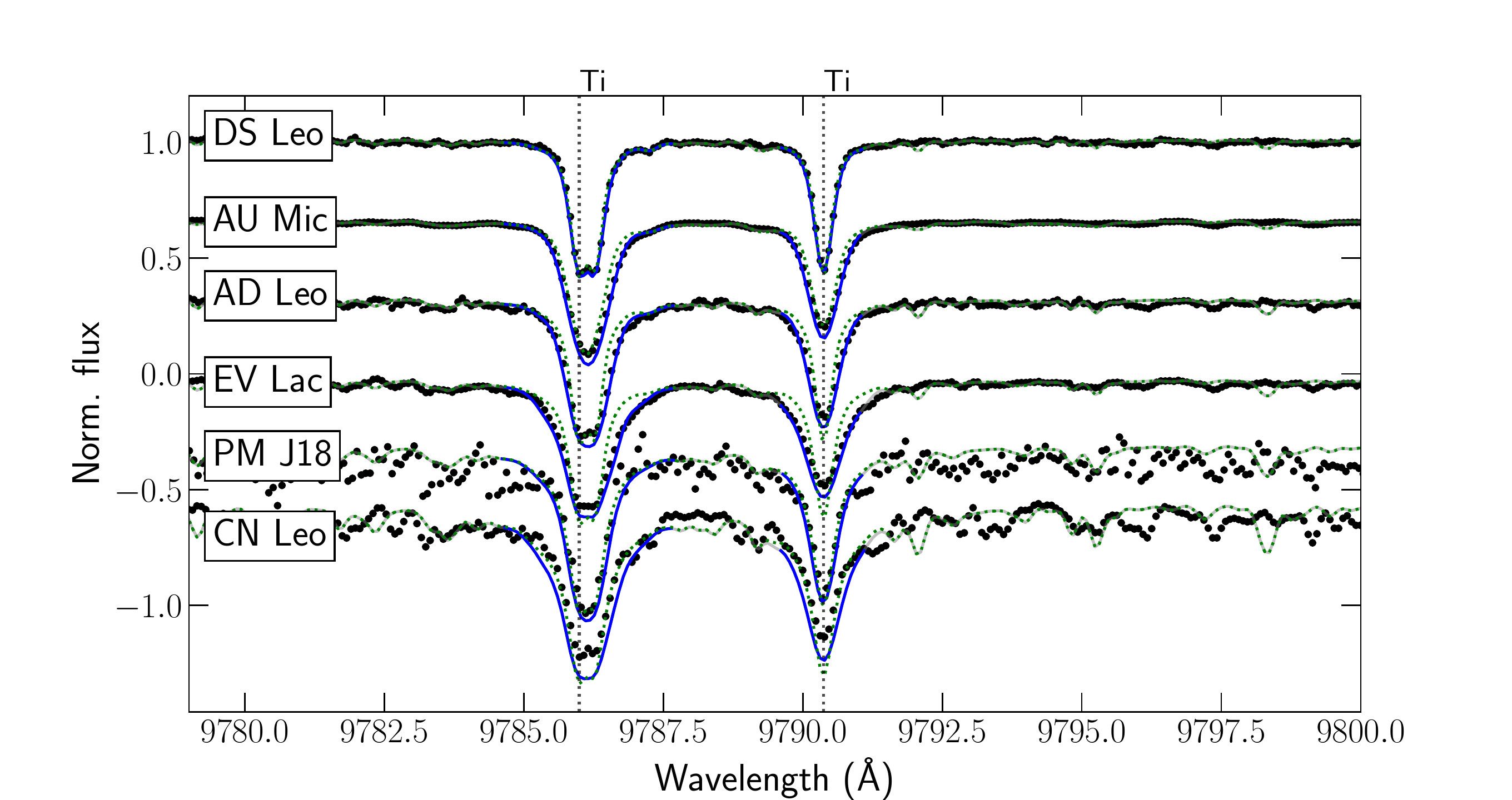}
    \caption*{\textbf{Figure B1} -- \textit{continued}}
\end{figure*}

\begin{figure*}
    \centering
    \includegraphics[scale=.47]{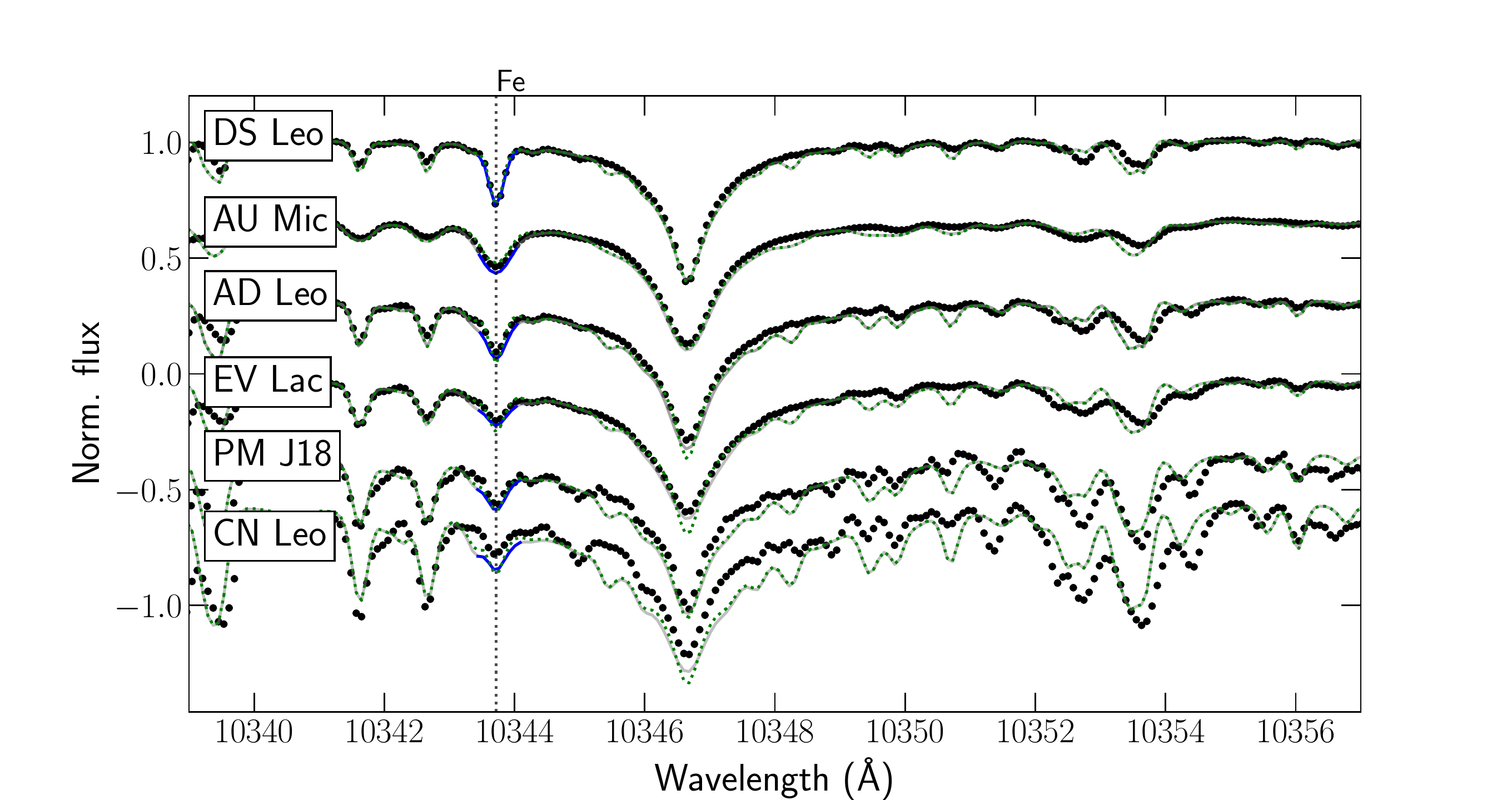}
    \caption*{\textbf{Figure B1} -- \textit{continued}}
\end{figure*}

\begin{figure*}
	\centering
	\includegraphics[scale=.47]{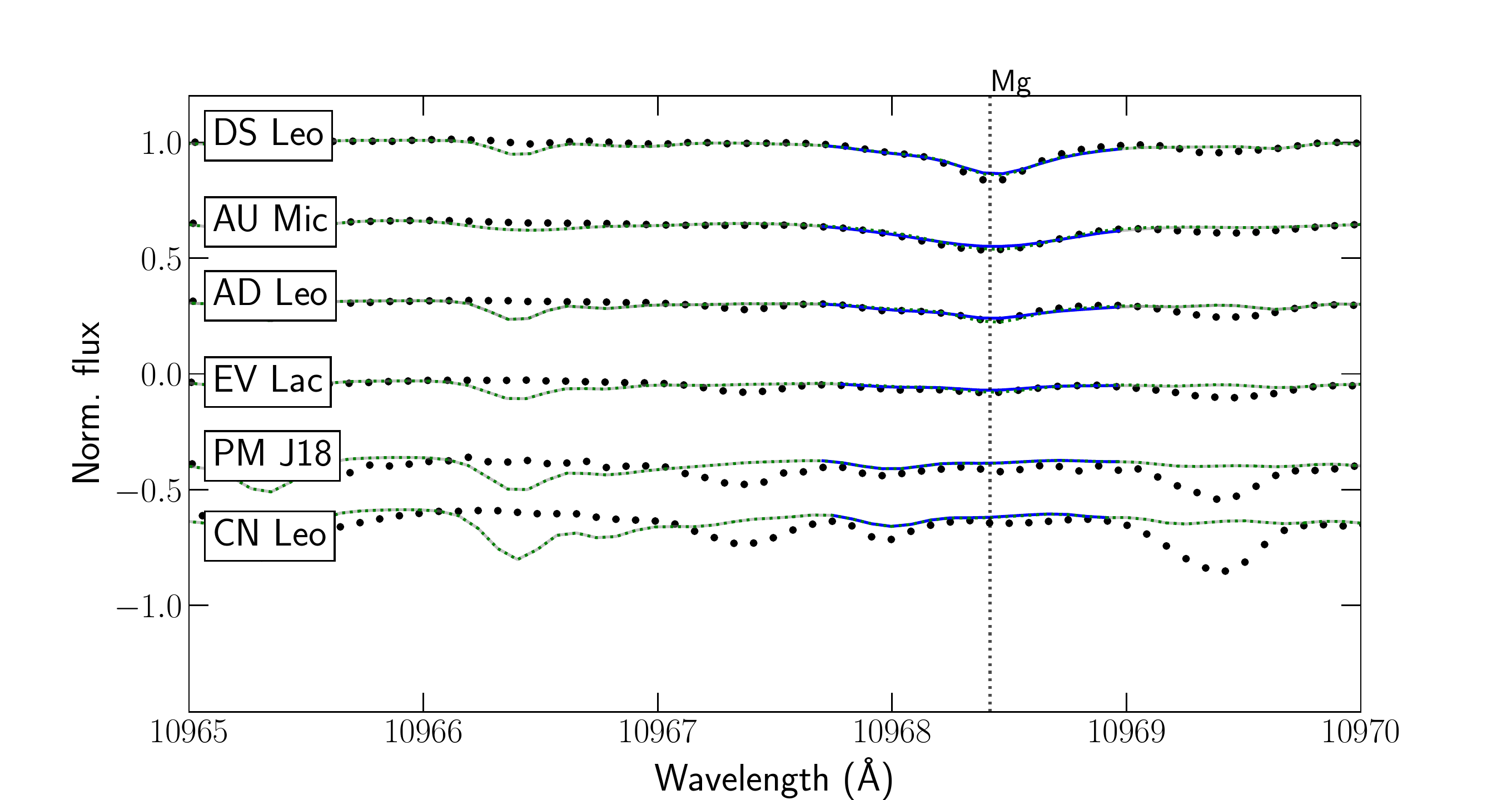}
    \caption*{\textbf{Figure B1} -- \textit{continued}}
\end{figure*}

\begin{figure*}
    \centering
    \includegraphics[scale=.47]{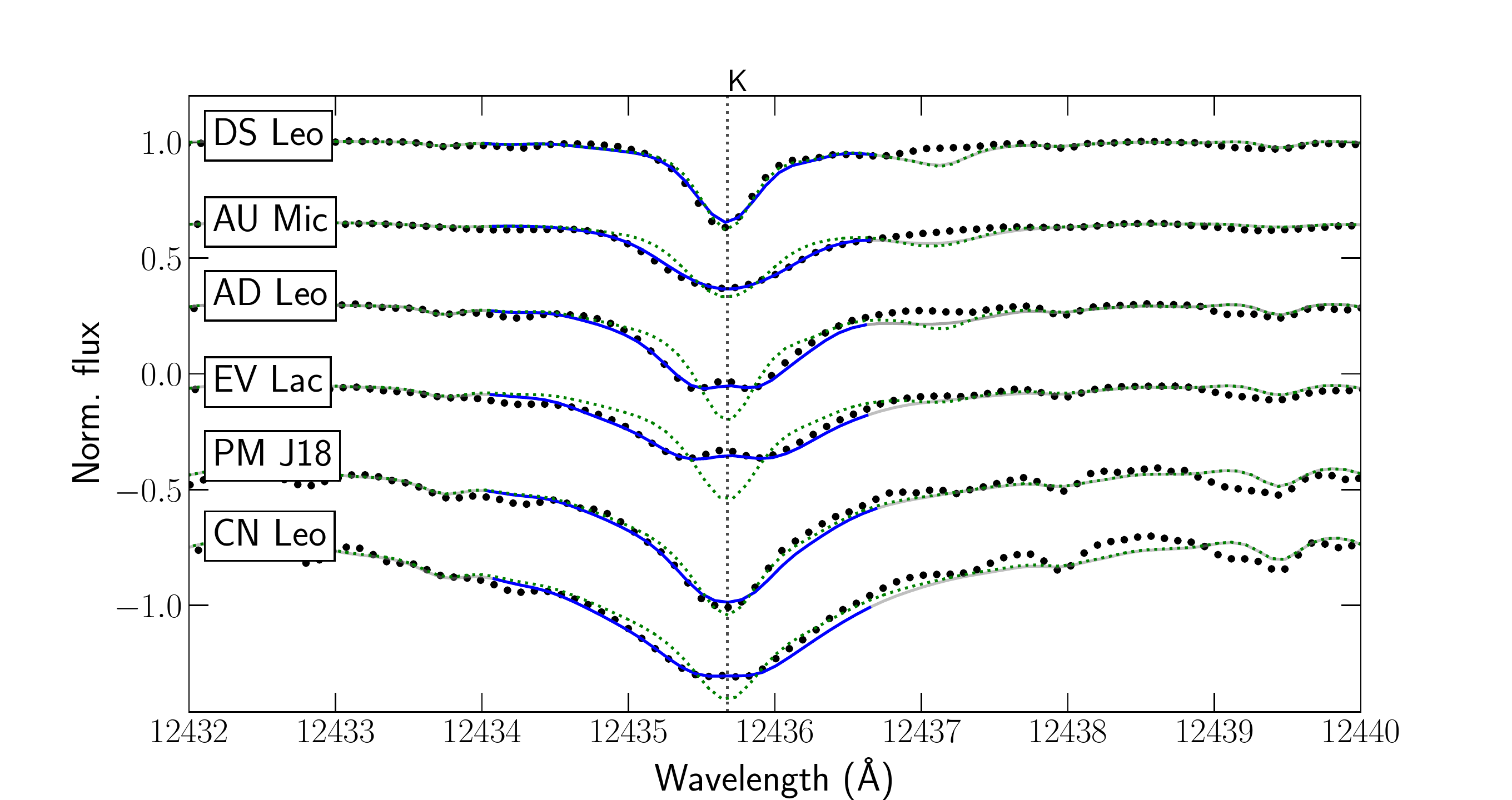}
    \caption*{\textbf{Figure B1} -- \textit{continued}}
\end{figure*}

\begin{figure*}
    \centering
    \includegraphics[scale=.47]{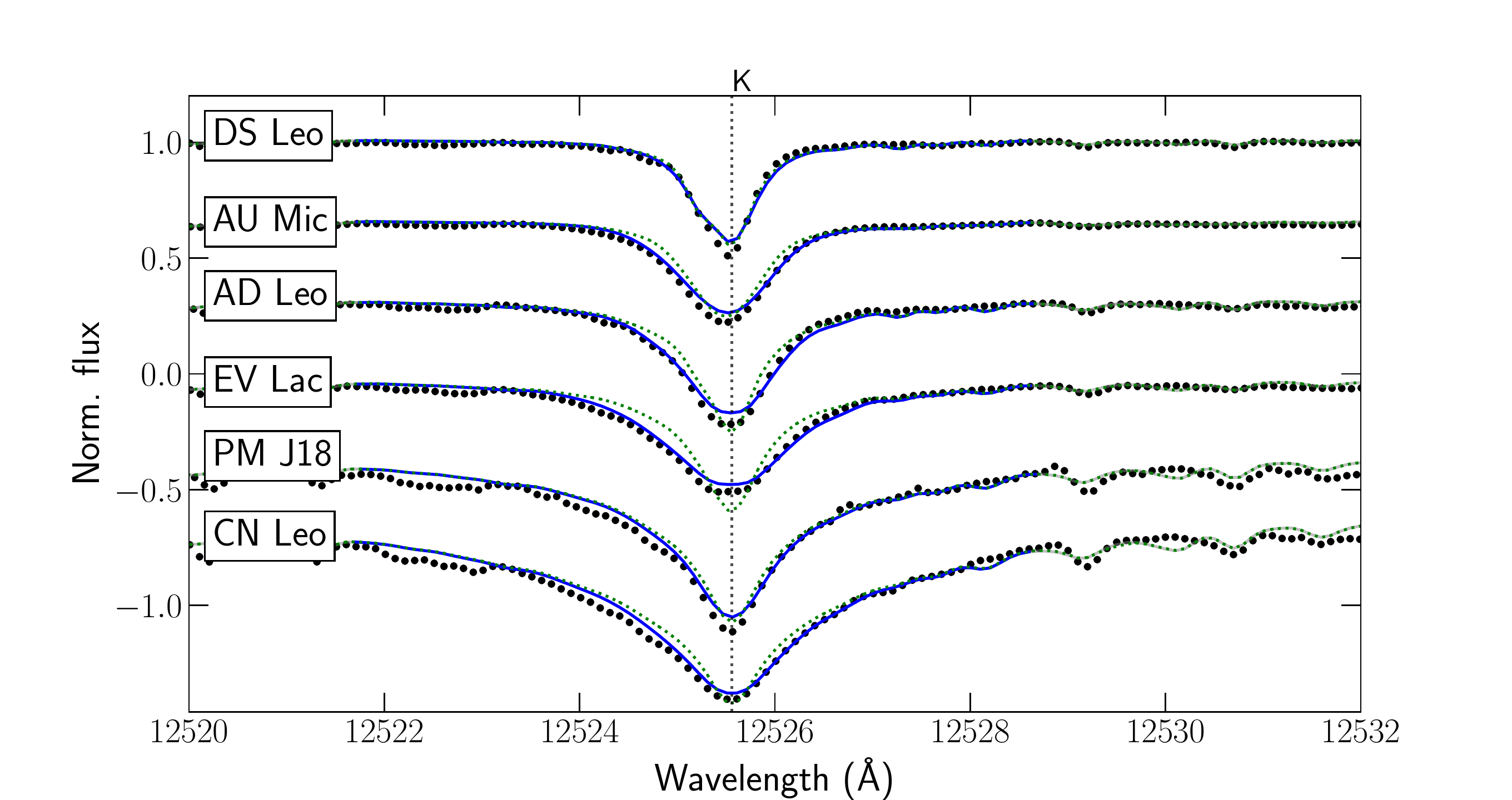}
    \caption*{\textbf{Figure B1} -- \textit{continued}}
\end{figure*}

\begin{figure*}
    \centering
    \includegraphics[scale=.47]{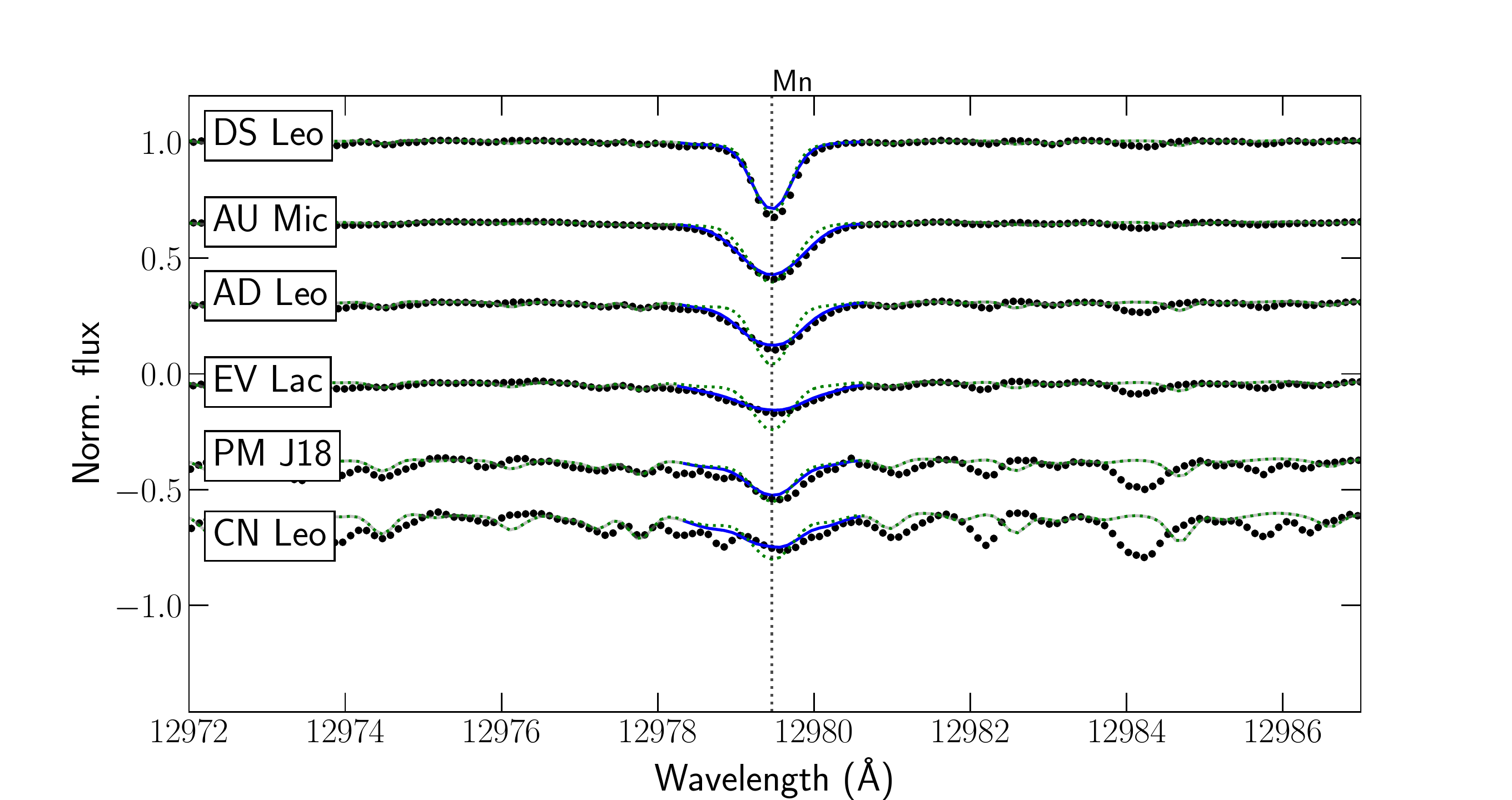}
    \caption*{\textbf{Figure B1} -- \textit{continued}}
\end{figure*}

\begin{figure*}
    \centering
    \includegraphics[scale=.47]{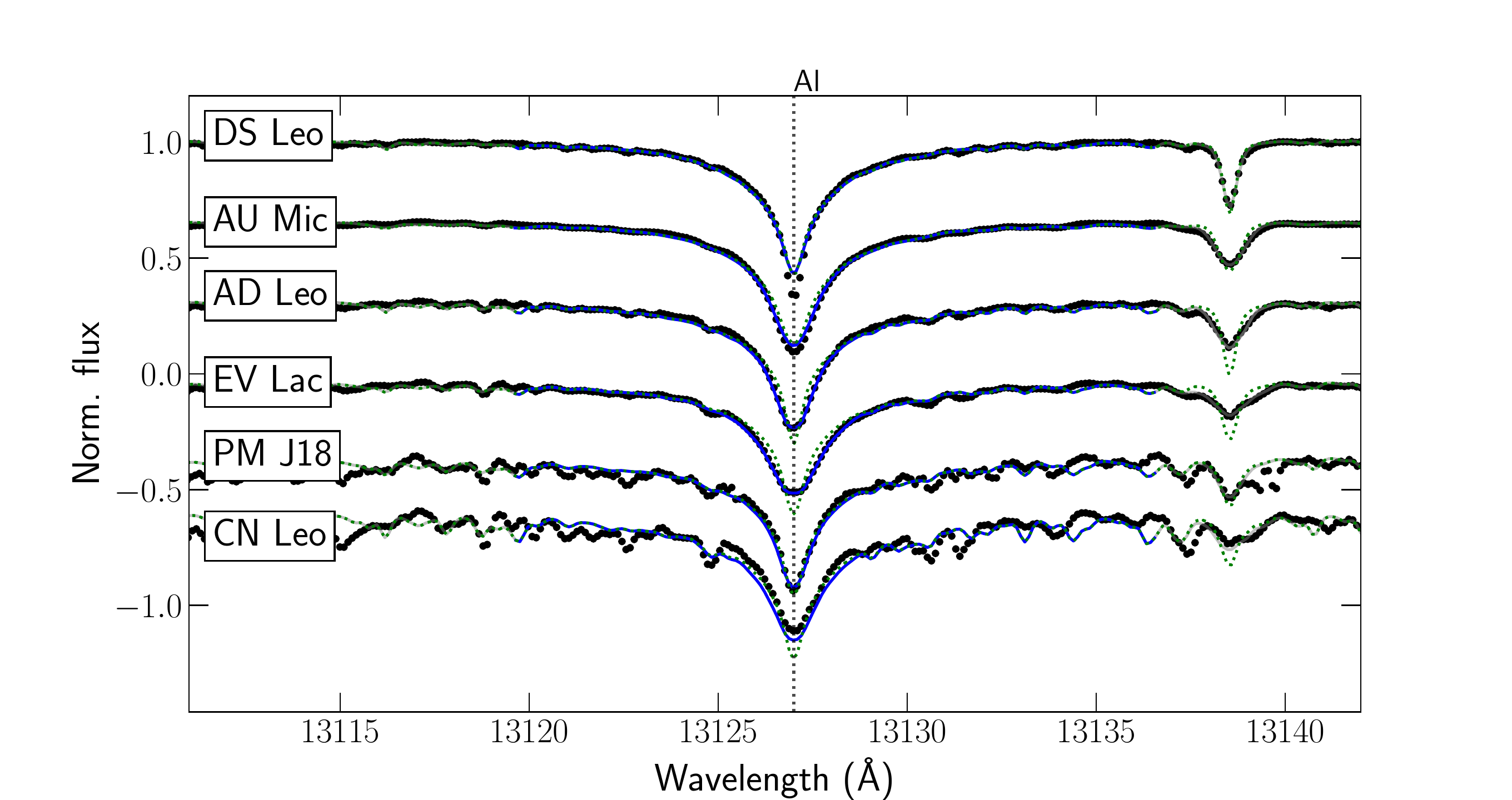}
    \caption*{\textbf{Figure B1} -- \textit{continued}}
\end{figure*}

\begin{figure*}
    \centering
    \includegraphics[scale=.47]{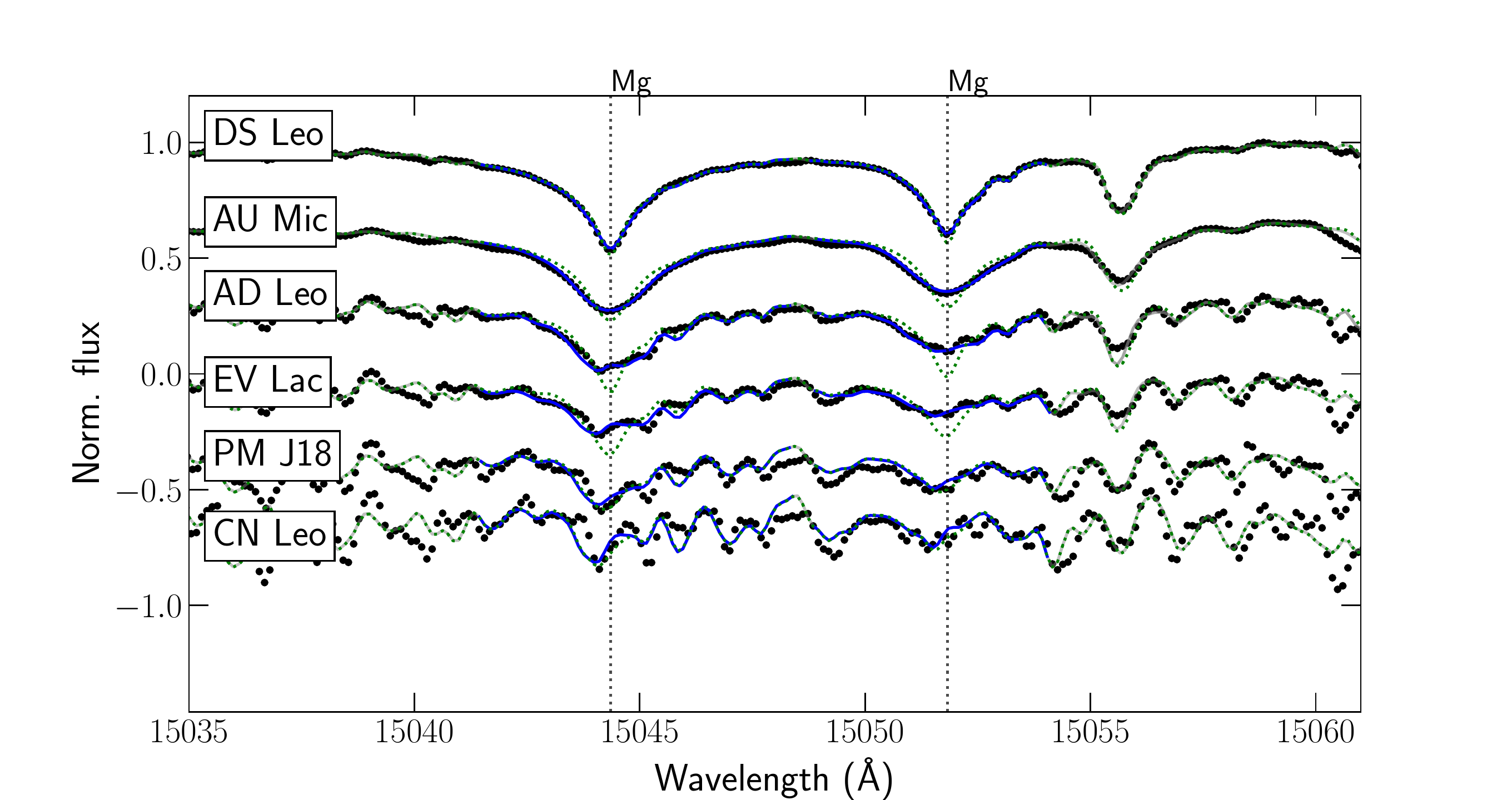}
    \caption*{\textbf{Figure B1} -- \textit{continued}}
\end{figure*}

\begin{figure*}
    \centering
    \includegraphics[scale=.47]{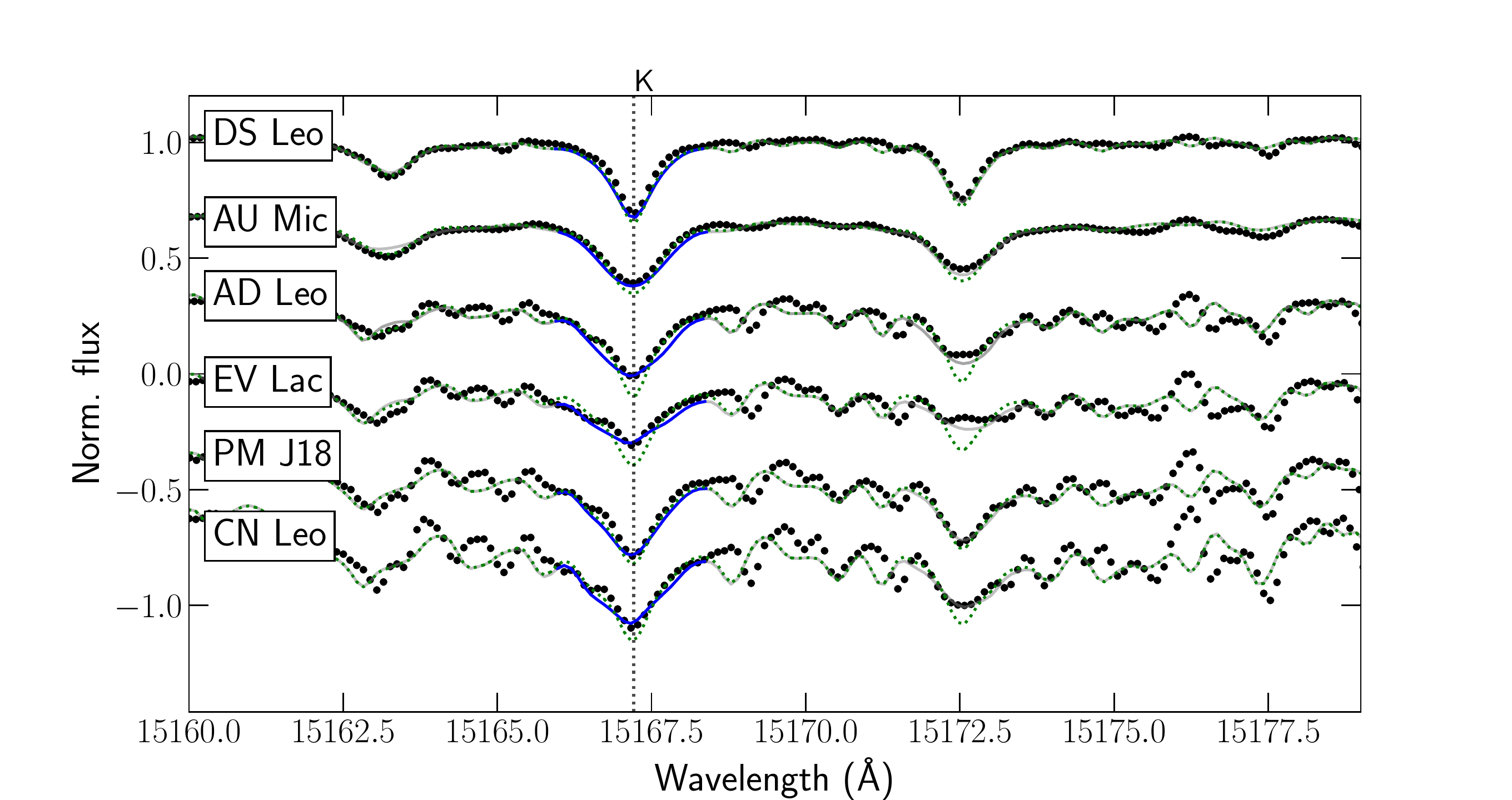}
    \caption*{\textbf{Figure B1} -- \textit{continued}}
\end{figure*}

\begin{figure*}
    \centering
    \includegraphics[scale=.47]{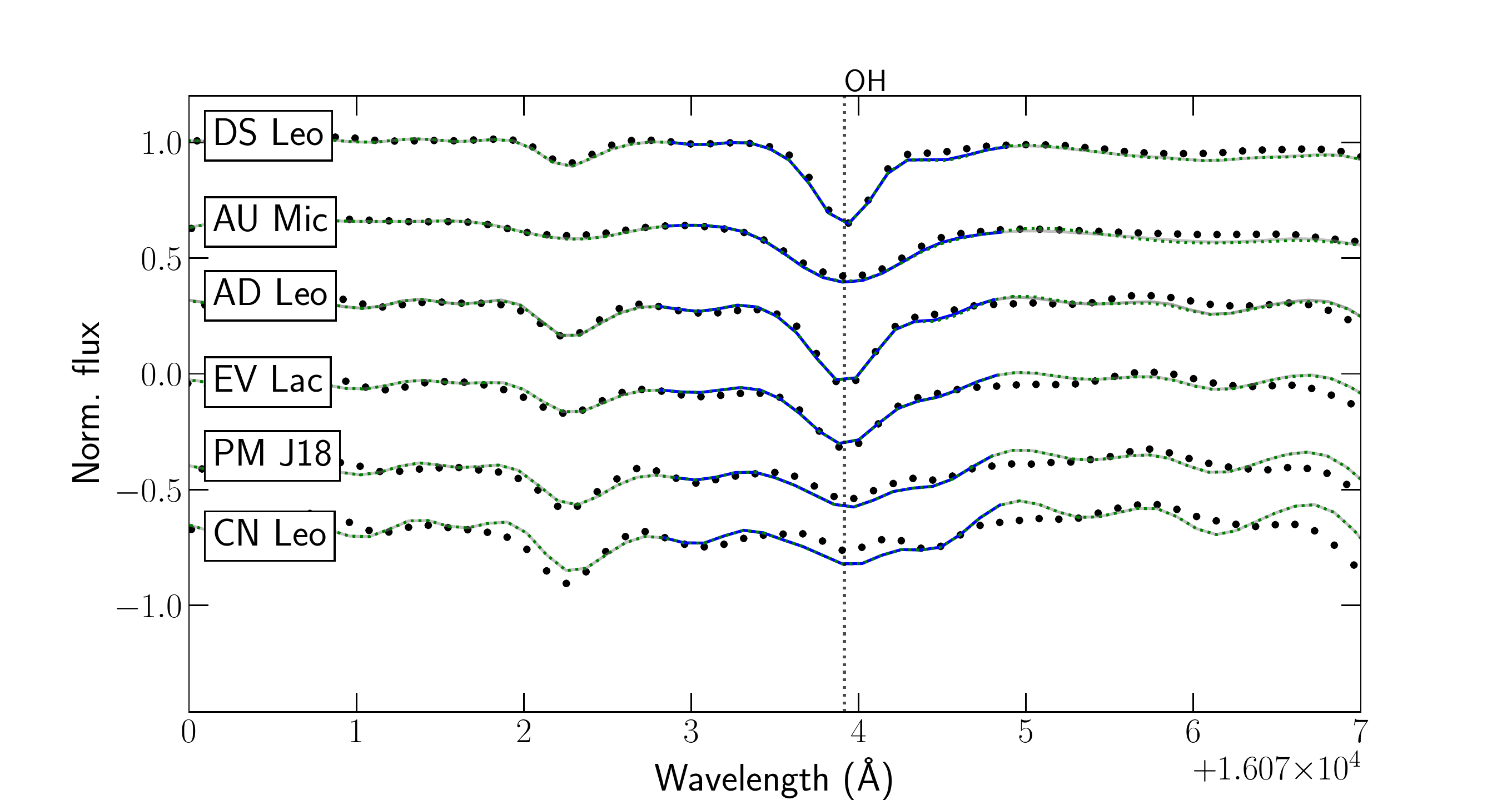}
    \caption*{\textbf{Figure B1} -- \textit{continued}}
\end{figure*}

\begin{figure*}
    \centering
    \includegraphics[scale=.47]{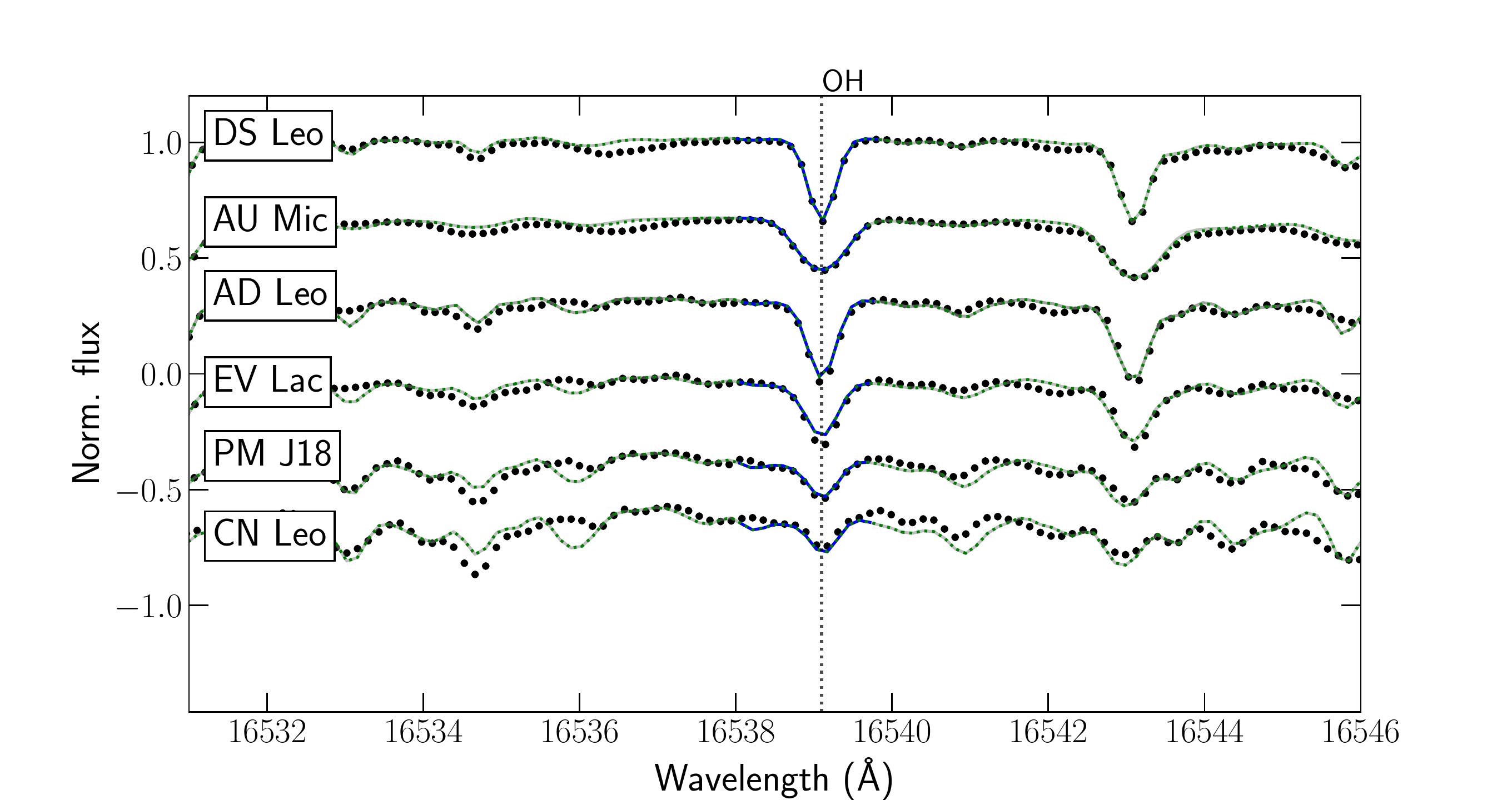}
    \caption*{\textbf{Figure B1} -- \textit{continued}}
\end{figure*}

\begin{figure*}
    \centering
    \includegraphics[scale=.47]{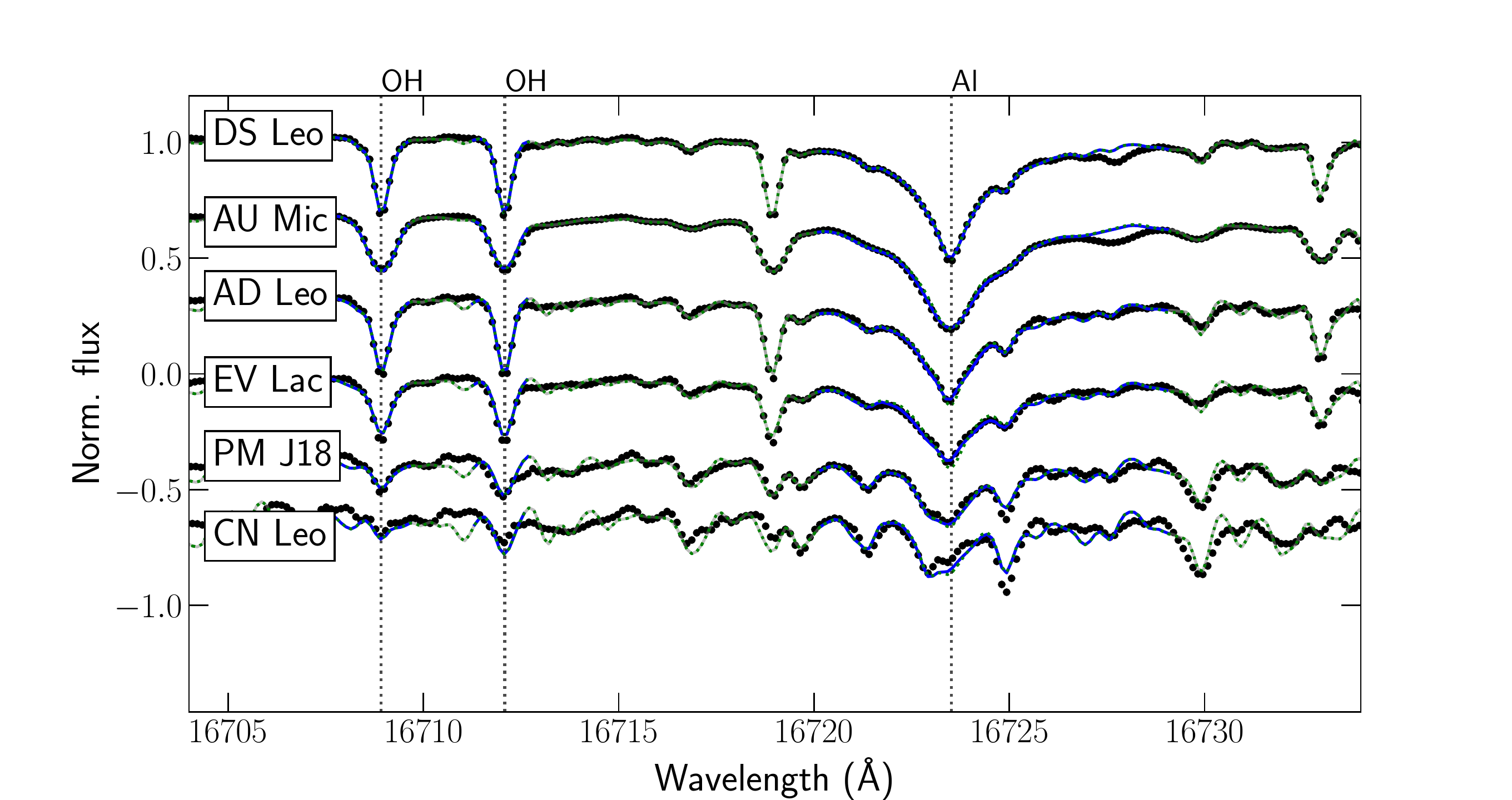}
    \caption*{\textbf{Figure B1} -- \textit{continued}}
\end{figure*}

\begin{figure*}
    \centering
    \includegraphics[scale=.47]{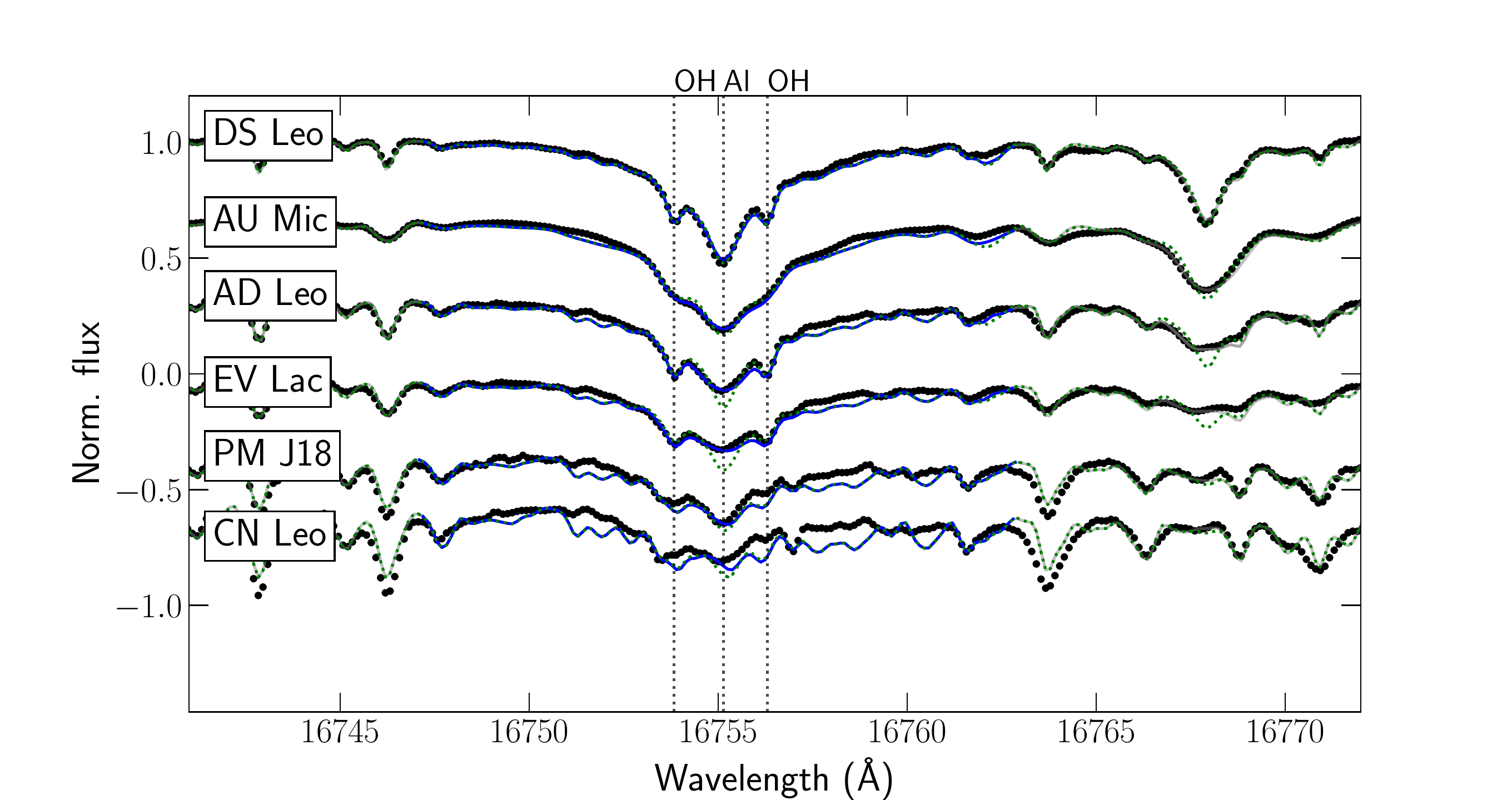}
    \caption*{\textbf{Figure B1} -- \textit{continued}}
\end{figure*}

\begin{figure*}
    \centering
    \includegraphics[scale=.47]{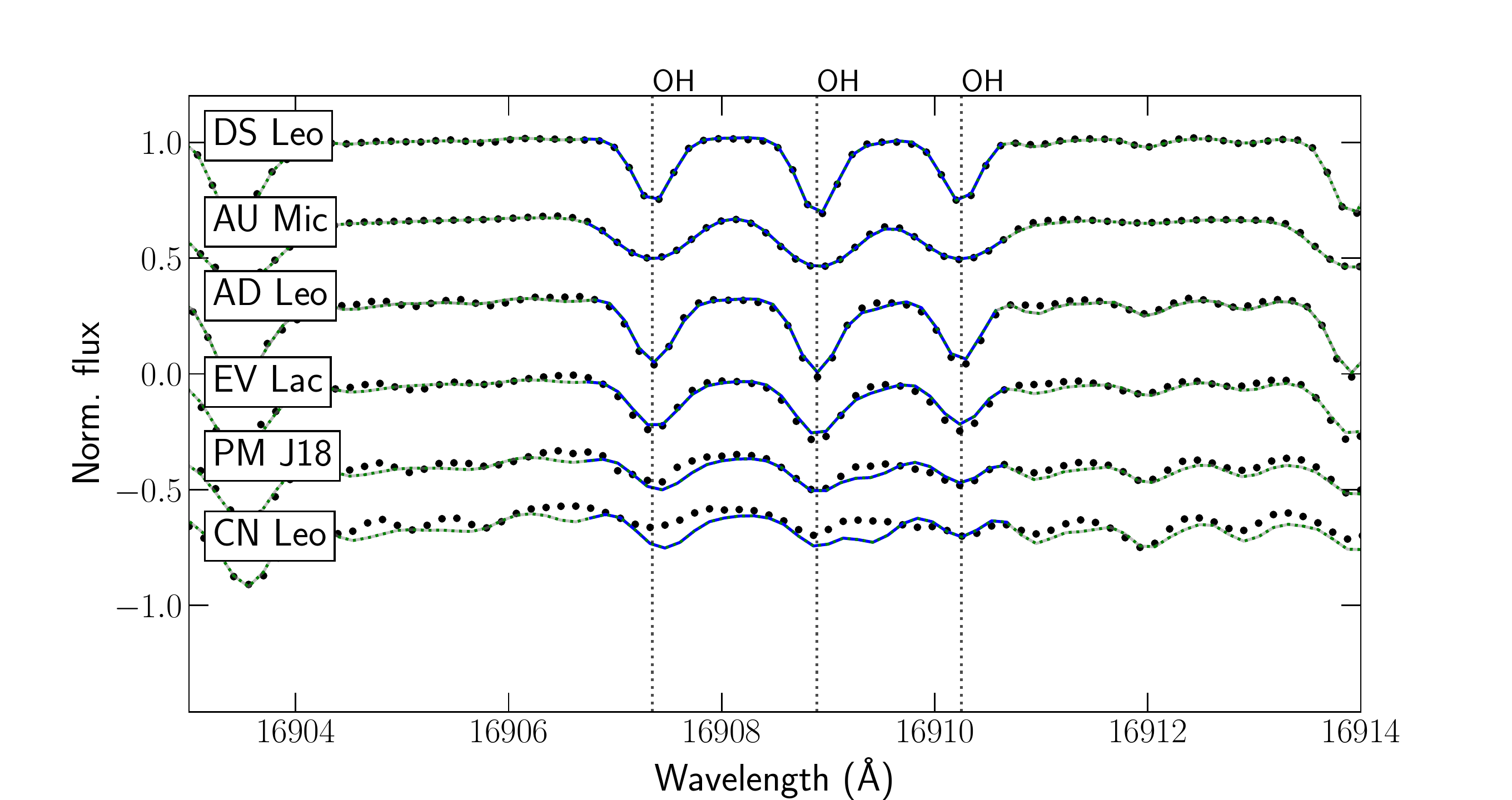}
    \caption*{\textbf{Figure B1} -- \textit{continued}}
\end{figure*}

\begin{figure*}
    \centering
    \includegraphics[scale=.47]{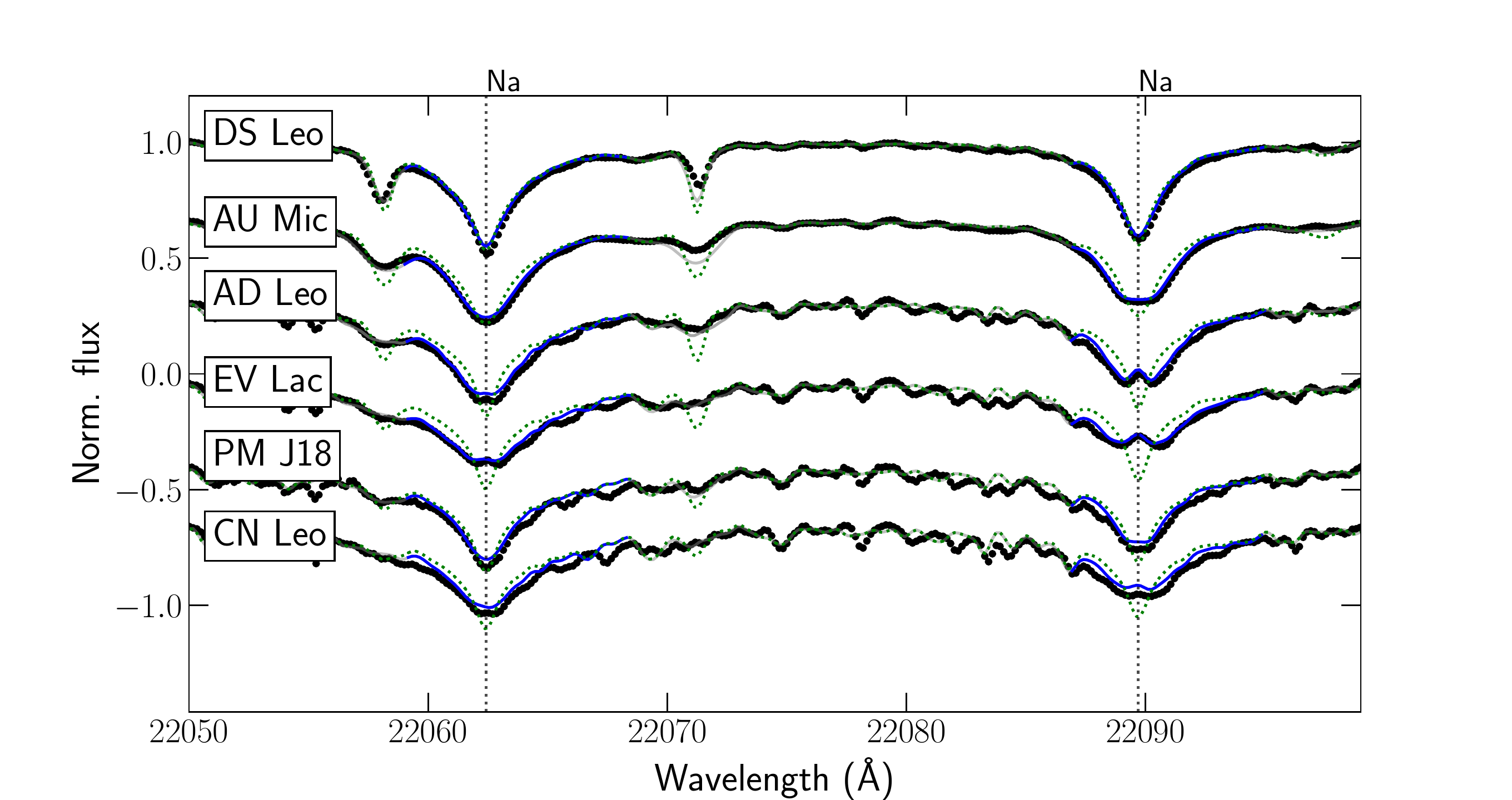}
    \caption*{\textbf{Figure B1} -- \textit{continued}}
\end{figure*}

\begin{figure*}
    \centering
    \includegraphics[scale=.47]{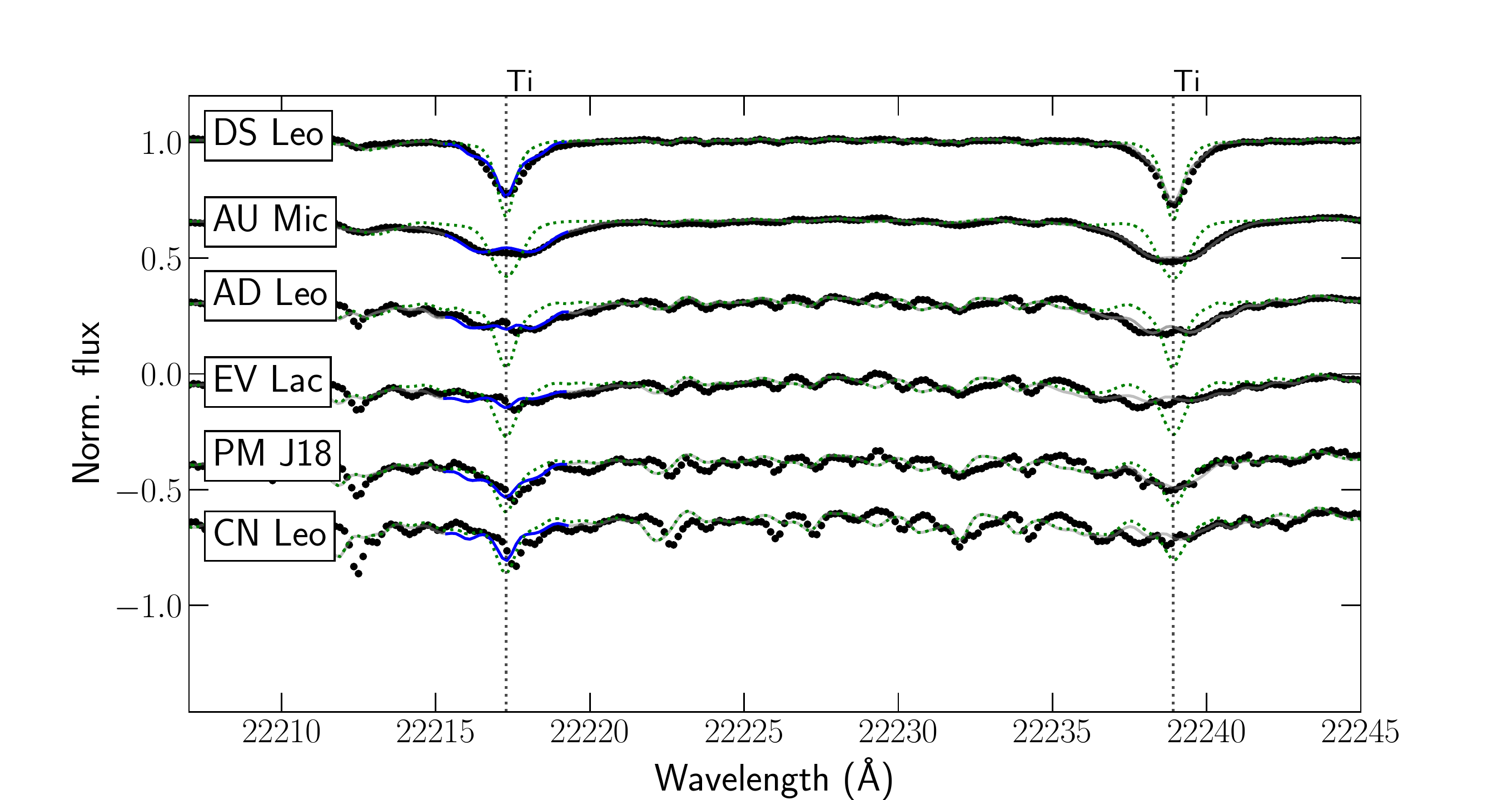}
    \caption*{\textbf{Figure B1} -- \textit{continued}}
\end{figure*}

\begin{figure*}
	\centering
	\includegraphics[scale=.47]{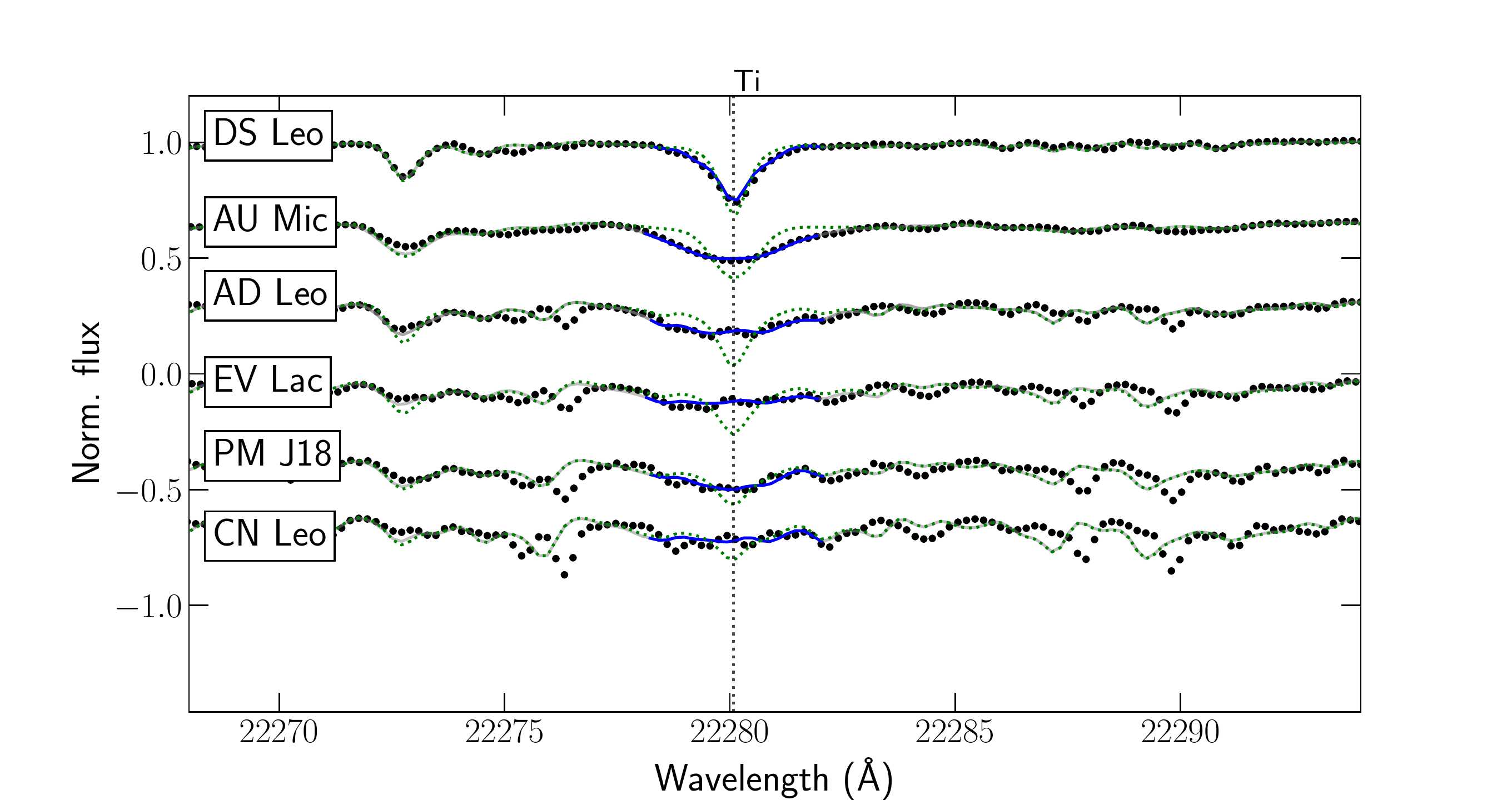}
    \caption*{\textbf{Figure B1} -- \textit{continued}}
\end{figure*}

\begin{figure*}
	\centering
	\includegraphics[scale=.47]{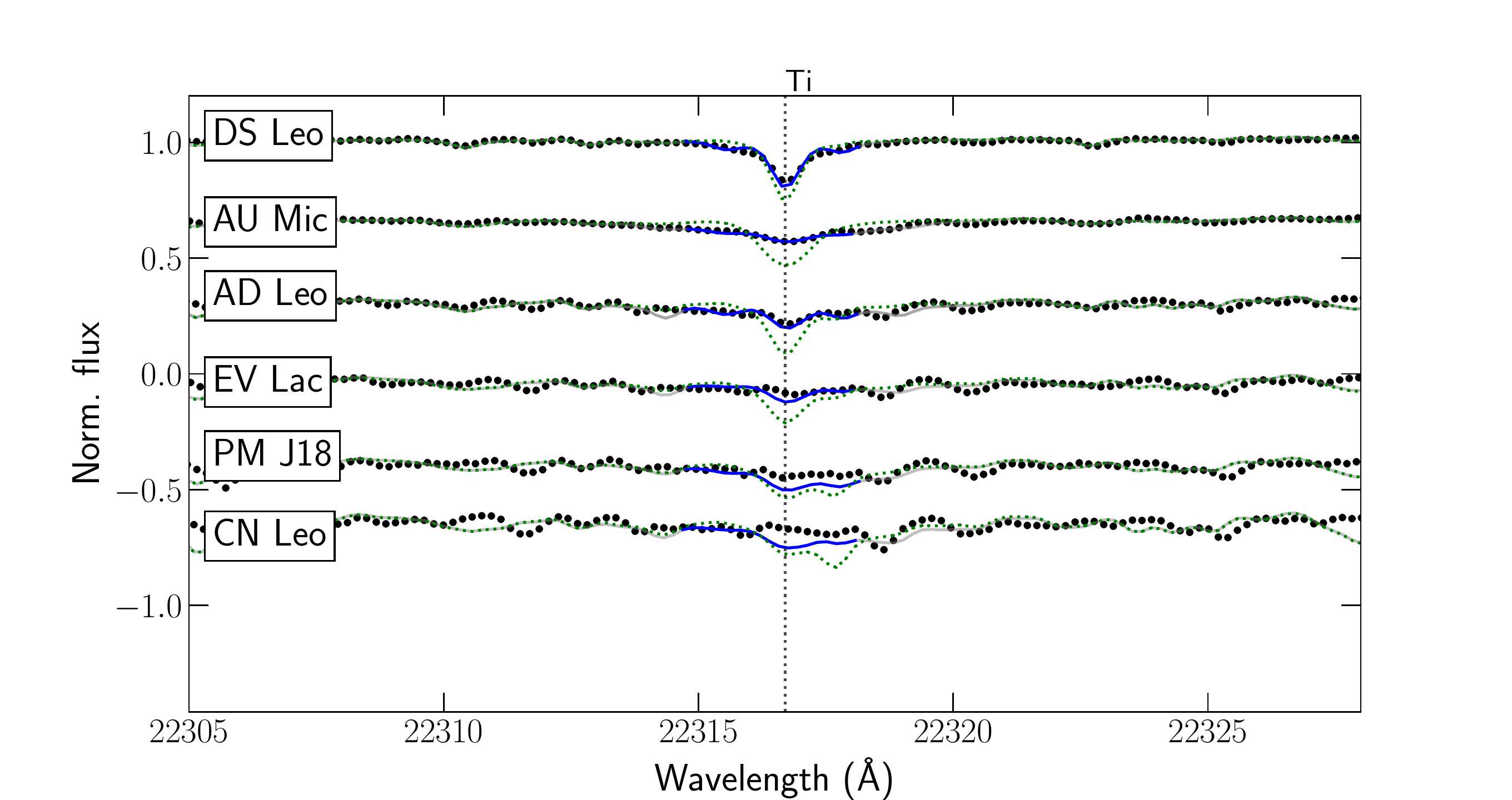}
    \caption*{\textbf{Figure B1} -- \textit{continued}}
\end{figure*}

\begin{figure*}
	\centering
	\includegraphics[scale=.47]{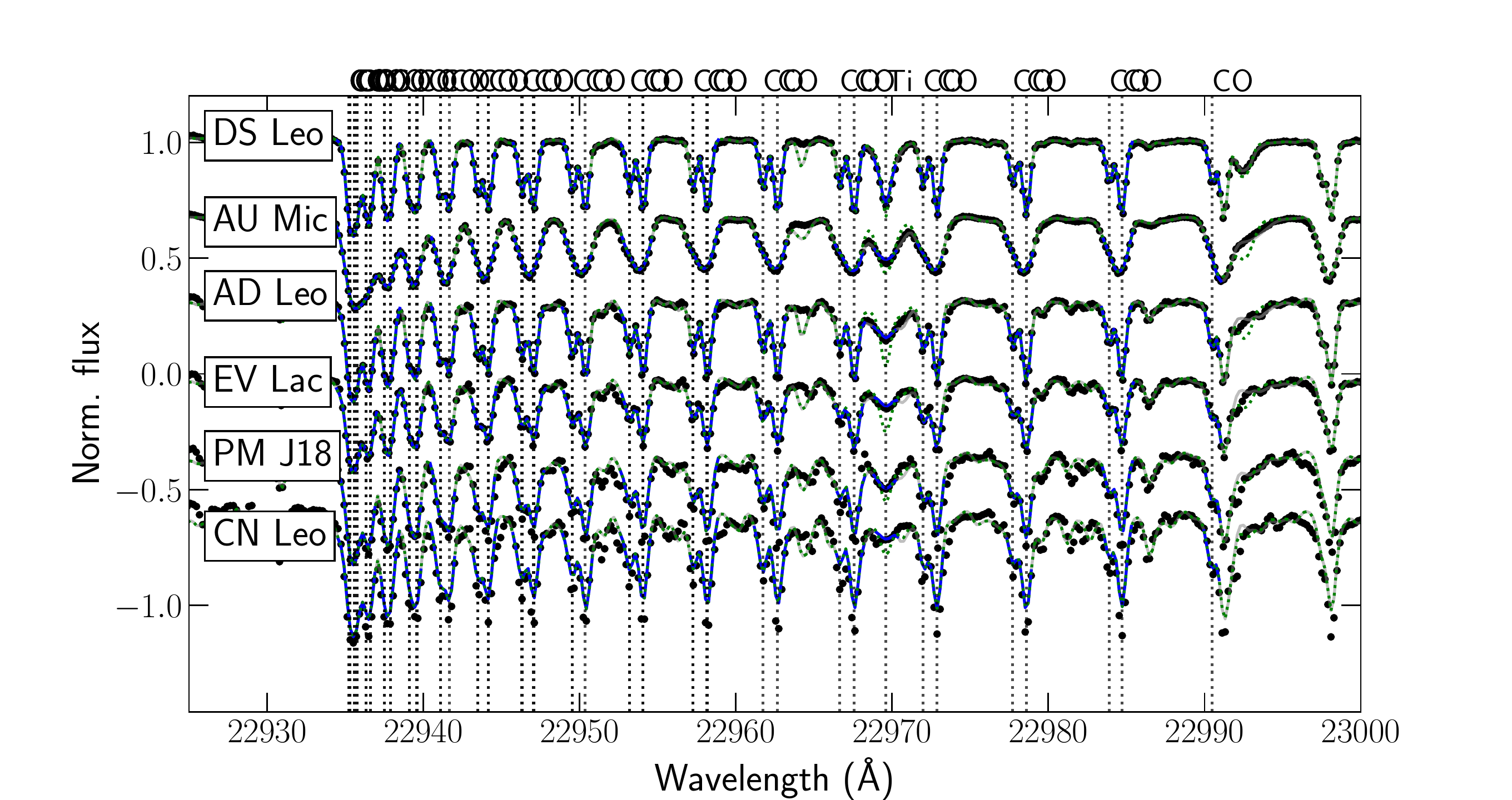}
    \caption*{\textbf{Figure B1} -- \textit{continued}}
\end{figure*}

\section*{APPENDIX C: CORNER PLOTS}
Figure~C1 presents the corner plots obtained for AU Mic, AD Leo, EV Lac, CN Leo, PM~J18482+0741 and DS Leo.

\begin{figure*}
    \centering
    \includegraphics[scale=.27]{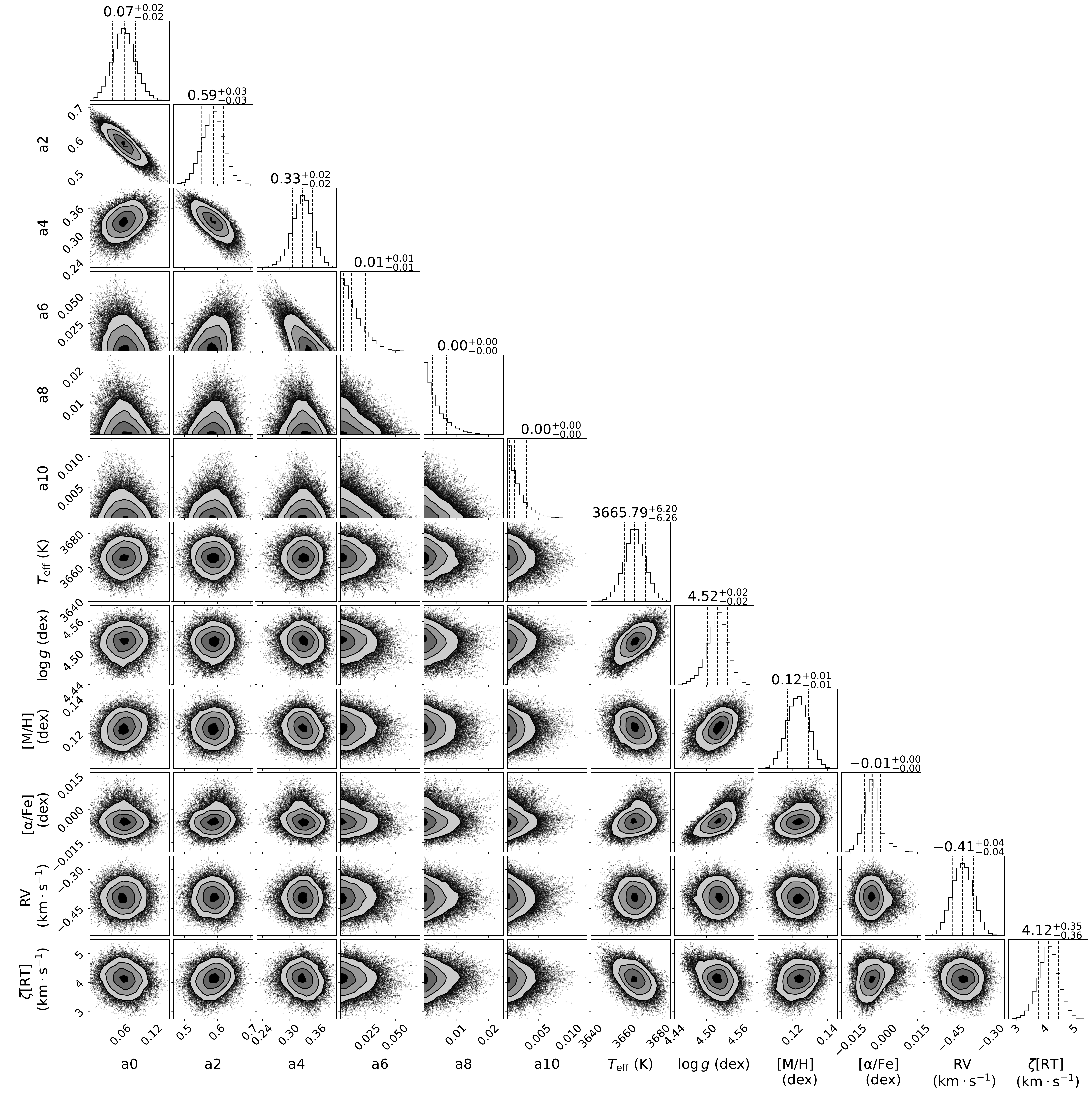}
    \caption*{\textbf{Figure C1.} Corner plot presenting the posterior distribution for filling factors and atmospheric parameters obtained for AU~Mic.}
    \label{fig:example_lines_annex}
\end{figure*}

\begin{figure*}
    \centering
    \includegraphics[scale=.27]{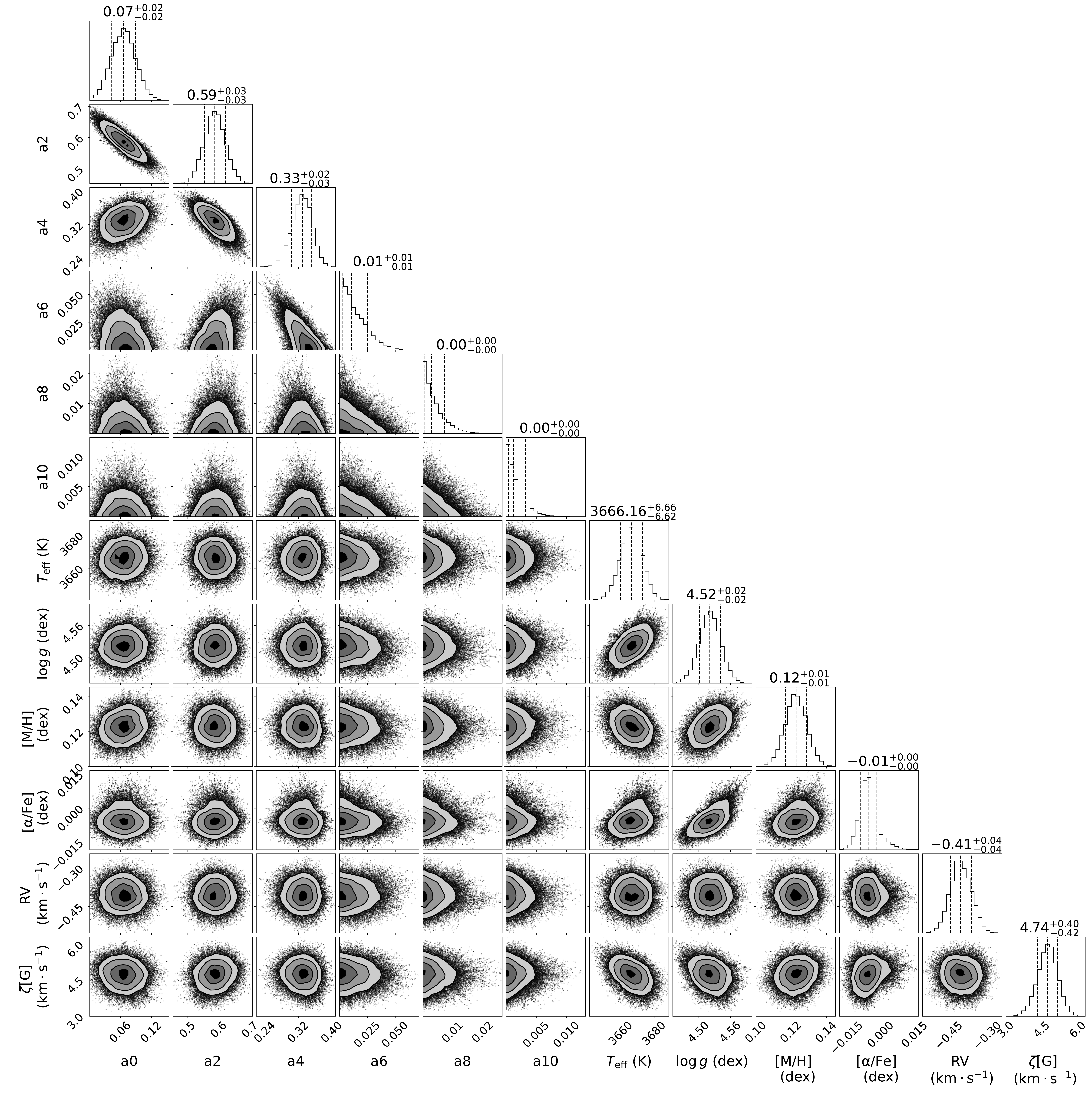}
    \caption*{\textbf{Figure C2. } Same as Fig.~C1 but with a Gaussian macroturbulence model.}
\end{figure*}



\begin{figure*}
    \centering
    \includegraphics[scale=.27]{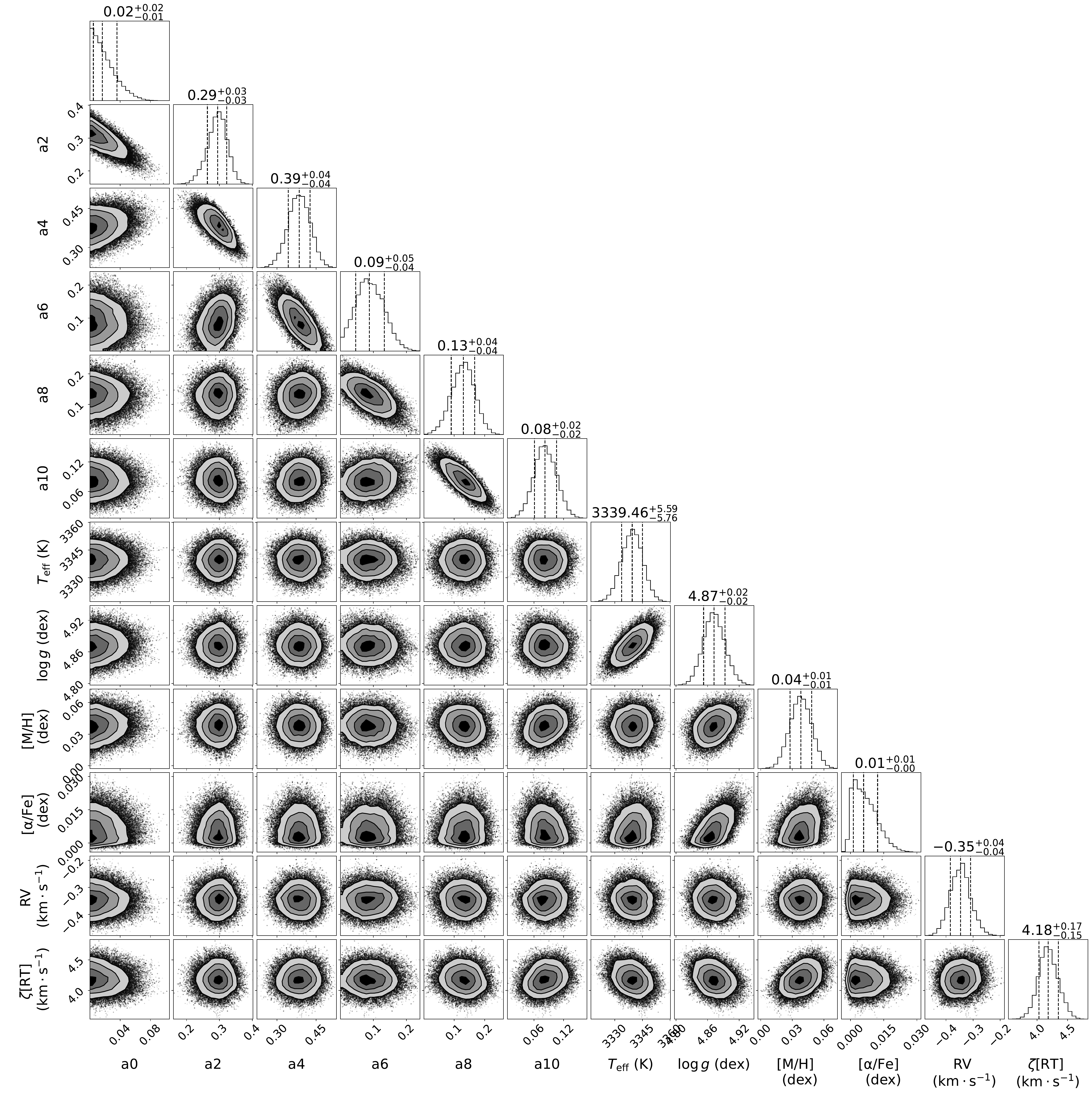}
    \caption*{\textbf{Figure C5. } Same as Fig.~C1 but for EV~Lac.}
\end{figure*}

\begin{figure*}
    \centering
    \includegraphics[scale=.27]{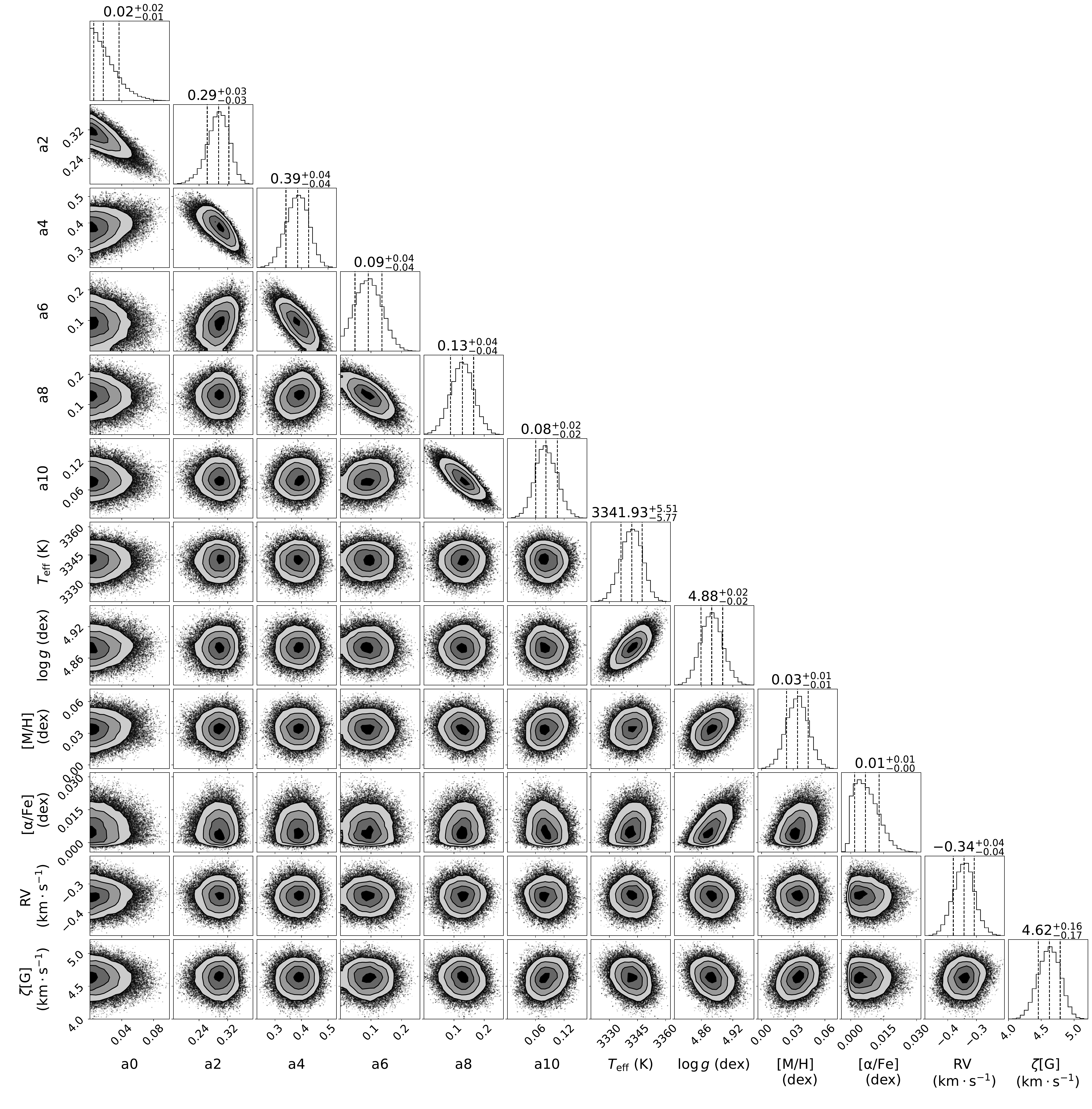}
    \caption*{\textbf{Figure C6. } Same as Fig.~C5  but with a Gaussian macroturbulence model.}
\end{figure*}

\begin{figure*}
    \centering
    \includegraphics[scale=.27]{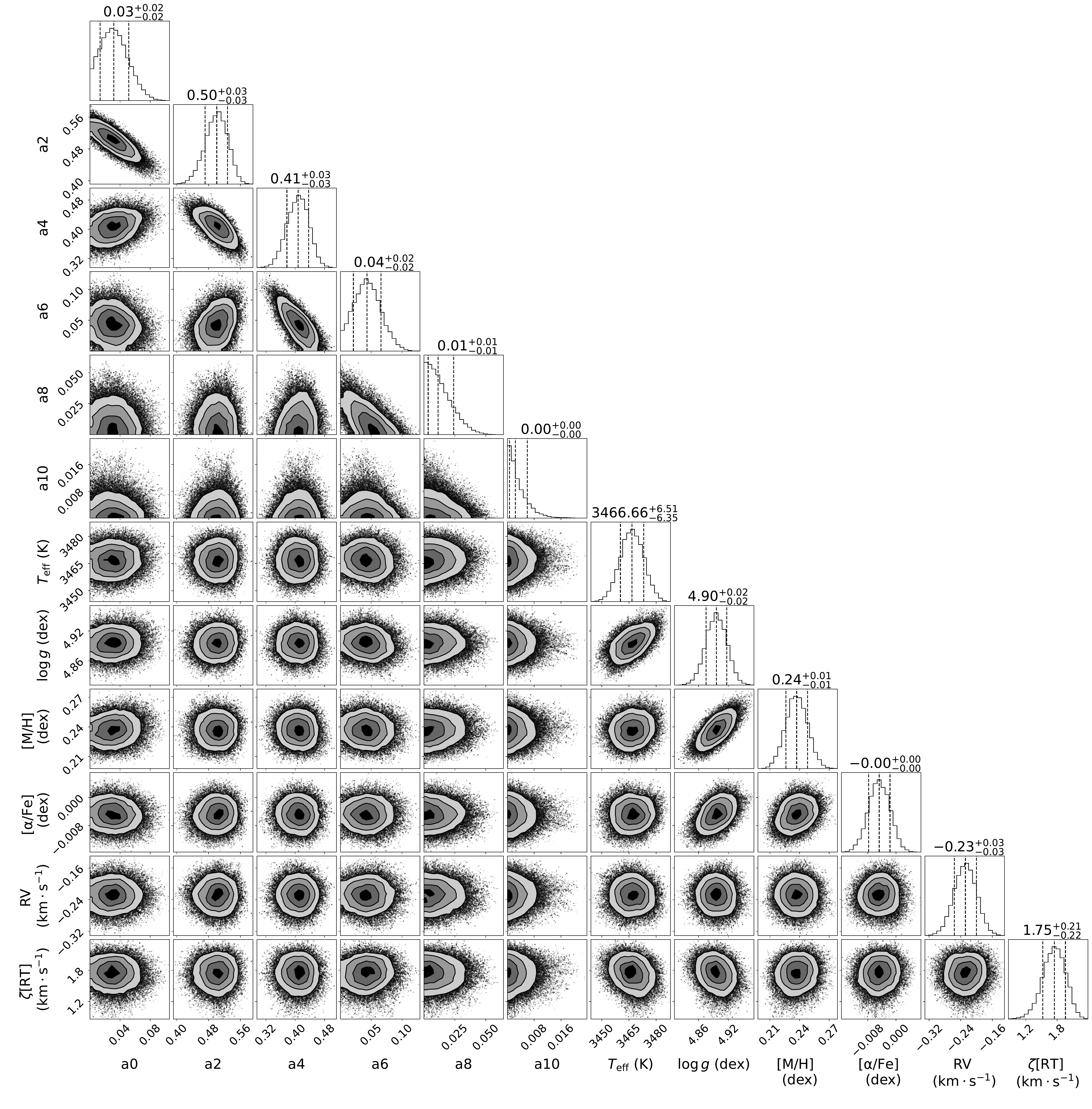}
    \caption*{\textbf{Figure C7. } Same as Fig.~C1 but for AD~Leo.}
\end{figure*}

\begin{figure*}
    \centering
    \includegraphics[scale=.27]{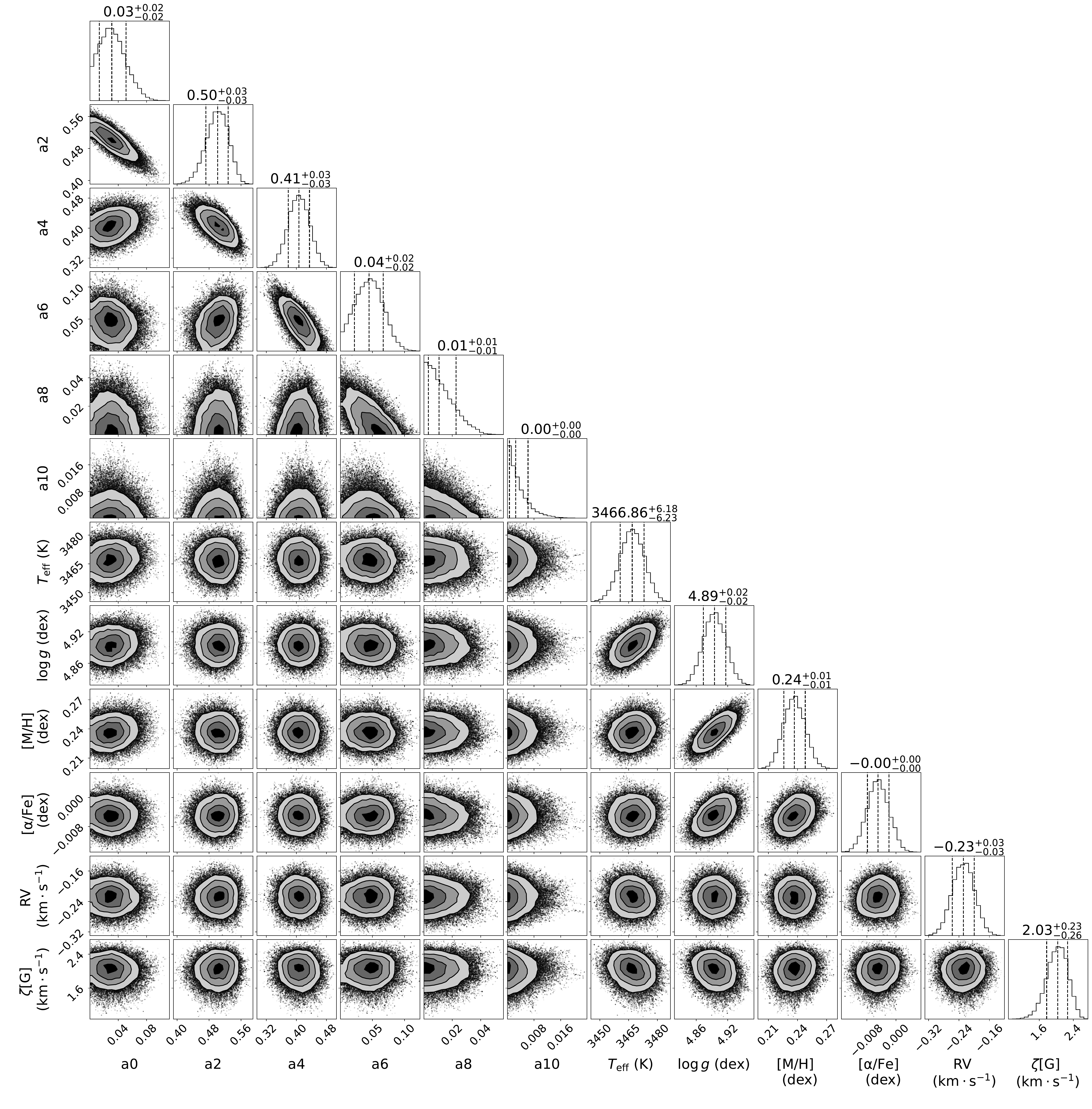}
    \caption*{\textbf{Figure C8. } Same as Fig.~C7  but with a Gaussian macroturbulence model.}
\end{figure*}

\begin{figure*}
    \centering
    \includegraphics[scale=.27]{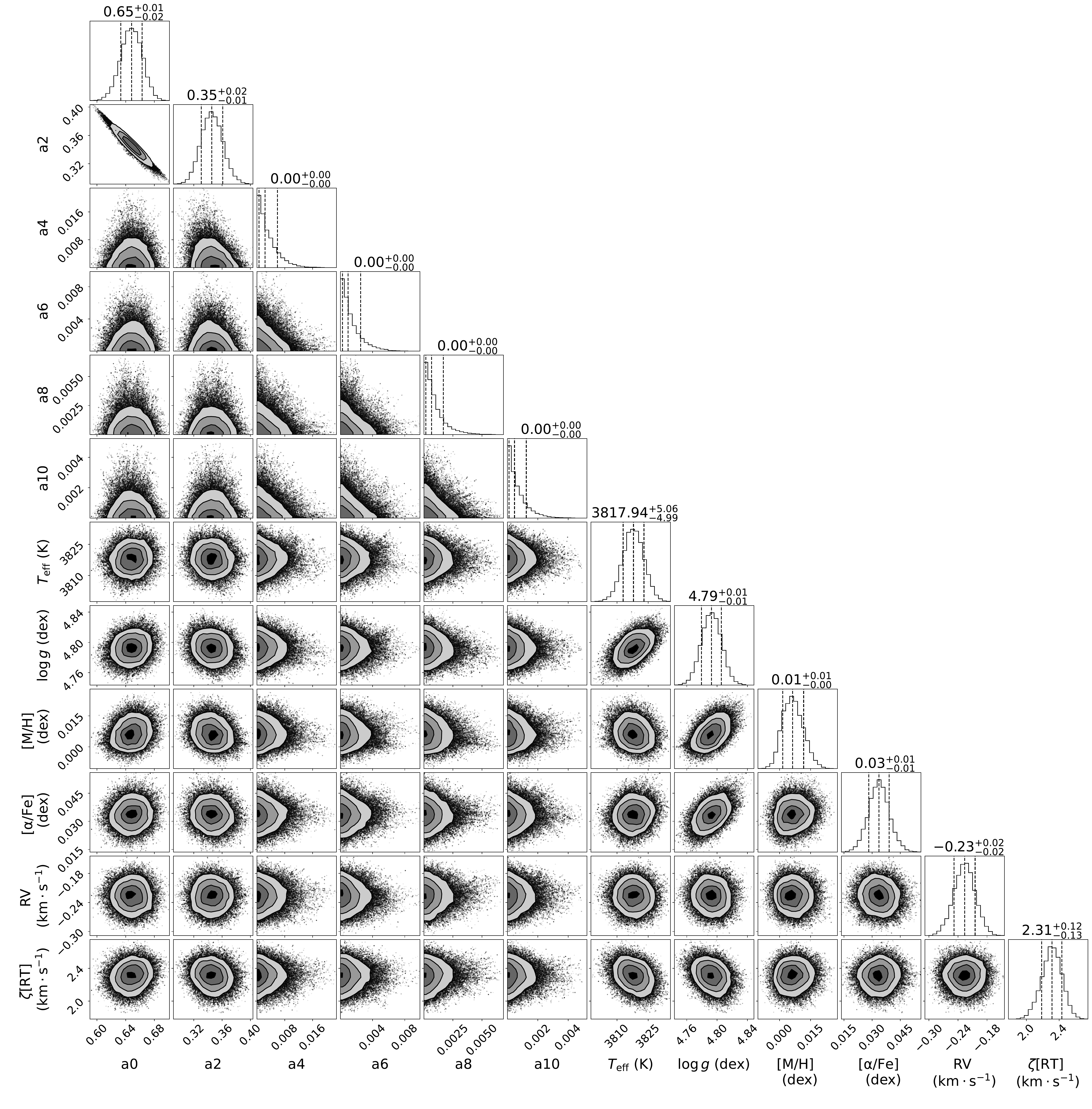}
    \caption*{\textbf{Figure C9. } Same as Fig.~C1 but for DS~Leo.}
\end{figure*}

\begin{figure*}
    \centering
    \includegraphics[scale=.27]{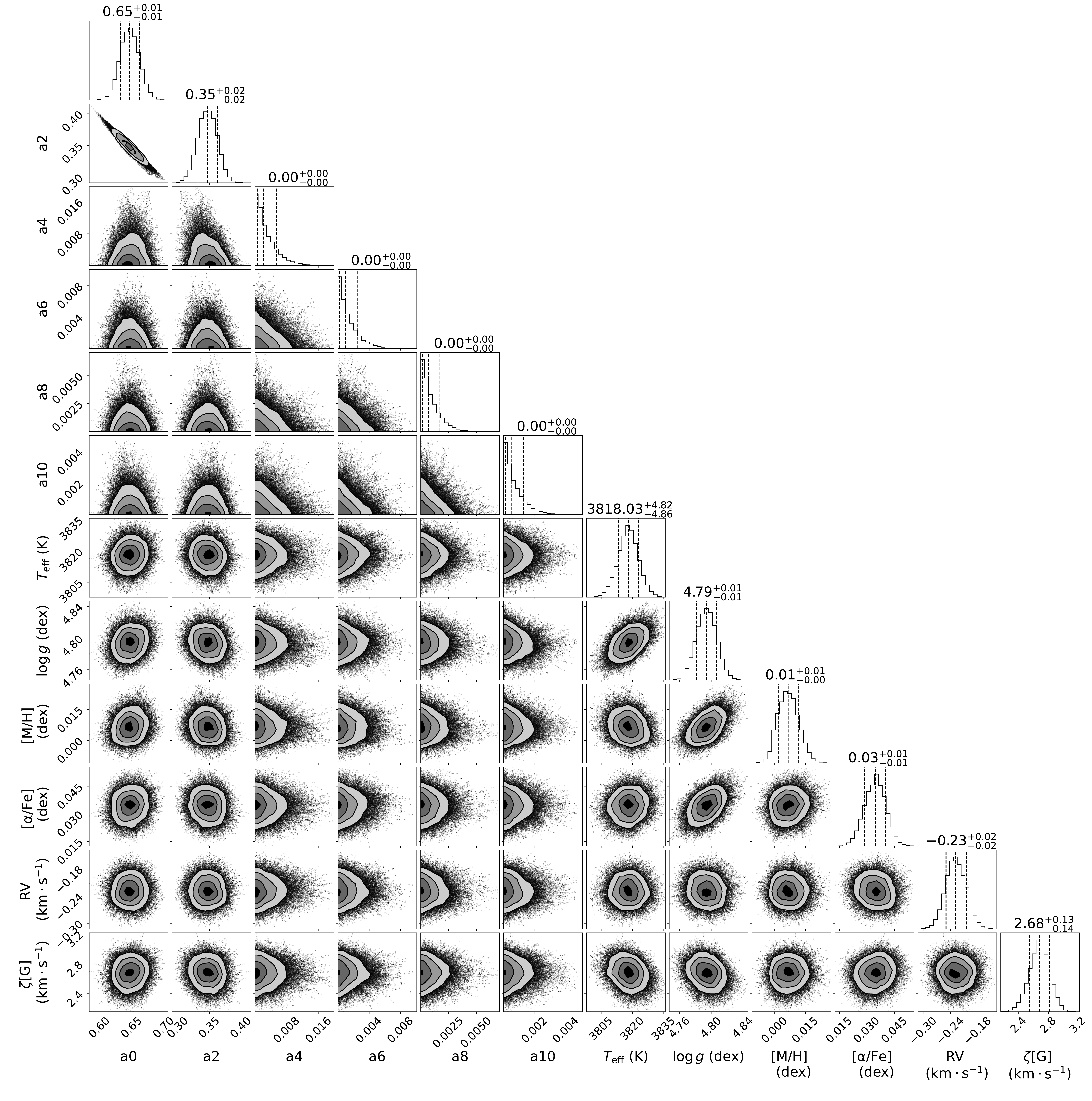}
    \caption*{\textbf{Figure C10. } Same as Fig.~C9  but with a Gaussian macroturbulence model.}
\end{figure*}

\begin{figure*}
    \centering
    \includegraphics[scale=.27]{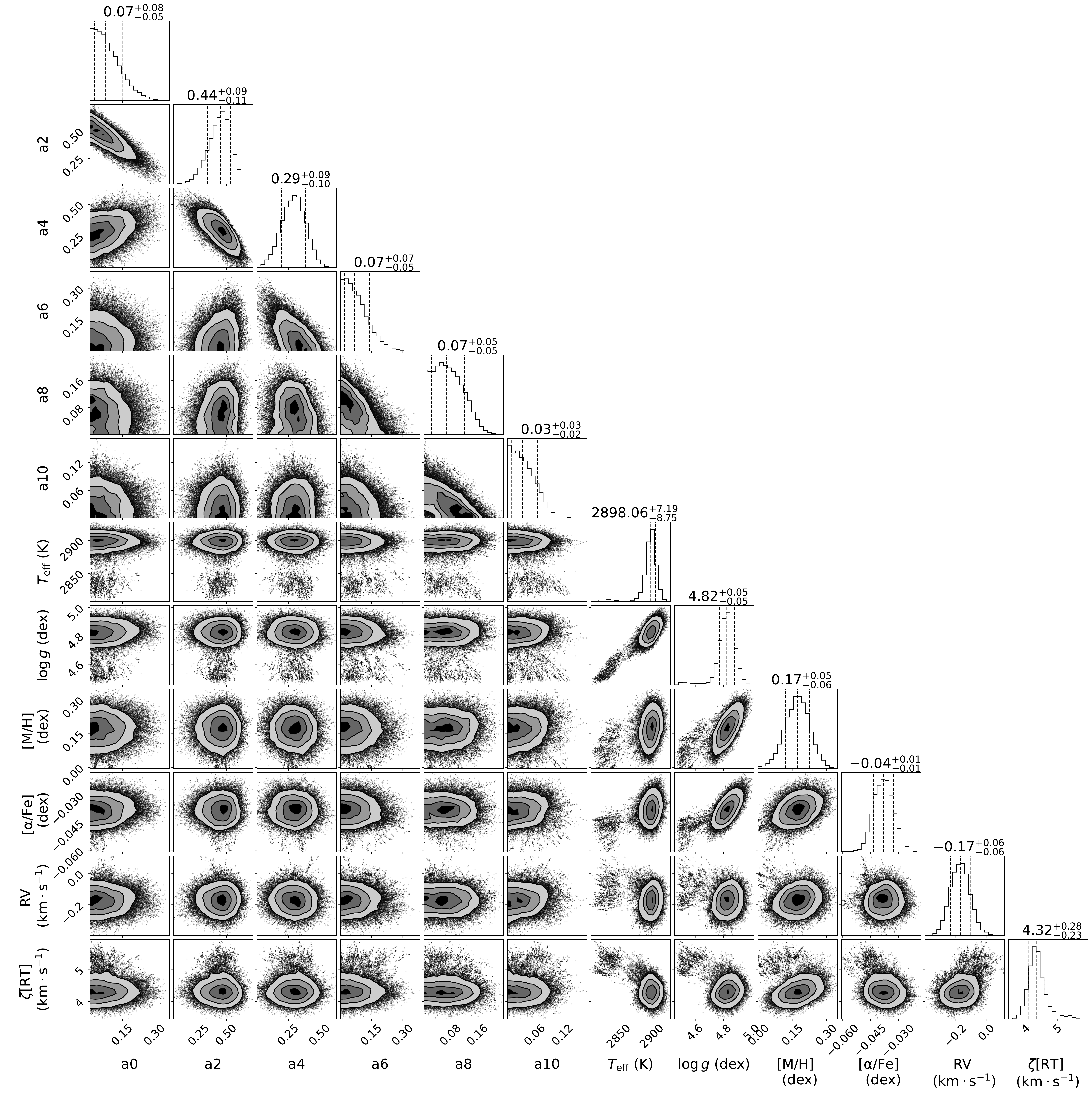}
    \caption*{\textbf{Figure C11. } Same as Fig.~C1 but for CN~Leo.}
\end{figure*}

\begin{figure*}
    \centering
    \includegraphics[scale=.27]{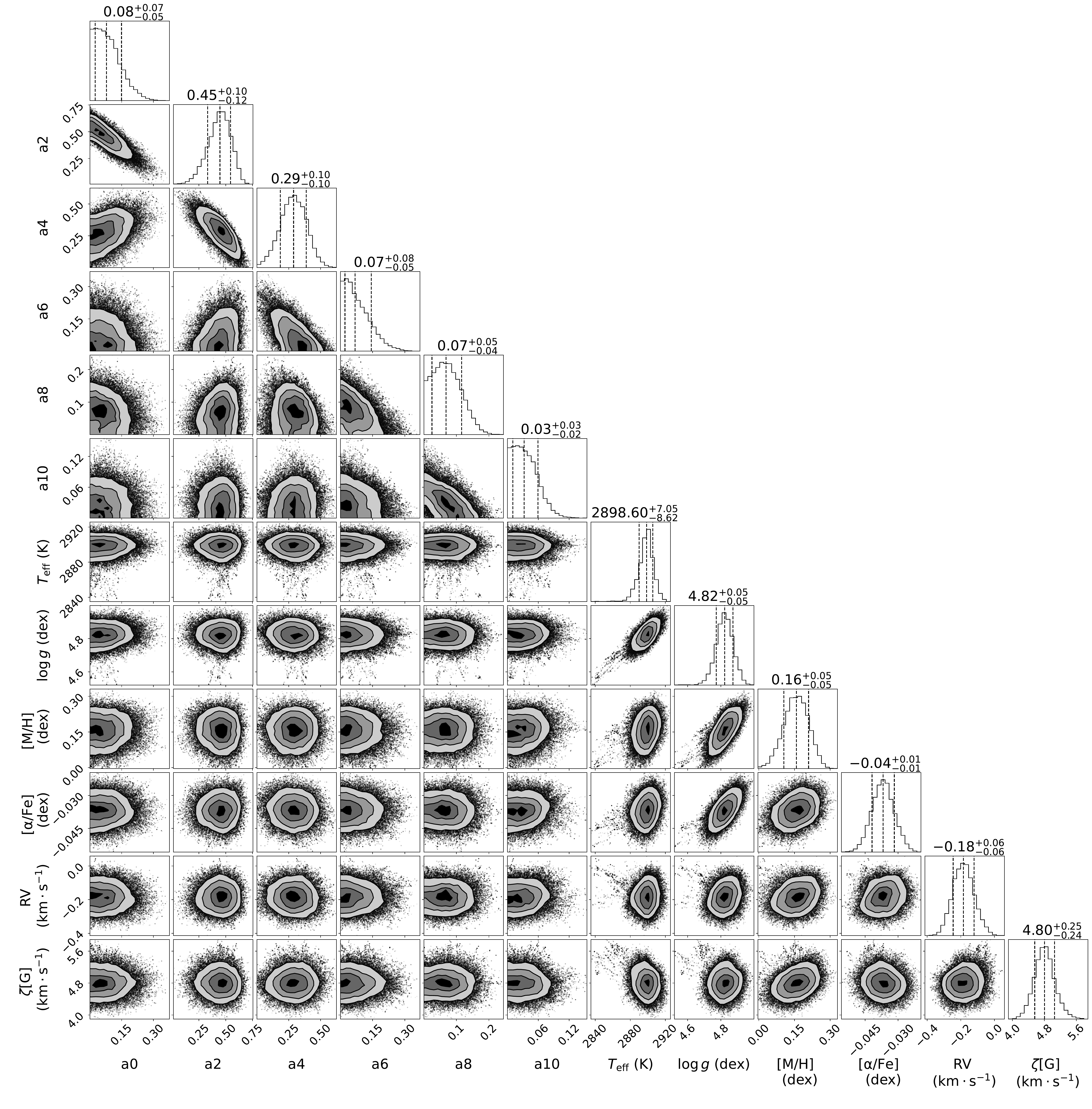}
    \caption*{\textbf{Figure C12. } Same as Fig.~C11  but with a Gaussian macroturbulence model.}
\end{figure*}

\begin{figure*}
    \centering
    \includegraphics[scale=.27]{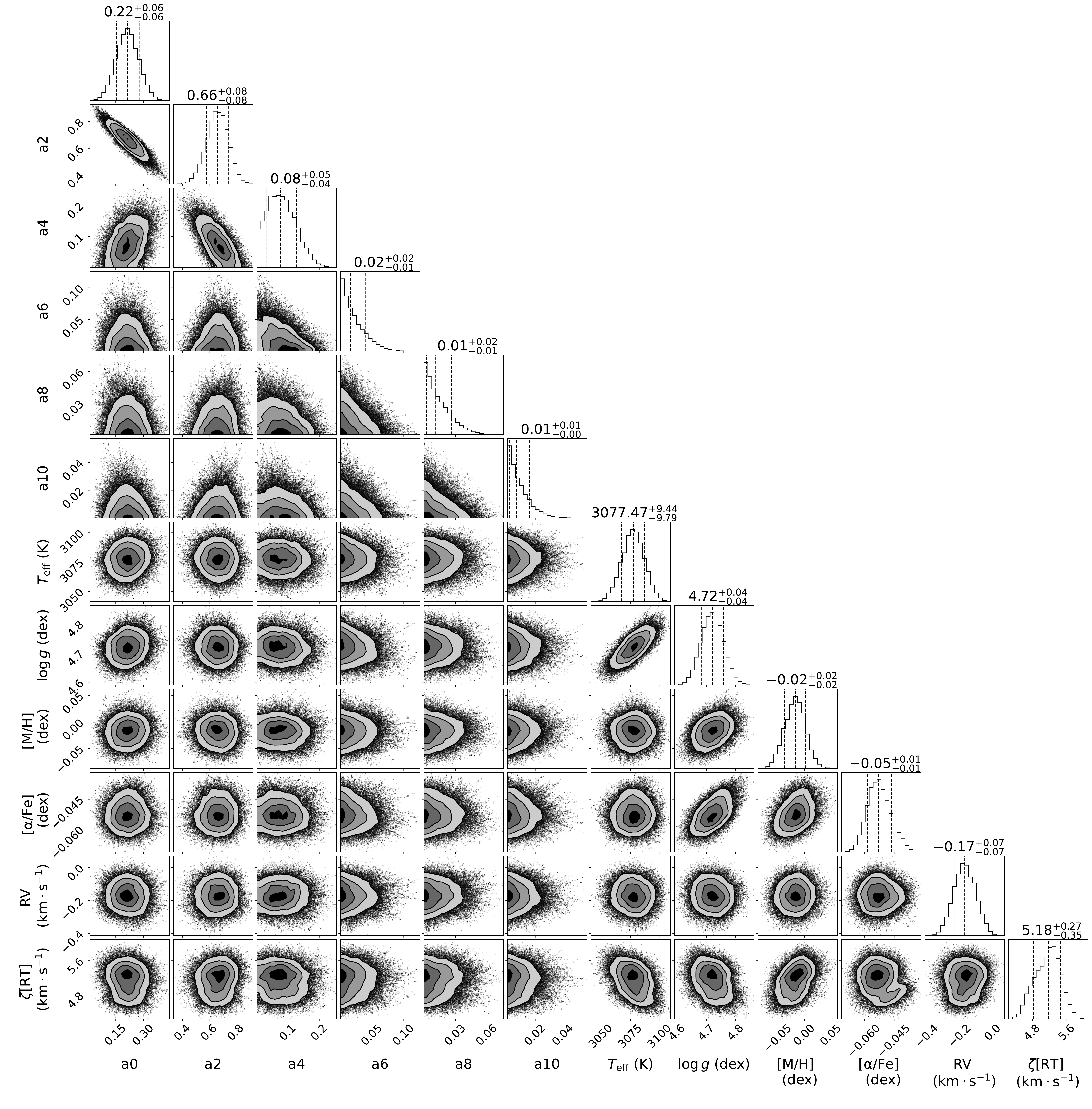}
    \caption*{\textbf{Figure C13. } Same as Fig.~C1 but for PM~J18482+0741.}
\end{figure*}

\begin{figure*}
    \centering
    \includegraphics[scale=.27]{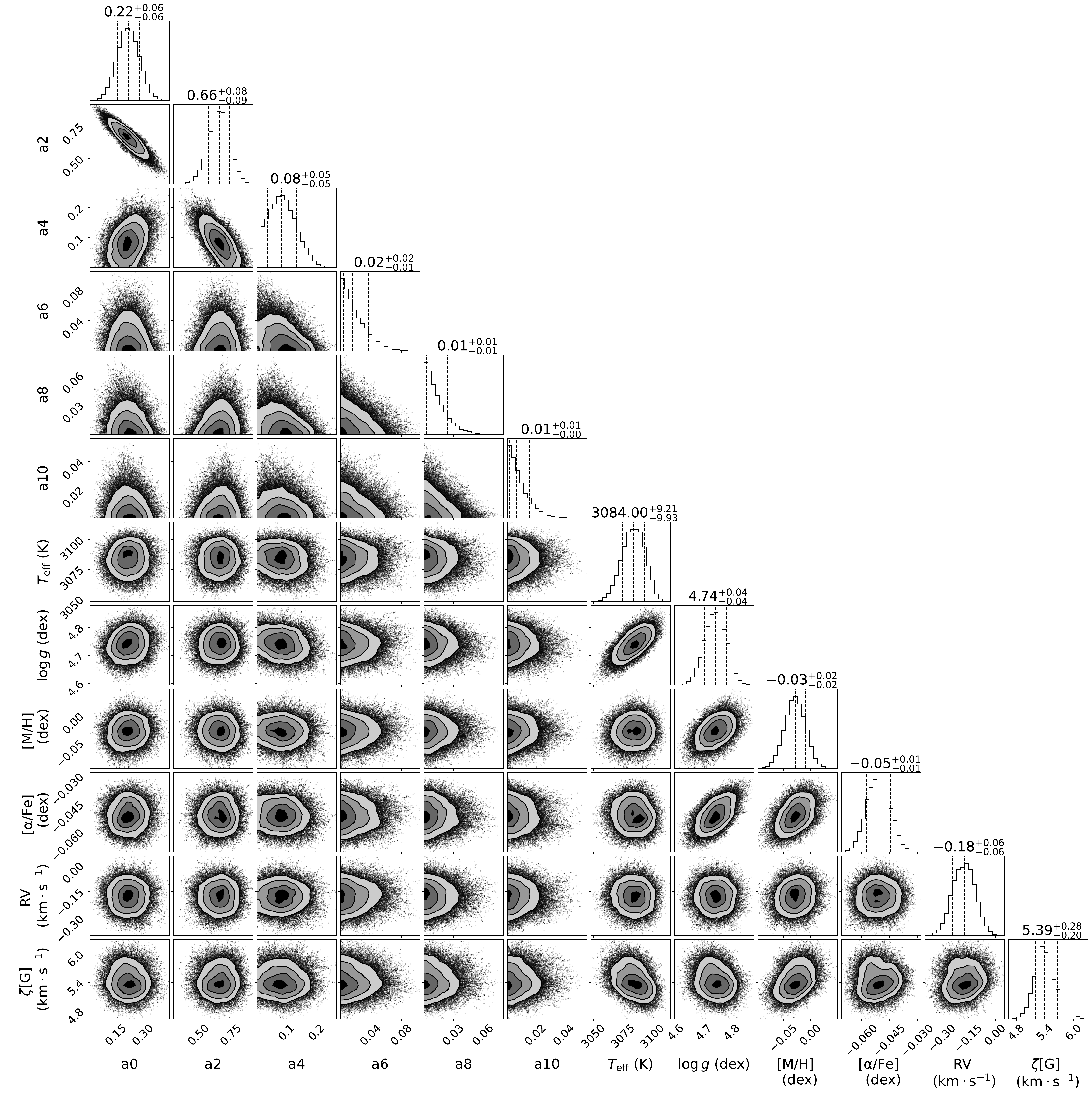}
    \caption*{\textbf{Figure C14. } Same as Fig.~C13  but with a Gaussian macroturbulence model.}
\end{figure*}